\documentclass[11pt]{article}

\usepackage[latin1]{inputenc}

\usepackage{amstext,amsmath,amssymb,amsfonts,bbm,amsmath}
\usepackage{extarrows}
\usepackage{xcolor} 
\usepackage{hyperref}
\hypersetup{
    colorlinks=false,
    menubordercolor=red,
    linkbordercolor=blue
}
\usepackage{authblk}
\usepackage{caption} 
\usepackage{epsfig} 
\usepackage{color} 
\usepackage{graphicx} 
\usepackage{braket} 
\usepackage{slashed} 
\usepackage{dsfont} 
\usepackage{amsthm}  

\usepackage{cite}

\allowdisplaybreaks[4]

\if 0
\usepackage{psfrag}
\usepackage{tikz} 
\usetikzlibrary{arrows}
\usetikzlibrary{plotmarks}
\usetikzlibrary{external}
\usetikzlibrary{patterns}
\usepackage{pgfplots}
\pgfplotsset{compat=1.9}
\usetikzlibrary{patterns}
\usetikzlibrary{calc}
\usepackage[compat=1.1.0]{tikz-feynman}
\usetikzlibrary{automata,positioning}
\fi

\usepackage{float}  
\usepackage{subfig}
\usepackage[lmargin=50pt,rmargin=60pt,tmargin=60pt,bmargin=65pt]{geometry}

\numberwithin{equation}{section}

\newcommand{\be}{\begin{equation}}
\newcommand{\ee}{\end{equation}} 
\newcommand{\Tr}{{\rm Tr}}
 
\DeclareMathOperator{\im}{\mathrm{i}}

\newcommand{\mba}{\mathbf{a}}
\newcommand{\mbb}{\mathbf{b}}
\newcommand{\mbc}{\mathbf{c}}
\newcommand{\mbd}{\mathbf{d}}

\newcommand{\cG}{\mathcal{G}}

\allowdisplaybreaks[4]

\theoremstyle{remark}

\begin{document}

\title{\bf Conformal Symmetry and Composite Operators \\ in the $O(N)^3$ Tensor Field Theory}

\author[1]{Dario Benedetti}
\author[1,2]{Razvan Gurau}
\author[1]{Kenta Suzuki}

\affil[1]{\normalsize \it 
 CPHT, CNRS, Ecole Polytechnique, Institut Polytechnique de Paris, Route de Saclay, \authorcr 91128 PALAISEAU, 
 France
 \authorcr
emails: dario.benedetti@polytechnique.edu, rgurau@cpht.polytechnique.fr, %sabine.harribey@polytechnique.edu,
 kenta.suzuki@polytechnique.edu
 \authorcr \hfill }

\affil[2]{\normalsize\it 
Perimeter Institute for Theoretical Physics, 31 Caroline St. N, N2L 2Y5, Waterloo, ON,
Canada
 \authorcr \hfill}

\date{}

\maketitle

\hrule\bigskip

\begin{abstract}
We continue the study of the bosonic $O(N)^3$ model with quartic interactions and long-range propagator.
The symmetry group allows for three distinct invariant $\phi^4$ composite operators, known as tetrahedron, pillow and double-trace. 
As shown in  \cite{Benedetti:2019eyl, Benedetti:2019ikb}, the tetrahedron operator is exactly marginal in the large-$N$ limit
and for a purely imaginary tetrahedron coupling a line
of real infrared fixed points (parametrized by the absolute value of the tetrahedron coupling) is found for the other two couplings. These fixed points have real critical exponents and a real spectrum of bilinear operators, satisfying unitarity constraints.
This raises the question whether at large-$N$ the model is unitary, despite the tetrahedron coupling being imaginary.

In this paper, we first rederive the above results by a different regularization and renormalization scheme.
We then discuss the operator mixing for composite operators and we give a perturbative proof of conformal invariance of the model at the infrared fixed points
by adapting a similar proof from the long-range Ising model.
At last, we identify the scaling operators at the fixed point and compute the two- and three-point functions of $\phi^4$ and $\phi^2$ composite operators. The correlations have the expected conformal behavior and the OPE coefficients are all real, reinforcing the claim that the large-$N$ CFT is unitary.

\end{abstract}

\hrule\bigskip

%\newpage
\tableofcontents

%%%%%%%%%%%%%%%%%%%%%%%%%%%%%%%%%%%%%%%%%%%%%%%%%%%%%%%%
%%%%%%%%%%%%%%%%%%%%%%%%%%%%%%%%%%%%%%%%%%%%%%%%%%%%%%%%
\section{Introduction}
\label{sec:introduction}
%%%%%%%%%%%%%%%%%%%%%%%%%%%%%%%%%%%%%%%%%%%%%%%%%%%%%%%%
%%%%%%%%%%%%%%%%%%%%%%%%%%%%%%%%%%%%%%%%%%%%%%%%%%%%%%%%

Finding and studying non-supersymmetric interacting conformal field theories (CFT) in $d>2$ dimensions is a very challenging task with a long history.
A young and successful approach is the conformal bootstrap \cite{Poland:2018epd}, working completely within the framework of CFT and seeking to identify CFT's by imposing consistency conditions. 
The most standard and historical approach is however based on the renormalization group: here the challenge is to find non-trivial (interacting) fixed points. They correspond by construction to scale-invariant theories, but very commonly invariance under the full conformal group arises as well \cite{Nakayama:2013is}.
The great hurdle faced by the renormalization group is that our main tool of investigation, perturbation theory, demands that we approach the interacting fixed point by perturbing the free theory. Clearly, the task becomes more daunting the further the two theories are, and therefore it is extremely important from the theoretical point of view to have adjustable parameters that allow one to bring the interacting fixed point closer to the free theory or to tame the perturbative series.
There are two widely exploited parameters of this sort \cite{Wilson:1972cf}: one is the analytically continued spacetime dimension, as typically the interacting fixed point collapses into the non-interacting one at some critical dimension; the other is the number of field components $\cal N$, when their interactions are constrained by a (global or local) symmetry group, as the large-$\cal N$ limit can lead to drastic simplifications of the perturbative expansion.

Tensor models, in which ${\cal N}=N^r$ with $r$ the rank of the tensor, are a recent entry in the menu of field theories admitting an interesting large-$N$ limit. They typically admit a large $N$ limit dominated by \emph{melonic} diagrams \cite{Bonzom:2011zz,RTM,Klebanov:2018fzb,Ferrari:2017jgw,Prakash:2019zia} different from both the large $N$ limit of vector models ($r=1$, dominated by bubble diagrams \cite{Moshe:2003xn}) and the one of matrix models ($r=2$,  dominated by planar diagrams \cite{DiFrancesco:1993nw}).
The melonic dominance translates into a perturbative expansion which is richer than the one of vectors and more manageable than that of matrices. It is therefore interesting to construct models of tensor field theories in $d$ dimensions and look for their fixed points at large $N$\cite{Giombi:2017dtl,Prakash:2017hwq,Benedetti:2017fmp,Giombi:2018qgp,Benedetti:2018ghn,Benedetti:2019eyl,Benedetti:2019ikb,Popov:2019nja}.\footnote{Tensor models were initially studied in zero dimension in the context of quantum gravity and random geometry \cite{Ambjorn:1990ge,Sasakura:1990fs,color,RTM,review,Oriti:2011jm}. They were then studied in one dimension \cite{Witten:2016iux,Gurau:2016lzk,Klebanov:2016xxf,Peng:2016mxj,Krishnan:2016bvg,Krishnan:2017lra,Bulycheva:2017ilt,Choudhury:2017tax,Halmagyi:2017leq,Klebanov:2018nfp,Carrozza:2018psc,Klebanov:2019jup,Ferrari:2019ogc} (see also \cite{Delporte:2018iyf,Klebanov:2018fzb} for reviews) as a generalization of the Sachdev-Ye-Kitaev model \cite{Sachdev:1992fk,Kitaev2015,Maldacena:2016hyu,Polchinski:2016xgd,Jevicki:2016bwu,Gross:2016kjj} without quenched disorder.}
We call the resulting conformal field theories \textit{melonic}.

In this paper we study the bosonic tensor field theory introduced in  \cite{Benedetti:2019eyl}, that is the Carrozza-Tanasa-Klebanov-Tarnopolsky \cite{Carrozza:2015adg,Klebanov:2016xxf} (CTKT) model with long-range propagator. The model has the following features:
\begin{enumerate}
\item The global symmetry group is $O(N)^3$ with fields transforming in the tri-fundamental representation, that is the fields are rank-3 tensors. The
$O(N)^3$  models were introduced in zero dimensions in \cite{Carrozza:2015adg}, and studied in dimension  one and higher in \cite{Klebanov:2016xxf} and later in \cite{Giombi:2017dtl,Giombi:2018qgp}. The existence of a large-$N$ expansion (with melonic dominance at leading order for the case of a tetrahedron interaction) was proved in \cite{Carrozza:2015adg} by adapting the methods of \cite{Bonzom:2011zz,Bonzom:2012hw}.
\item The action contains only quartic interactions, and there are three of them which are allowed by the symmetry; they are known as tetrahedron, pillow, and double-trace invariants. The same set of interactions were considered in Ref.~\cite{Giombi:2017dtl}, but the two models are distinguished by the following two features.
\item The propagator is long-range, $C(p)=1/p^{d/2}$ in $d<4$ dimensions, such that the quartic couplings are dimensionless in any dimension $d$. From this point of view the model can be thought of as a generalization of the long-range Ising model \cite{Fisher:1972zz,Sak:1973}, which has been studied extensively with various methods, including constructive methods  \cite{Brydges:2002wq,Abdesselam:2006qg}, large-$N$ expansion \cite{Brezin:2014}, functional renormalization group \cite{Defenu:2014}, and CFT methods \cite{Paulos:2015jfa,Behan:2017emf}.
\item The tetrahedron coupling is purely imaginary. As the tetrahedron invariant is unbounded (from above and below) the choice of imaginary coupling is reminiscent of the Lee-Yang model with an $\im \lambda \phi^3$ interaction \cite{Fisher:1978pf,Cardy:1985yy}.
\end{enumerate}

The main result of Ref.~\cite{Benedetti:2019eyl} is that in the large-$N$ limit the tetrahedron coupling $g$ is exactly marginal (its beta function is identically zero) while the beta functions of the other two couplings are quadratic in the couplings themselves with $g$-dependent coefficients. The model has four $g$-dependent fixed points, which are real for $g$  purely imaginary and below some critical value. One of them is IR attractive for both the pillow and the double-trace couplings.
Notice that from the perspective of the long-range Ising model ($N=1$, only one quartic interaction) this infrared fixed point is surprising. In fact, 
our kinetic term corresponds to the transition point between  the long-range behavior and  the mean field theory one, in which the infrared fixed point disappears in the Ising case.
The existence of a non-trivial IR fixed point in our $O(N)^3$ model is exclusively due to the tensor structure and the large-$N$ limit.
The critical exponents at the IR fixed point, that is the scaling dimensions of the pillow and double-trace operators, are real and above the unitarity bounds. 
The spectrum of dimensions of bilinear operators with arbitrary spin, as well as their  OPE coefficients with two fundamental fields, has been computed in Ref.~\cite{Benedetti:2019ikb}, where it was found again to be real and above the unitarity bounds.

Given the fact that the model has an imaginary coupling, it is to be expected that it is non-unitary, but the results of  \cite{Benedetti:2019eyl, Benedetti:2019ikb} raise the tantalizing possibility that in the large-$N$ limit we could find a unitary theory.\footnote{Here and occasionally in the rest of the paper, by an abuse of terminology, we talk about unitarity rather than reflection positivity despite working exclusively in Euclidean signature. It should be clear that we have in mind unitarity of the Wick-rotated theory.} For example, non-unitarity could manifest itself in some dimensions or OPE coefficients having imaginary parts which are suppressed in $1/N$.

In this paper we address the following two questions: {\it (i)} does the large-$N$ infrared fixed point of Ref.~\cite{Benedetti:2019eyl} define a conformal field theory? {\it (ii)} Does it define a unitary theory?

In order to tackle the first question, we will adapt to our model the methods of Ref.~\cite{Paulos:2015jfa}, which gives a proof of conformal invariance to all orders in perturbation theory for the infrared fixed-point of the long-range Ising model with propagator $C(p)=1/p^{(d+\epsilon)/2}$.
Most of that proof is built on standard ideas (e.g.\ from Ref.~\cite{Brown:1979pq}), except that the non-local propagator of the long-range model implies the absence of a local energy-momentum tensor. The main point is to use the Caffarelli-Silvestre trick \cite{Caffarelli} of localizing the kinetic term by means of an embedding of the theory in $d+p$ dimensions, with $p=2-(d+\epsilon)/2$. 

The main differences in our case are that: first, we are interested in $\epsilon=0$, and second, we must deal with multiple quartic interactions which mix under renormalization. We thus need to revisit  the results of Ref.~\cite{Benedetti:2019eyl} and \cite{Benedetti:2019ikb} with an analytic regularization (rather than using a momentum cutoff), and construct renormalized composite operators. 
Once this is done we conclude along the lines of \cite{Paulos:2015jfa} that the fixed-point theory of our $O(N)^3$ model is indeed conformally invariant.

We then compute two and three-point functions among the renormalized composite operators, which we use to further test conformal invariance and to address our second question.
Conformal invariance greatly constraints these correlators \cite{Osborn:1993cr}: the two-point functions between operators of different scaling dimension are zero, the three-point functions among three operators are completely fixed by their dimensions (up to an overall OPE coefficient), and so on. Such constraints are respected by all the correlators we have computed. 
%{\color{red} with a caveat in one case on which we comment further below.}

For special values of the dimensions of operators some subtle issues appear \cite{Bzowski:2015pba}. For instance one should be careful to distinguish between the $d$-dimensional Dirac delta $\delta(x)$, which is a homogeneous distribution under conformal transformations, and $|x|^{-d}$ which has a singular Fourier transform and upon regularization does not transform homogeneously \cite{Todorov:1985xs}. We need to deal with this issue as in our model we have an exactly marginal operator, the tetrahedron, which by definition has dimension $d$. We find contact terms (i.e.\ terms including a delta function) in its three-point function with itself Eq.~\eqref{eq:ttt}, as expected \cite{Seiberg:1988pf,Nakayama:2019mpz}. These contact terms 
do not lead to an anomaly as their coefficients are finite. However, an anomaly can arise from  its two-point function \cite{Gomis:2015yaa}, which has the functional form $1/|x|^{2d}$, and which in two dimensions has a singular distributional limit: $1/|x|^{2d-\epsilon} \sim \frac{1}{\epsilon} (\partial^{2}) \delta(x)$. We therefore expect a conformal anomaly in $d=2$. Such a conformal anomaly in two dimensions would not be a big surprise, but in the absence of a local energy-momentum tensor it is not obvious how it should be interpreted. It should also be noted that in Ref.~\cite{Benedetti:2019ikb} a puzzling discontinuity was found in the spectrum of bilinear operators: the computation at $d=2$ differs from that at $d=2+\epsilon$ in the limit of vanishing $\epsilon$. We hope to come back in future work to the two-dimensional case to clarify these issues. 

The computation of  two- and three-point functions allows us to address also the question about unitarity.
Unitarity constrains the correlators \cite{Poland:2018epd}: we can check whether our two-point functions satisfy reflection positivity, whether the OPE coefficients appearing in our three-point functions are real, and so on. It turns out that these constraints are satisfied.

\paragraph{Melons vs fishnets.} We conclude this overview of results by a remark. There are a number of intriguing similarities between our model and the conformal fishnet theory introduced in Ref.~\cite{Gurdogan:2015csr}, or more precisely with its  generalization to dimension $d<4$ \cite{Kazakov:2018qbr}, which  requires a long-range propagator like ours.
The conformal fishnet theory is a model of two complex matrices, with a single-trace chiral quartic interaction, without its hermitian conjugate. The interaction is therefore complex, as in our model. Furthermore this interaction is exactly marginal in the large-$N$ limit, as in our model. 
Moreover, renormalizability requires the introduction of double-trace interactions, and the four-point function renormalizing them is built out of ladders and bubbles in a similar fashion to what was found in Ref.~\cite{Benedetti:2019eyl} for our model (see also Sec.~\ref{sec:quartic} below). The resulting beta functions for the running couplings are therefore quadratic in both models, with coefficients parametrically depending on the exactly marginal coupling (compare Eq.~(13) of Ref.~\cite{Grabner:2017pgm}, with our beta functions \eqref{eq:beta1a} below).

There are of course also important differences. The different names that have been attached to the two conformal theories are not an accident: whereas at large-$N$ our model is dominated by melonic diagrams, the model of G\"urdogan and Kazakov is dominated by fishnet diagrams. What is accidental is the fact that for the four-point functions of fundamental fields both types of diagrams reduce to ladders: indeed a ladder can be thought either as melonic graph which has been open on two edges (and with resummed propagators), or as a fishnet with periodicity of length two in one direction. However, all the other $n$-point functions of fundamental fields are different. In particular, while the two-point function in a melonic CFT is given by a sum over melonic two-point diagrams, in the fishnet CFT there is no  correction to the bare propagator (and hence no mass or wave function renormalization) at leading order. Such a difference is not very important for our long-range model in $d<4$, which has no wave function renormalization anyway, but it becomes relevant for models with a standard short-range propagator, including ours at $d=4$. Last but not least, the conformal fishnet theory is a logarithmic CFT \cite{Gromov:2017cja}, while so far we have not found logarithmic correlators in our model.
We will study in future work whether this property survives in other correlators or at sub-leading order in $1/N$.

\paragraph{Plan of the paper.}
In section \ref{sec:model}, we introduce and review the model.
In section \ref{sec:renorm}, we discuss renormalization and fixed points of the model, both in the Wilsonian picture and in the minimal subtraction scheme.
Then, in section \ref{sec:composite operator}, we discuss the mixing of the $\phi^4$ composite operators under the renormalization group flow.
In section \ref{sec:embedding}, we give a proof of the conformal symmetry at the infrared fixed point of a class of correlations, based on the $D=d+p$ dimensional embedding method of Ref.~\cite{Paulos:2015jfa}.
Lastly, in section \ref{sec:3pt}, we use the perturbative expansion at the fixed-point (our small parameter being the exactly marginal tetrahedron coupling)
in order to compute the two- and three-point functions (and hence the OPE coefficients) among the $\phi^4$ and $\phi^2$ composite operators.

In the appendices we include some detailed computations and additional remarks. Appendix \ref{app:int} contains detailed computations for several integrals we use in the main text.
The beta functions for the spin zero bilinear operators are presented in appendix \ref{app:bilinear}.
In appendix \ref{app:1/N}, we discuss the large-$N$ scaling of the maximally single-trace (MST) and maximally multi-trace (MMT) operators,
and in appendix \ref{app:pillow}, we discuss correlators containing the pillow operator, which is neither of MST nor MMT.
% {\color{green}
% Finally, in appendix \ref{app:N=1}, we give a comparison with the long-range Ising model, that loosely corresponds to $N=1$.}

%%%%%%%%%%%%%%%%%%%%%%%%%%%%%%%%%%%%%%%%%%%%%%%%%%%%%%%%
%%%%%%%%%%%%%%%%%%%%%%%%%%%%%%%%%%%%%%%%%%%%%%%%%%%%%%%%
\section{Overview of the Model}
\label{sec:model}
%%%%%%%%%%%%%%%%%%%%%%%%%%%%%%%%%%%%%%%%%%%%%%%%%%%%%%%%
%%%%%%%%%%%%%%%%%%%%%%%%%%%%%%%%%%%%%%%%%%%%%%%%%%%%%%%%
We study the tensor model of \cite{Benedetti:2019eyl, Benedetti:2019ikb}, that is,  the $O(N)^3$ tensor model of Klebanov and Tarnopolsky \cite{Klebanov:2016xxf}
and Carrozza and Tanasa \cite{Carrozza:2015adg} (CTKT model) with a long-range covariance.
The fundamental field is a real tensor field of rank $3$, $\phi_{a^1a^2 a^3}(x)$,
transforming under $O(N)^3$ with indices distinguished by the position, and we denote $\mba = (a^1,a^2,a^3)$. The action of the model is:
	\begin{align}
	\label{eq:action}
		S[\phi] &=  \frac{1}{2} \int d^dx \, \phi_{\mba}(x) (  - \partial^2)^{\zeta}\phi_{\mba}(x) + S^{\rm int}[\phi] \,,  \\
		S^{\rm int}[\phi]  &= \frac{ m^{2\zeta}}{2} \int d^dx \, \phi_{\mba}(x)  \phi_{\mba}(x) 
		+ \frac{1}{4} \int d^d x \, \left[ \im\lambda \hat{\delta}^t_{\mba \mbb\mbc\mbd} + \lambda_1 \hat{P}^{(1)}_{\mba\mbb; \mbc\mbd}
		+ \lambda_2 \hat{P}^{(2)}_{\mba\mbb; \mbc\mbd } \right] \phi_{\mba}(x) \phi_{\mbb}(x) \phi_{\mbc}(x) \phi_{\mbd}(x) \, , \nonumber
	\end{align}
where repeated tensor indices are summed over $a^i = 1, \cdots, N$ and we introduced the projectors:
	\begin{equation}
		\hat{P}^{(1)}_{\mba\mbb; \mbc\mbd} \, = \, 3 (\hat{\delta}^p_{\mba\mbb;\mbc\mbd} - \hat{\delta}^d_{\mba\mbb;\mbc\mbd}) \, , \qquad 
		\hat{P}^{(2)}_{\mba\mbb; \mbc\mbd} \, = \, \hat{\delta}^d_{\mba\mbb;\mbc\mbd} \, .
	\end{equation}
and the rescaled operators:
\be \label{eq:deltas}
\hat{\delta}^t_{\mba\mbb\mbc\mbd}=\frac{1}{N^{3/2}} \, \delta^t_{\mba\mbb\mbc\mbd} \,,
\quad \hat{\delta}^p_{\mba\mbb;\mbc\mbd}=\frac{1}{N^{2}} \, \delta^p_{\mba\mbb;\mbc\mbd}\,,
\quad \hat{\delta}^d_{\mba\mbb;\mbc\mbd}=\frac{1}{N^{3}} \, \delta^d_{\mba\mbb;\mbc\mbd}\, ,
\ee
with
\be
\begin{split} \label{eq:deltas-nohat}
&\delta^t_{\mba \mbb\mbc\mbd}  = \delta_{a^1 b^1}  \delta_{c^1 d^1} \delta_{a^2 c^2}  \delta_{b^2 d^2 } \delta_{a^3 d^3}   \delta_{b^3 c^3} \, , \\
	\delta^p_{\mba\mbb; \mbc\mbd } &= \frac{1}{3} \sum_{i=1}^3  \delta_{a^ic^i} \delta_{b^id^i} \prod_{j\neq i}  \delta_{a^jb^j}  \delta_{c^jd^j} \,,
	\qquad  \delta^d_{\mba\mbb; \mbc\mbd }  = \delta_{\mba \mbb}  \delta_{\mbc \mbd} \,.
\end{split}
\ee
Here $t$ stands for \emph{tetrahedron},  $p$ for \emph{pillow}, and $d$ for \emph{double-trace}. Such names refer to the graphical representation of the respective pattern of contraction of indices, as recalled below. 

We use the following shorthand notations for the quadratic invariant:
	\begin{equation}
		\phi^2(x) \, \equiv \, \phi_\mba(x) \phi_\mba(x) \, ,
	\label{eq:qadrainv}
	\end{equation}
and for the quartic invariants:
\be\label{eq:quartinv}
\begin{split}
& \phi^4_t(x) \equiv \im \hat{\delta}^t_{\mba \mbb\mbc\mbd} \, \phi_{\mba}(x) \phi_{\mbb}(x) \phi_{\mbc}(x) \phi_{\mbd }(x) \,, \qquad \\
& \phi^4_i(x) \equiv \hat{P}^{(i)}_{\mba\mbb; \mbc\mbd} \, \phi_{\mba}(x) \phi_{\mbb}(x) \phi_{\mbc}(x) \phi_{\mbd }(x) \, , \qquad (i=1,2)\,.
\end{split}
\ee

The difference between this model and the CTKT model is that the Laplacian is allowed to have a non integer power $0<\zeta \leq 1$. 
This modification preserves the reflection positivity of the propagator: the free theory is unitary for any $\zeta \le 1$.
The choice $\zeta = d/4$ renders the quartic invariants marginal in any $d$ \cite{Benedetti:2019eyl}. It is this value of $\zeta$ that interests us in this paper. Moreover, we will restrict to $d<4$ in the following, in order to avoid a wave function renormalization (see Sec.~\ref{sec:2pt}, and footnote~\ref{foot:melon-int} in particular).

We have not assigned any subscript to the coupling of the tetrahedral invariant, as it plays a special role in the model. 
Observe also that the infrared fixed point  found in  \cite{Benedetti:2019eyl, Benedetti:2019ikb}, and  that we aim to study, occurs for a purely imaginary tetrahedral coupling, hence we have chosen here to make that explicit  from the onset, by writing the coupling as  $\im \lambda$, with $\lambda\in \mathbb{R}$. 

\begin{figure}[ht]
\begin{center}
\includegraphics[width=0.5\textwidth]{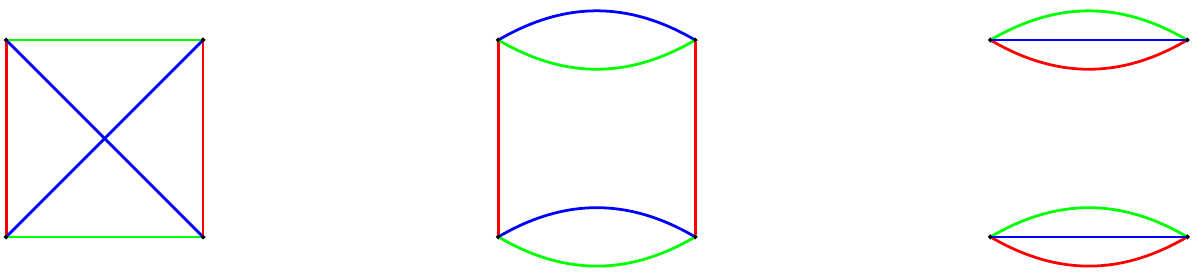}
 \caption{Graphical representation of the quartic $O(N)^3$ invariants. From left to right: the tetrahedron, the pillow, and the double-trace (there are three pillow contractions, distinguished by the color of the vertical edge).} \label{fig:interactions}
 \end{center}
\end{figure}

As usual, it is convenient to introduce a graphical representation of the $O(N)^3$ invariants, which also justifies their names.
We represent every tensor ($\phi_{\mba}$, $\phi_{\mbb}$ and so on) as a three-valent node and every contraction of two indices ($a^i$ and $b^i$ for instance)
as an edge with a color $i=1, 2$, or 3 (red, green, or blue) corresponding to the position $i$ of the indices.
As a result, $O(N)^3$ invariants are represented by 3-colored graphs.
The graphs corresponding to the quartic invariants of Eq.~\eqref{eq:action} are depicted in Fig.~\ref{fig:interactions}.

\begin{figure}[ht]
\begin{center}
\includegraphics[width=0.5\textwidth]{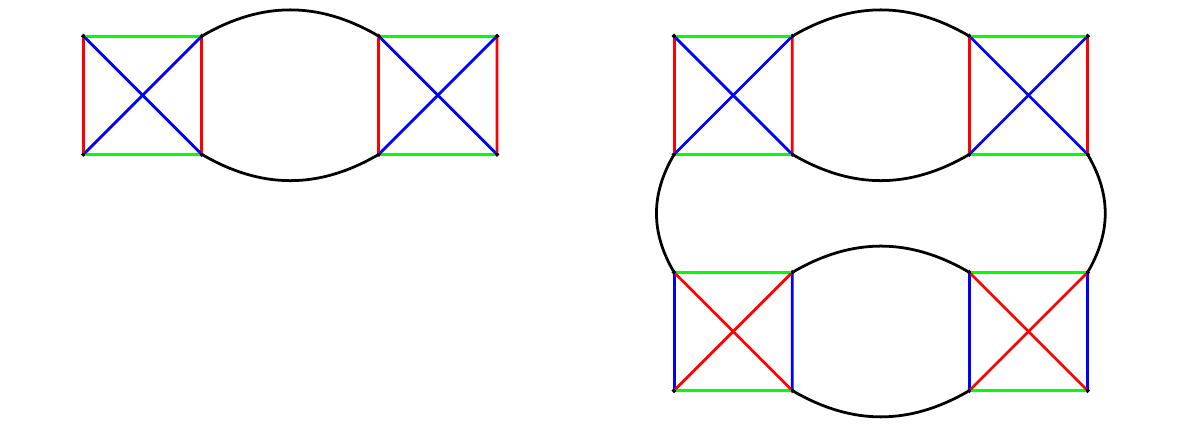}
 \caption{Two $4$-colored graphs, with external tensor contractions equivalent to the pillow (left) and double-trace (right) invariants.} \label{fig:graph}
 \end{center}
\end{figure}

We can expand the free energy and the connected $n$-point functions perturbatively around the Gaussian theory. We introduce two graphical representations for the terms in the perturbative expansion.
In the first, each interaction invariant is a $3$-colored graph, and the Feynman propagators are represented as edges with a new color, which we call 0 (pictured in black), connecting the tensors. This leads to a representation of the perturbative expansion in terms of 4-colored graphs, as for example in Fig.~\ref{fig:graph}. 
In the second representation, we simplify the graphs by shrinking each interaction invariant (i.e. all its edges with colors from 1 to 3) to a point. We call the resulting object Feynman diagrams, as they represent in a more straightforward way the spacetime integrals associated to the amplitude.
An important class of graphs and diagrams are the melonic ones, which however have very different features depending on whether it is the 4-colored graph which is melonic, or the (single color) diagram. The pillow and double-trace interactions are examples of melonic 3-colored graphs, and models based on such type of interactions are known  to be dominated by melonic 4-colored graphs at leading-order in $1/N$ \cite{Bonzom:2012hw}. The corresponding Feynman diagrams are cactus diagrams, as in vector models. On the contrary, the tetrahedron interaction is not a melonic 3-colored graph, but it leads to melonic Feynman diagrams in the large-$N$ limit \cite{Carrozza:2015adg,Klebanov:2016xxf}.

\begin{figure}[ht]
\begin{center}
\vspace{-10pt}
\includegraphics[width=0.25\textwidth]{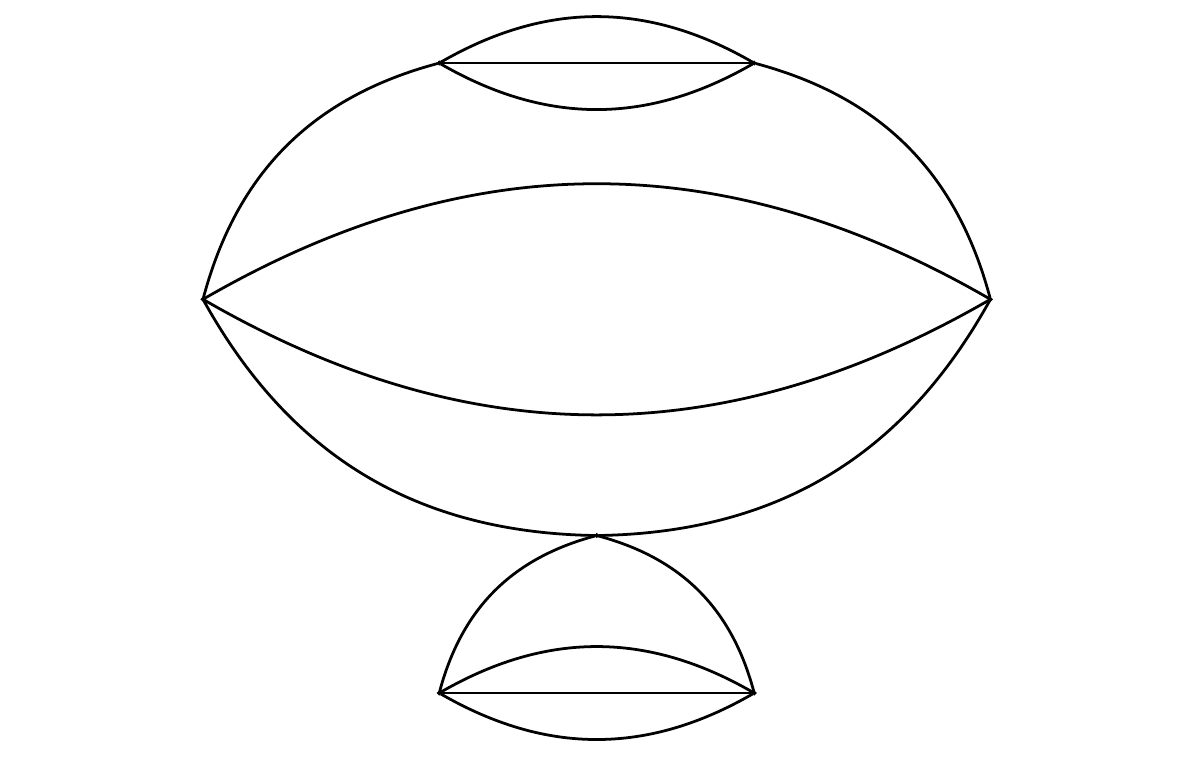}
 \caption{A vacuum melon-tadpole Feynman diagram, where all the invariants have been shrunk to point-like vertices.} \label{fig:melontadpoles}
 \end{center}
\end{figure}

As a result of the combination of pillow, double-trace, and tetrahedron interactions, our model has a $1/N$ expansion dominated by \emph{melon-tadpole} diagrams  \cite{Benedetti:2019eyl} with melons based on couples of tetrahedral vertices and tadpoles based on either pillow or double-trace vertices (see Fig.\ref{fig:melontadpoles}). 

\begin{figure}[ht]
\begin{center}
\includegraphics[width=0.5\textwidth]{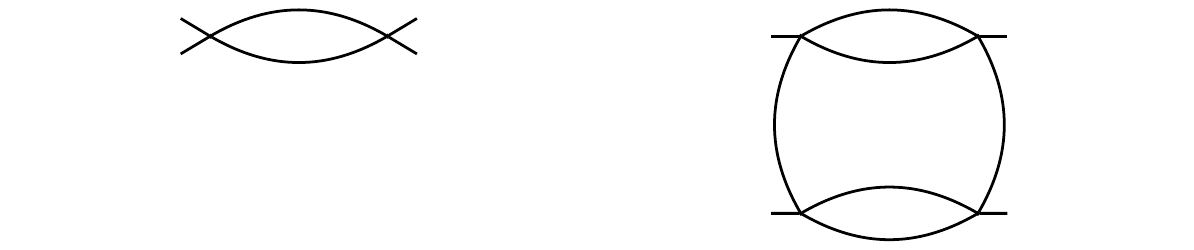}
 \caption{Two Feynman diagrams obtained from the two $4$-colored graphs of Fig.~\ref{fig:graph} by shrinking the colored edges. Half-edges are also added to keep track of the external fields.} \label{fig:FeynDiagr}
 \end{center}
\end{figure}

%%%%%%%%%%%%%%%%%%%%%%%%%%%%%%%%%%%%%%%%%%%%%%%%%%%%%%%%
%%%%%%%%%%%%%%%%%%%%%%%%%%%%%%%%%%%%%%%%%%%%%%%%%%%%%%%%
\section{Renormalization and Fixed Points}
\label{sec:renorm}
%%%%%%%%%%%%%%%%%%%%%%%%%%%%%%%%%%%%%%%%%%%%%%%%%%%%%%%%
%%%%%%%%%%%%%%%%%%%%%%%%%%%%%%%%%%%%%%%%%%%%%%%%%%%%%%%%
We consider the $d$-dimensional theory in Euclidean signature, and our aim is to study the case $\zeta = d/4$.

We introduce an infrared regulator $\mu$ by modifying the free covariance of the theory to: 
	\begin{equation} \label{eq:reg-C}
		C_{\mu}(p) = \frac{1}{(p^2 + \mu^2)^{\zeta}} \, = \, \frac{1}{\Gamma(\zeta)} \int_0^{\infty} da \; a^{\zeta-1} e^{-(p^2 + \mu^2) a} \,,
	\end{equation}
and we regulate the UV divergences by setting:
	\begin{equation}
		\zeta \, = \, \frac{d+\epsilon}{4} \, , \qquad (0 \le \epsilon \ll 1) \,.
	\label{eq:zeta}
	\end{equation}
This implies that the ultraviolet dimension of the field is
\be \label{eq:Delta_phi}
\Delta_{\phi} = ( d-2\zeta)/2 = ( d-\epsilon)/4 \,.
\ee
The reader should keep in mind that for us $\epsilon$ is just a regulator and we always intend to send $\epsilon \to 0$ in the end. 
In particular, we are not looking for Wilson-Fisher type of fixed points at small but finite $\epsilon$.

 We will need the following Fourier transform, which holds for $d/2>\gamma>0$:
\be\label{eq:Fourier1}
\begin{split}
& \int \frac{d^dp}{(2\pi)^d} \; \frac{1}{|p|^{2\gamma} } \; e^{ \im p(x-y)}    
    = \; \frac{c(\gamma)}{|x-y|^{d-2\gamma}}  \,,\qquad
  c(\gamma)  = \frac{\Gamma(\frac{d}{2} -\gamma )}{2^{2\gamma} \pi^{d/2}\Gamma(\gamma)}  \,.
\end{split}
\ee 

We sometimes rescale the couplings to $\lambda = (4\pi)^{d/2} \Gamma(\zeta)^{ 2} \tilde \lambda $ and so on, and
we denote the dimensionless running couplings at scale $\mu$ by $g,g_1,g_2$ (respectively $\tilde g, \tilde g_1, \tilde g_2$).

%%%%%%%%%%%%%%%%%%%%%%%%%%%%%%%%%%%%%%%%%%%%%%%%%%%%%%%%
\subsection{The two-point function}
\label{sec:2pt}
%%%%%%%%%%%%%%%%%%%%%%%%%%%%%%%%%%%%%%%%%%%%%%%%%%%%%%%%

We now discuss the bare and the full two-point functions of the model.

\paragraph{The bare propagator.}
The bare propagator  in the direct space, with infrared regulator set to zero, is obtained from Eq.~\eqref{eq:Fourier1} by simply setting $\gamma=\zeta$:
	\begin{equation}
C(x-y) = \int \frac{d^dp}{(2\pi)^d} \, \frac{e^{ \im p(x-y)}}{|p|^{2\zeta}} \, = \, \frac{c(\zeta)}{|x-y|^{d-2\zeta}} \,.
	\label{eq:fourier}
	\end{equation}
We will encounter below the convolution of the cube of the bare propagators with another propagator. Using repeatedly the Fourier transform in Eq.~\eqref{eq:Fourier1}, we obtain the formal result:
\be\label{eq:bbaremelon}
 \int d^d z \, C(x-z) C(z-y)^3 = \frac{c(\zeta)^3}{c(3\zeta-d)} \; \frac{c(4\zeta-d)}{|x-y|^{3d-8\zeta}}  \,.
\ee
The problem with this formula is that we used Eq.~\eqref{eq:Fourier1} for $\gamma = 3\zeta -d  =-d/4+3\epsilon/4<0$, hence the result is only formal. In fact this convolution hides an ultraviolet divergence which needs to be subtracted. Taking this into account, we obtain:\footnote{\label{foot:melon-int}In Fourier space the subtracted melon contribution is:
\be\label{eq:mmelosub}\nonumber
\begin{split}
 &\int \frac{d^dq_1}{(2\pi)^d}\frac{d^dq_2}{(2\pi)^d} \;
   \;  C(q_1) C(q_2) \bigg[ C(p+q_1+q_2) - C(q_1+q_2) \bigg]    = -
   \frac{p^{2d-6\zeta}}{ (4\pi)^d \Gamma(\zeta)^3}   \int_{0}^{\infty} da_1da_2da_3 \crcr
& \qquad \qquad \frac{(a_1a_2a_3)^{\zeta-1}}{
  (a_1a_2+a_1a_3+a_2a_3)^{d/2}}\; \bigg(1- e^{ - \frac{a_1a_2a_3}{a_1a_2+a_1a_3+a_2a_3} }  \bigg) = 
   \frac{p^{2d-6\zeta}}{ (4\pi)^d \Gamma(\zeta)^3}
\frac{\Gamma(1+3\zeta -d)}{ 3\zeta -d } \; \frac{\Gamma(\frac{d}{2}-\zeta)^3}{\Gamma( \frac{3}{2} d - 3\zeta )}    
  \,,
\end{split}
\ee
where the integral over the  $a_i$ parameters is convergent for $d<4$ and $\epsilon<d/3$. We computed it in Appendix~\ref{app:int}. Multiplying Eq.~\eqref{eq:mmelosub} by $p^{-2\zeta}$, Fourier transforming back to the direct space (which is allowed for $\epsilon \ge 0$) and using the analytic continuation of the $\Gamma$ function we obtain Eq.~\eqref{eq:convolution}}
\be	\label{eq:convolution}
\begin{split}
 \int d^d z \, C(x-z)  
 \bigg( C(z-y)^3  -\delta(z-y) \int d^du \, C(u)^3  \bigg) 
  &= \frac{c(\zeta)^3}{c(3\zeta-d)} \; \frac{c(4\zeta-d) }{|x-y|^{8\Delta_{\phi} -d }} \\
  &= - \frac{4 \Gamma(1-\frac{d}{4})}{d (4\pi)^d \Gamma(\frac{3d}{4})} \delta(x-y)  +O(\epsilon) \,,
\end{split}
\ee
where this time the integral is convergent, and we have used the distributional limit $\lim_{\epsilon\to 0} c(\epsilon) / |x-y|^{d-2\epsilon} = \delta(x-y)$ (alternatively one can take the limit in Fourier space to reach the same conclusion). 

%%%%%%%%%%%%%%%%%%%%%%%%%%%%%
\paragraph{The full two-point function.}
%%%%%%%%%%%%%%%%%%%%%%%%%%%%%
We now discuss the full two-point function of the model. 
In the absence of spontaneous symmetry breaking,\footnote{Spontaneous symmetry breaking in tensor field theories has been so far not much explored, but see Ref.~\cite{Diaz:2018eik,Benedetti:2018ghn,Benedetti:2019sop}.} the full two-point function is diagonal in the tensor indices. That is, %denoting $A=(\mba,x)$ and so on, 
it can be written as: 
\be
G_{\mba \mbb}(x,y)=\braket{\phi_{\mba}(x) \phi_{\mbb}(y)} 
=\delta_{\mba \mbb} \, G(x-y) \,.
\ee
Due to the infrared regulator, the full two-point function also acquires a $\mu$-dependence, hence we write its regulated diagonal component as  $G_{\mu}(x-y)$.

Following \cite{Benedetti:2019eyl}, we observe that $G_{\mu}(p)$ at leading order in $N$ respects the melonic Schwinger-Dyson (SD) equation:
\be
 G_{\mu}(p)^{-1} = (p^2+\mu^2)^{\zeta} \, + m^{2\zeta} +  \lambda_2  \int 
 \frac{d^dq}{(2\pi)^d} \, G_{\mu}(q) \;  + \, \lambda^2 \int 
 \frac{d^dq_1}{(2\pi)^d}\frac{d^dq_2}{(2\pi)^d} \;
   \;  G_{\mu}(q_1) G_{\mu}(q_2)  G_{\mu}(p+q_1+q_2)  \,,
\ee
where the sign of the last term is changed with respect to  \cite{Benedetti:2019eyl} due to the explicit $\im$ factor in our action.
Similar to Eq.~\eqref{eq:bbaremelon}, this equation exhibits a power divergence in the ultraviolet. The critical theory is obtained by tuning the bare mass to exactly cancel this divergence
such that the renormalized mass is zero. The melonic SD equation in the limit $\mu\to 0$ for the critical theory is:
\be
 G(p)^{-1} = p^{2\zeta} \, + \, \lambda^2 \int 
 \frac{d^dq_1}{(2\pi)^d}\frac{d^dq_2}{(2\pi)^d} \;
   \;  G(q_1) G(q_2) \bigg[ G(p+q_1+q_2) - G(q_1+q_2) \bigg] \,,
\ee
where $G(p)$ denotes the full two-point function with no cutoff. It is solved self consistently at $\epsilon=0$, for any $d<4$ by the ansatz $G(p)^{-1}= Zp^{2\zeta}$: 
\be
 \begin{split}
   Z p^{2 \zeta} =  
  p^{2\zeta} \, - \, \frac{p^{2d-6\zeta}   }{Z^3}  \; \; 
  \lambda^2  \left( - \frac{c(\zeta)^3}{c(3\zeta - d)} \right)
 \,, \qquad  \zeta = \frac{d}{4} \,, 
\end{split}
\ee
provided that the constant $Z$ is chosen as:
\be\label{eq:wf}
   Z =1 \, - \, \frac{ \lambda^2}{Z^3} 
 \left( - \frac{ c(d/4)^3 }{c(-d/4)} \right) \,.
\ee
This in turn fixes the bare mass to $m^{2\zeta} =  m_c^{2\zeta}$, where:
\be\label{eq:mbare}
 m_c^{2\zeta} = - \frac{\lambda_2}{Z} \int \frac{d^dq}{(2\pi)^d} 
  \frac{1}{q^{2\zeta}} - \frac{\lambda^2}{Z^2} \int\frac{d^dq_1}{(2\pi)^d}\frac{d^dq_2}{(2\pi)^d} \; \frac{1}{q_1^{2\zeta} } \, \frac{1}{q_2^{2\zeta} } \,
  \frac{1}{(q_1+q_2)^{2\zeta}} \,.
\ee

Summarizing, in the critical theory, the net effect of all the Feynman diagrams on the two-point function is a just multiplicative factor: $G(p)= Z^{-1} C(p)$.
The constant $Z$ can be seen as a finite wave function renormalization which resums all the melonic insertions.
By resumming an infinite series of diagrams, $Z$ carries non-perturbative information on the radius of convergence of the perturbative series \cite{Benedetti:2019eyl} at large $N$.

The fact that we can perform such a resummation is an important property of our tensor model at $\epsilon=0$.
Nevertheless, in order to regulate the UV divergences, below we will consider $\epsilon>0$. 
In this case, the resummation of the leading-order two-point function can not be performed explicitly. 
Furthermore, for theories with a non-local propagator, due to the locality of counterterms, there is no need to renormalize the fields. Therefore, it is common to  set the the wave function renormalization to $1$ (e.g.\ \cite{Paulos:2015jfa}), and we will do the same here.
We thus seem to have a discontinuous $Z$ at $\epsilon=0$, but in fact this is not the case. It is all only a matter of repackaging of diagrams:
the two-point melonic insertions contribute also at $\epsilon>0$, but as we let $\epsilon\to 0$, we can conveniently resum them by working with melon-free diagrams and propagators divided by $Z$. Differences appear if we decide to rescale or not the fields by $\sqrt{Z}$: rescaling is a natural choice at $\epsilon=0$, as it brings the two-point back function to its original form, but it is not justified at $\epsilon>0$ as $G(p)$ and $C(p)$ have different functional forms. In any case, rescaling by a finite constant is just a choice of renormalization scheme like others, and as such it will not affect the universal part of the beta functions.

For $\epsilon=0$ the melonic SD equation in direct space is:
\be	\label{eq:exact prop}
 ( G^{-1})_{xy }  = (C^{-1})_{xy}  + \lambda^2 \bigg[ G_{xy}^3 - \delta_{xy}\int d^d u \, G(u)^3\bigg]\,, \quad
 		G(x-y) \, = \, \frac{c(d/4) }{ Z \, |x-y|^{d/2}} \, ,
\ee
which can also be written as:
\be	\label{eq:SD-eq}
\int d^d z \,  G(x-z) \bigg( G(z-y)^3 - \delta_{zy}\int d^d u \, G(u)^3 \bigg) = \frac{Z-1}{Z \, \lambda^2} \,\delta(x-y) \,,\quad 
\frac{Z-1}{Z \, \lambda^2} =
 \frac{1}{Z^4}  \; \frac{c(d/4)^3}{c(-d/4)}  \,.
\ee
The SD equation simplifies if the integral over $z$ is understood in the sense of dimensional regularization. In this case the local part of the melonic correction and the tadpoles are set to zero. In the Wilsonian picture the melon integral combines with the explicit mass counterterms which provide the subtraction. This works as long as the position $y$ is an internal position. If $y$ is an external argument, that is when this is a contribution to a correlator $\Braket{\dots \phi^4_t(y)}$
one needs to replace the tetrahedral operator by the renormalized tetrahedral operator:
\be
 \Braket{\dots [\phi^4_t(y)]} \,, \qquad [\phi^4_t(y)] = \phi^4_t(y) - 2 \phi^2(y) \int d^du \,G^3(u) + \dots \,.
\ee

%%%%%%%%%%%%%%%%%%%%%%%%%%%%%%%%%%%%%%%%%%%%%%%%%%%%%%%%
\subsection{The beta functions}
%%%%%%%%%%%%%%%%%%%%%%%%%%%%%%%%%%%%%%%%%%%%%%%%%%%%%%%%

We review the $\beta$ functions of the theory defined by the action \eqref{eq:action} tuned to criticality.

%%%%%%%%%%%%%%%%%%%%%%%%%%%%%%%%%%
\subsubsection{Quartic couplings}
\label{sec:quartic}
%%%%%%%%%%%%%%%%%%%%%%%%%%%%%%%%%%

We denote $ \im \Gamma^R_t , \Gamma^R_1$, and $\Gamma^R_2$ the appropriately normalized one-particle irreducible four-point functions at zero external momentum. They are computed using the bare expansion in terms of connected amputated one-particle irreducible four-point diagrams $\cG$ with amplitude: 
\begin{align}
\label{eq:ampli1}
 &A(\cG) = \mu^{(d - 4 \Delta_\phi)[1-n(\cG)]}  \hat A(\cG) = \mu^{\epsilon -n(\cG)\epsilon} \hat A(\cG) \,,\\ 
& \hat A(\cG)   =\int_{0}^{\infty} 
 \left( \prod_{e\in \cG} d\alpha_e \;  \alpha_e^{\zeta-1}\right) \; 
 \frac{ 1 }{ \big[ \sum_{ {\cal T} \subset \cG} \prod_{e\notin {\cal T} } \alpha_e \big]^{d/2} } \;  e^{-\sum_{e\in \cG} \alpha_e}\,, 
\end{align}
where $n(\cG)$ denotes the number of vertices of $\cG$, $e\in \cG$ denotes the edges of $\cG$, and ${\cal T}$ runs over the spanning trees in $\cG$ (each having $n(\cG)-1$ edges).
The tetrahedral four-point function is trivial, as it receives no radiative corrections at large $N$:
\be
  \Gamma^R_t = \tilde \lambda \,.
\ee

At leading order in $1/N$ the remaining four-point functions $\Gamma^R_1$ and $\Gamma^R_2$ are identical up to replacing $\tilde \lambda^2$ by $3\tilde \lambda^2$ and $\tilde\lambda_1$ by $\tilde \lambda_2$. We discuss $\Gamma^R_1$.
Only chain diagrams \cite{Benedetti:2019eyl} contribute to $\Gamma^R_1$ at leading order in $1/N$.  A chain diagram $\cG$ is a sequence of irreducible pieces connected by pairs of parallel horizontal edges. The irreducible pieces are either vertical ladder rungs with two tetrahedral couplings, or bare vertices $\lambda_1$. There are $2^{n}$ chain diagrams with $n$ irreducible parts (that is vertical rungs or bare vertices). The edges of a chain diagram are decorated by arbitrary melonic insertions, but we do not include tadpoles, as they are assumed to have been taken care of (by dimensional regularization or mass subtraction).

The chain diagram consisting in a bare vertex has amplitude $1$. 
We denote  $n_t(\cG)$ the number of tetrahedral vertices of $\cG$
(which is always even), $n_1(\cG)$ the numbers of vertices $\lambda_1$ of $\cG$, and $\mathfrak{G}$ the set of connected chain diagrams with at least two internal vertices. We have:
\begin{equation}\label{eq:bare}
    \Gamma^R_1  =   \tilde \lambda_1   
  +  \mu^{\epsilon}  \sum_{\cG\in \mathfrak{G}} (-1)^{1 + n_1(\cG) +n_t(\cG) } 
      \left( \mu^{-\epsilon } \tilde \lambda\right)^{n_{t}(\cG)} 
 \bigg( \mu^{-\epsilon } \tilde  \lambda_1   \bigg)^{n_1(\cG)} 
 \hat A(\cG)\,.
\end{equation}

The case $\epsilon=0$ is special. The series in \eqref{eq:bare} can be further simplified: all the melonic insertions can be analytically resummed at the price of multiplying the bare propagator by $Z^{-1}$ from Eq.~\eqref{eq:wf}. Furthermore, in this case the four-point function itself should be divided 
by $Z^2$. The overall effect is that Eq.~\eqref{eq:bare} can be rewritten at $\epsilon=0$ by dividing all the couplings by $Z^{2}$ and reducing the sum to chain diagrams with no melonic insertions.

The chain diagrams can be analyzed in terms of their one-vertex irreducible components.
Adapting the notation of \cite{Benedetti:2019eyl} to the case $\epsilon>0$, we denote:
\begin{itemize}
\item $\hat{U}_r$ the sum of dimensionless amplitudes of the ladders with $r\geq 1$ rungs, and with melonic insertions;
 we include in $\hat{U}_r$ the $\tilde\lambda$-dependence due to melonic insertions, but not that due to the pairs of vertices in a rung. Therefore, we write the sum over the ladders of arbitrary length as:
\be \label{eq:U-def}
U(x) = \sum_{r\ge 1} (-1)^r x^{2r} \hat{U}_r(x) \,,
\ee
with $x=\mu^{-\epsilon} \tilde \lambda$. Notice that unlike in \cite{Benedetti:2019eyl}, $U(x)$ is not the generating function of the amplitudes $\hat{U}_r(x)$, due to the $x$-dependence of the latter. We also define $U_r \equiv \hat{U}_r(0)$, for the amplitudes without melonic insertions (Fig.~\ref{fig:pureladders}).
\begin{figure}[H]
\begin{center}
\includegraphics[scale=1]{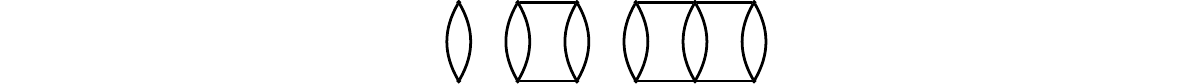} 
 \caption{The pure ladders $U_1$, $U_2$, and $U_3$.} \label{fig:pureladders}
 \end{center}
\end{figure}

\item $\hat{S}_r$ the sum of dimensionless amplitudes of the caps with $r\geq 1$ rungs, i.e.\ ladders with $r$ rungs closed on a $ \lambda_1$ vertex on one side, with melonic insertions; we write the sum over caps as:
\be \label{eq:S-def}
S(x) = \sum_{r\ge 1} (-1)^r x^{2r} \hat{S}_r(x) \,,
\ee
with $x=\mu^{-\epsilon} \tilde \lambda$.  We also define $S_r \equiv \hat{S}_r(0)$, for the amplitudes without melonic insertions (Fig.~\ref{fig:pladders}).
\begin{figure}[H]
\begin{center}
\includegraphics[scale=1]{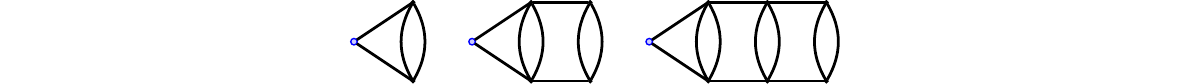} 
 \caption{The caps $S_1$, $S_2$, and $S_3$. The blue vertex represents $\lambda_1$.} \label{fig:pladders}
 \end{center}
\end{figure}

\item $\hat{T}_r$ the sum of dimensionless amplitudes of the double-caps with $r\geq 0$ rungs, i.e.\ ladders with $r$ rungs closed on a $ \lambda_1$ vertex on each side, with melonic insertions;  we write the sum over double-caps as:
\be \label{eq:T-def}
T(x) = \sum_{r\ge 0} (-1)^r x^{2r} \hat{T}_r(x) \,,
\ee
with $x=\mu^{-\epsilon} \tilde \lambda$.  We also define $T_r \equiv \hat{T}_r(0)$, for the amplitudes without melonic insertions (Fig.~\ref{fig:pladdersp}).
\begin{figure}[H]
\begin{center}
\includegraphics[scale=1]{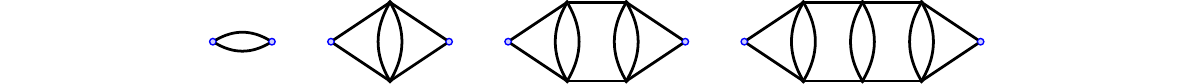} 
 \caption{The caps $T_0$, $T_1$, $T_2$, and $T_3$. The blue vertices represent $\lambda_1$.} \label{fig:pladdersp}
 \end{center}
\end{figure}
\end{itemize}

Observe that for the caps and double-caps exactly one and  two vertices, respectively,  correspond to couplings $\lambda_1$.
The leading $1/\epsilon$ behavior of these amplitudes is
$
 {U}_r\sim \epsilon^{- ( 2r-1) },\,{S}_r \sim \epsilon^{-2r},
 \, {T}_r\sim \epsilon^{-( 2r+1)}\,.
$
In terms of the resummed amplitudes, the bare expansion writes \cite{Benedetti:2019eyl}:
\be
  \mu^{-\epsilon}\Gamma^R_1 = -U \left( \mu^{-\epsilon} \tilde \lambda  \right)
 + \mu^{-\epsilon} \tilde \lambda_1  
 \frac{ \big[ 1 +S\left( \mu^{-\epsilon} \tilde \lambda \right) \big]^2 }{ 1 +  \mu^{-\epsilon}  \tilde \lambda_1  
 T\left( \mu^{-\epsilon} \tilde \lambda \right) } \,,
\ee
which exhibits an ascending series of poles in $1/\epsilon$.
In the limit $\epsilon\to 0$, as discussed above, we can restrict to ladders, caps, and double-caps with no melonic insertions (that is, we can remove hats from $\hat U_r$, $\hat S_r$, and $\hat T_r$), provided that the couplings and the amplitude are rescaled by $Z^{-2}$.

\paragraph{Wilsonian beta functions.} In the Wilsonian picture we identify the four-point functions  with the running couplings:
\be
 \tilde g =  \mu^{-\epsilon}\Gamma^R_t \,,\qquad 
 \tilde g_1 = \mu^{-\epsilon} \Gamma^R_1 \,,\qquad 
 \tilde g_2 =  \mu^{-\epsilon}\Gamma^R_2 \,,
\ee
and for the theory at $\epsilon=0$ we would further rescale the right-hand side  by $Z^{-2}$. The beta functions are the scale derivative of the running couplings at fixed bare couplings $\tilde \lambda,\tilde \lambda_1,\tilde \lambda_2$. The beta functions of $\tilde g_1$ and $\tilde g_2$ decouple.
In order to compute them, we invert the bare series:
\be\label{eq:renseries}
		\tilde \lambda \, = \, \mu^{\epsilon} \,  \tilde g   \,, \qquad
		\tilde \lambda_1 =\mu^{\epsilon}  \; \frac{\tilde g_1 + U(\tilde g)}
 { [1+S(\tilde g)]^2 - [ \tilde g_1 + U(\tilde g) ]  T(\tilde g)  } \,.
\ee
We first observe that the $\beta$ function of the tetrahedral coupling is trivial:
\be\label{eq:betaWilson_t}
\beta_t(\tilde g)  = \mu \partial_{\mu} \tilde g =  
- \epsilon \tilde g  \,.
\ee
In particular $\beta_t$ is identically zero at $\epsilon =0$ indicating a line of fixed points in that case \cite{Benedetti:2019eyl}. For the other couplings, the beta functions are obtained by taking $\mu\frac{d}{d\mu}$ in equation \eqref{eq:renseries}, leading to:
\be
\beta_1(\tilde g_1,\tilde g)  =  \mu\partial_{\mu}\tilde g_1 = \epsilon  \bigg[    \tilde g U' - 
     \bigg( \frac{2\tilde gS'}{1+S} +1 \bigg)(\tilde g_1+U) 
     +   \frac{  \tilde gT' +T }{(1+S)^2}  \; (\tilde g_1 + U)^2  \bigg] \,,
\ee
where $U = - \tilde g^2 \hat{U}_1 + \dots $, 
$S=- \tilde g^2\hat{S}_1 + \dots $ and $T=\hat{T}_0 - \tilde g^2\hat{T}_1 +\dots$ and their derivatives are evaluated at the effective tetrahedral coupling $\tilde g$. Comparing with the results of \cite{Benedetti:2019eyl}, we see that the beta functions remain a quadratic polynomial in $\tilde g_1$ and $\tilde g_2$ also at $\epsilon >0$:
\begin{align}
&\beta_1 (\tilde g_1,\tilde g) = - \epsilon  \tilde g_1 + 
       \beta^{(0)} (\tilde{g}^2)  -2 \beta^{(1)}( \tilde g^2) \; \tilde g_1 + 
        \beta^{(2)} (\tilde g^2) \; \tilde g_1^2 \,, \label{eq:beta1} \\
&\beta_2 (\tilde g_2,\tilde g) = - \epsilon  \tilde g_2 + 
       \beta^{(0)} (3\tilde g^2)  -2 \beta^{(1)}( 3\tilde g^2) \; \tilde g_2 + 
        \beta^{(2)} (3\tilde g^2) \; \tilde g_2^2 \,, \label{eq:beta2}
\end{align}
where the coefficients are:
\be
\begin{split}
\beta^{(0)}( \tilde g^2) & =  \epsilon   \bigg[ \tilde g U' -U - 2\frac{ U }{1+S} \tilde gS' + \frac{U^2}{(1+S)^2} ( \tilde g T' +T) \bigg] \,,
\crcr 
\beta^{(1)}( \tilde g^2) & =   \epsilon  \bigg[ \frac{1}{1+S}\tilde gS' - \frac{U}{(1+S)^2} (\tilde gT' +T) \bigg] \,, 
\qquad \beta^{(2)}( \tilde g^2)  = \epsilon  \; \frac{\tilde gT'+T}{(1+S)^2} \,. 
\label{eq:betacoeffW} 
\end{split}
\ee
Up to cubic order in the couplings (that is, two loops), and using the integrals $D\equiv T_0(0)= U_1(0) $ and $S_1(0)$ discussed in Appendix \ref{app:int}, we find:
 \be\label{eq:betaWilson}
    \beta_1( \tilde g_1,\tilde g) = -\epsilon \tilde g_1 
   - \epsilon D \tilde g^2 
    + 2 \epsilon(2S_1-D^2) \tilde g^2 \tilde g_1  + \epsilon D \tilde g_1^2 + O(g^4,g_1^2 g^2) \,,
 \ee
and at first order in $\epsilon$ the coefficients $\beta^{(0)},\beta^{(1)}$ and $\beta^{(2)}$ are:
\begin{align}
 \beta^{(0)} = & - \epsilon \, \tilde g^2 D =- 2 \frac{\Gamma(d/4)^2}{\Gamma(d/2)} \tilde g^2
  + O(\epsilon,g^4,g_1^2 g^2) \,, 
 \crcr
 \beta^{(1)} = & -\epsilon \, (2 S_1 - D^2 )\tilde g^2
 = -
\frac{2\Gamma(d/4)^4}{\Gamma(d/2)^2} \bigg[ \psi(1) + \psi(d/2) - 2 \psi(d/4) \bigg] \tilde g^2 + O(\epsilon,g^4,g_1^2 g^2) \,,\\
  \beta^{(2)} = & \; \epsilon \, D= 2 \frac{\Gamma(d/4)^2}{\Gamma(d/2)}
  + O(\epsilon,g^4,g_1^2 g^2) \,, \nonumber
\end{align}
 which reproduce the results of \cite{Benedetti:2019eyl} upon dividing all the couplings by $Z^2$.
 
\paragraph{Minimal subtraction.}
The minimal subtraction consists in fixing a series of counterterms for each coupling,
having ascending series of poles in 
$1/\epsilon$, with residues chosen such that the four-point 
functions
expressed in terms of the renormalized couplings $\tilde g,\tilde g_1,\tilde g_2$
do not have any poles in $1/\epsilon$.
The tetrahedron coupling is still trivial, as before. For the others we have:
\be \label{eq:lambdaMS}
\tilde \lambda_1 = \mu^{\epsilon}   \left( \tilde g_1 + \sum_{k\geq 1} \frac{B^{(k)}(\tilde g_1,\tilde g)}{\epsilon^k}   \right) \,,
\ee
where the $B^{(k)}(\tilde g_1,\tilde g)$ are $\epsilon$-independent.
A standard manipulation \cite{zinnjustin} then leads to:
\be \label{eq:betaMS-def}
\beta_1( \tilde g_1, \tilde g^2)  = -\epsilon \tilde g_1 + (\tilde g \partial_{\tilde g}+ \tilde g_1 \partial_{\tilde g_1} -1)B^{(1)}(\tilde g_1,\tilde g) \,.
\ee

The difference between minimal subtraction and Wilsonian scheme amounts to a mapping between renormalized couplings. 
In fact, in minimal subtraction we find:
\be
\mu^{-\epsilon} \Gamma^R_1 = -U^{(0)} \left( \tilde g \right)
 +  \tilde g_1   
 \frac{ \big[ 1 +S^{(0)}\left(  \tilde g \right) \big]^2 }{ 1 +  \tilde g_1  
 T^{(0)}\left(  \tilde g \right)} \equiv F(\tilde g_1,\tilde g)\,,
\ee
where we have used:
\be \label{eq:U-exp}
U \left( \tilde g \right) = \sum_{k\geq 0} \frac{U^{(k)}(\tilde g)}{\epsilon^k} \,,
\ee
and similar expansions for $S^{(0)}\left(  \tilde g \right)$ and $T^{(0)}\left(  \tilde g \right)$. The superscript here denotes the order of the pole in $\epsilon$, and each amplitude $U_r$ and so on has an expansion of the form \eqref{eq:U-exp}. Therefore, we have:
\be \label{eq:W-MS-map}
\tilde g_1^{\rm Wilson} = F(\tilde g_1^{\rm MS},\tilde g) \,.
\ee
We could in principle plug this transformation in Eq.~\eqref{eq:renseries}, find the expansion in Eq.~\eqref{eq:lambdaMS}, and write the beta function in terms of the one-vertex irreducible amplitudes. The result is rather cumbersome.

Let us instead compute directly the $\beta$ functions in minimal subtraction up to two loops (order $\lambda^3$). At this order, the beta functions of marginal couplings are scheme independent up to $O(\epsilon)$ terms because $F(\tilde g_1,\tilde g)= \tilde g_1+O(\tilde g^2)$. The bare series of $\Gamma^R_1$ at cubic order is:
\be
  \mu^{-\epsilon}\Gamma_1^R =  \mu^{-\epsilon} \tilde \lambda_1  + 
  (\mu^{-\epsilon} \tilde \lambda)^2 D - 
  ( \mu^{-\epsilon} \tilde \lambda_1)^2 D -
   2S_1 (\mu^{-\epsilon} \tilde \lambda_1)(\mu^{-\epsilon}\tilde \lambda)^2 +
   (\mu^{-\epsilon} \tilde \lambda_1)^3D^2  + O(\tilde \lambda^4,\tilde \lambda^2\tilde \lambda_1^2,\tilde \lambda_1^4) \,.
\ee
Its pole singularities arise from 
$D = \frac{D^{(1)}}{\epsilon} + D^{(0)} + O(\epsilon)$ 
and $S_1 = \frac{S_1^{(2)}}{\epsilon^2} + \frac{S_1^{(1)}}{\epsilon} + S_1^{(0)}+O(\epsilon) $, with 
$2S_1^{(2)} = (D^{(1)} )^2 $.
They are cancelled by choosing:
\be\label{eq:lambda1_MS}
\begin{split}
&\mu^{-\epsilon} \tilde \lambda  = \tilde g \,,  \\
&\mu^{-\epsilon} \tilde \lambda_1  =  \tilde g_1 - \frac{D^{(1)}}{\epsilon} \tilde g^2 +  \frac{D^{(1)}}{\epsilon}  \tilde g_1^2 -\left(\frac{2S_1^{(2)}}{\epsilon^2}-\frac{2S_1^{(1)} - 2D^{(1)} D^{(0) } }{\epsilon}\right) \tilde g^2 \tilde g_1
+   \frac{ ( D^{(1)} )^2}{ \epsilon^2 } \tilde g_1^3 + O( \tilde g^4) \,,
\end{split}
\ee
which in the limit $\epsilon\to 0$ leads to:
\be
F(\tilde g_1,\tilde g) = \tilde g_1 + D^{(0)} (\tilde{g}^2-\tilde{g_1}^2) - 2 S_1^{(0)} \tilde{g}^2 \tilde{g_1} + D^{(0)}{}^2 \tilde{g_1}^3
 + O(\tilde{g}^4,\tilde{g}^2\tilde{g_1}^2,\tilde{g_1}^4) \,.
\ee

From Eq.~\eqref{eq:lambda1_MS} we can read off $B^{(1)}(\tilde g_1,\tilde g)$, and plugging it into Eq.~\eqref{eq:betaMS-def}, we find up to two loops:
\be\label{eq:betaMS}
   \beta_t( \tilde g^2) = -\epsilon \tilde g \,,\qquad
   \beta_1( \tilde g_1, \tilde g^2)  = -\epsilon \tilde g_1 - 
   D^{(1)} \tilde g^2 + D^{(1)} \tilde g_1^2 +2 (2S_1^{(1)} - 2D^{(1)} D^{(0) } )
    \tilde g^2 \tilde g_1 \,.
\ee
The reader can check that these functions coincide with the Wilsonian beta functions in 
Eq.~\eqref{eq:betaWilson_t} and \eqref{eq:betaWilson}, up to terms that vanish for $\epsilon\to 0$.
At higher perturbative orders, even at $\epsilon=0$, the two schemes will differ. However, also in minimal subtraction, the $\beta$ functions at all orders have the form:
\begin{align} \label{eq:beta1a}
&\beta_1 (\tilde g_1,\tilde g) = - \epsilon  \tilde g_1 + 
       \bar\beta^{(0)} (\tilde{g}^2)  -2 \bar\beta^{(1)}( \tilde g^2) \; \tilde g_1 + 
        \bar\beta^{(2)} (\tilde g^2) \; \tilde g_1^2 \,,\\
&\beta_2 (\tilde g_2,\tilde g) = - \epsilon  \tilde g_2 + 
       \bar\beta^{(0)} (3\tilde g^2)  -2 \bar\beta^{(1)}( 3\tilde g^2) \; \tilde g_2 + 
        \bar\beta^{(2)} (3\tilde g^2) \; \tilde g_2^2 \,,
\end{align}
with $\bar\beta^{(i)} (\tilde g^2)$ the minimal subtraction versions of Eq.~\eqref{eq:betacoeffW}.

\paragraph{Fixed points and critical exponents.} Without loss of generality we consider $\tilde g>0$.
While the fixed point below exists at all orders in perturbation theory, we will restrict to the first non trivial order. The beta functions at one loop:
\be \label{eq:betas}
 \beta_t = -\epsilon \tilde g \,, \qquad 
 \beta_1 = -\epsilon \tilde g_1 + 
 2 \frac{\Gamma(d/4)^2}{\Gamma(d/2) }( \tilde g_1^2 - \tilde g^2) \,,\qquad
 \beta_2 = -\epsilon \tilde g_2 + 
 2 \frac{\Gamma(d/4)^2}{\Gamma(d/2) }( \tilde g_2^2 - 3\tilde g^2) \,,
\ee
admit at $\epsilon=0$, for any $\tilde g = \tilde g^{\star}$, a fixed point $\tilde g_1^{\star} = \tilde g^{\star}, \tilde g_2^{\star} = \sqrt{3} \, \tilde g^{\star}$
which is infrared attractive:
\be
 \partial_{\tilde g_1}\beta_1|_{\star} = 4 \frac{\Gamma(d/4)^2}{\Gamma(d/2) } \tilde g^{\star} >0 \,,\qquad
  \partial_{\tilde g_2}\beta_2|_{\star} = 4 \frac{\Gamma(d/4)^2}{\Gamma(d/2) }  \sqrt{3} \tilde g^{\star} >0 \,. 
\ee
The quartic operators acquire anomalous scaling $\Delta_a^{\star} = 4\Delta^{\star}_{\phi} + \delta h_a$ (with $a = t,1,2$).
Here $\Delta^{\star}_{\phi} = d/4 $ is the dimension of the field at the fixed point.
As the stability matrix $\partial_{\tilde g_a} \beta_b$ is triangular, $\delta h_a =  \partial_{\tilde g_a} \beta_a|_{\star}$ hence:
\be\label{eq:dim-an-quart}
 \delta h_t  = 0 \,, \qquad \delta h_1 = 4 \frac{\Gamma(d/4)^2}{\Gamma(d/2) } \tilde g^{\star}  \,,\qquad \delta h_2 =   4 \frac{\Gamma(d/4)^2}{\Gamma(d/2) }  \sqrt{3} \, \tilde g^{\star} \,.
\ee
We will come back to the corresponding scaling operators in Sec.~\ref{sec:scaling-op}, after having discussed composite operators and their mixing.

%%%%%%%%%%%%%%%%%%%%%%%%%%%%%%%%%%%%%%%%%%%%%%%%%%%%%%%%
\subsubsection{Quadratic couplings}
\label{sec:quadratic couplings}
%%%%%%%%%%%%%%%%%%%%%%%%%%%%%%%%%%%%%%%%%%%%%%%%%%%%%%%%
%\paragraph{The $\phi^2$ operator.}
Let us consider a $\phi^2$ perturbation of the critical theory. This comes to considering from the onset a mass parameter 
$m^{2\zeta} = m_{c}^{2\zeta} + \lambda_{\phi^2}$, with $m_c^{2\zeta}$ given by 
\eqref{eq:mbare}. With respect to our previous discussion we now need to add bi-valet vertices in the theory, corresponding to the insertion of $\lambda_{\phi^2}$. The back reaction of $\lambda_{\phi^2}$ on the flow of $g_1$ and $g_2$ can be neglected, as the ultraviolet behavior of the massive propagator 
$[(p^2+ \mu^2)^{\zeta} + \lambda_{\phi^2}]^{-1}$ is identical to that of the  massless one.

Let us consider the one-particle irreducible two-point function at zero external momentum $ \Gamma^R_{(2)}(0) =  G^{-1}_{\mu}(0)$. At one loop, only the tadpole with an insertion of a bi-valent vertex contributes:
\be
\Gamma^R_{(2)}(0)  = 
  \lambda_{\phi^2} - \lambda_2 \lambda_{\phi^2} \int\frac{d^dp}{(2\pi)^d} 
 \left(  \frac{1}{(p^2+\mu^2)^{\zeta}}\right)^2=   \lambda_{\phi^2} -
  \frac{ \lambda_2 \lambda_{\phi^2}  }{(4\pi)^{d/2}\Gamma(\zeta)^2} \mu^{-\epsilon} D \,,
\ee
with $D$ the integral of App.~\ref{app:int}. For both the Wilsonian prescription and in minimal subtraction, we get:
\be \label{eq:lambda_m}
  \tilde \lambda_{\phi^2}  = \mu^{d- 2 \Delta_{\phi}} 
  \bigg(   \tilde g_{\phi^2} + \frac{D^{(1)}}{\epsilon} \tilde g_2 \tilde g_{\phi^2}  \bigg) \,,
  \qquad  \beta_{\phi^2} = -(d-  2 \Delta_{\phi} ) \tilde g_{\phi^2}  
  +  D^{(1)} \tilde g_{\phi^2} \tilde g_2 \,,
\ee
which is valid for $\epsilon\ge 0$, and where $ 2\Delta_{\phi} = \frac{d - \epsilon}{2} $ is the classical dimension of the $\phi^2$ operator.
Observe that $\tilde g_{\phi^2}=0$ is always a fixed point of this equation. For $\epsilon=0$ (when the dimension of the field becomes $\Delta_{\phi}^{\star} =d/4$), the beta function becomes:
\be \label{eq:beta-mass}
 \beta_{\phi^2} \, = \, - ( d  - 2\Delta_{\phi}^{\star} ) \; \tilde g_{\phi^2}  
  +   2 \frac{\Gamma(d/4)^2}{\Gamma(d/2) }   \tilde g_{\phi^2} \tilde g_2 \,,
\ee
and therefore,  close to the fixed point $\tilde g_2^\star =  \sqrt{3} \, \tilde g^{\star}$ the operator $\phi^2$ acquires
the anomalous scaling
\be \label{eq:phi2-andim}
 \delta h_{\phi^2} =  2 \frac{\Gamma(d/4)^2}{\Gamma(d/2) }  \tilde g_2^{\star}
  =    2 \frac{\Gamma(d/4)^2}{\Gamma(d/2) }  \sqrt{3} \,  \tilde g^{\star} \,,
  \qquad \Delta_{\phi^2}^{\star} = 2\Delta^{\star}_{\phi} + \delta h_{\phi^2}  \,,
\ee
reproducing the anomalous dimension found in \cite{Benedetti:2019eyl} by diagonalizing the four-point kernel.\footnote{For the comparison, one should note that in this paper we explicitly separated the imaginary unit $\im$ from the tetrahedral coupling, see Eq.~\eqref{eq:action}.
This factor was instead included in the coupling in \cite{Benedetti:2019eyl}.}

The spin-zero bilinear operators of the type: $\phi(-\partial^2)^n \phi$ can be treated similarly.
We present a detail computation of the corresponding beta functions in Appendix \ref{app:bilinear}.
The result again reproduces the anomalous dimensions of this type of operators derived in \cite{Benedetti:2019eyl} by diagonalizing the four-point kernel.

%%%%%%%%%%%%%%%%%%%%%%%%%%%%%%%%%%%%%%%%%%%%%%%%%%%%%%%%
%%%%%%%%%%%%%%%%%%%%%%%%%%%%%%%%%%%%%%%%%%%%%%%%%%%%%%%%
\section{Composite Operators}
\label{sec:composite operator}
%%%%%%%%%%%%%%%%%%%%%%%%%%%%%%%%%%%%%%%%%%%%%%%%%%%%%%%%
%%%%%%%%%%%%%%%%%%%%%%%%%%%%%%%%%%%%%%%%%%%%%%%%%%%%%%%%

Consider a field theory described by the action:
\be \label{eq:pert-action}
 S [\phi]= S_0[\phi] + S^{\rm int}[\phi]\;, \qquad
  S^{\rm int}[\phi] = \sum_a \lambda_a \int d^d x \; {\cal O}_a(x) \,,
\ee
where $S_0$ is some free quadratic action and the bare perturbation $S^{\rm int}[\phi]$ is a sum over local operators ${\cal O}_a$ with associated bare couplings $\lambda_a$.
To simplify the discussion we assume that we do not have a wave function renormalization, which is in particular the case for our model. 
We regulate the logarithmic ultraviolet divergences by some dimensional continuation and we assume that the power ultraviolet divergences have already been taken care of. We denote $\mu$ the infrared regulator (which could be a cutoff on the covariance) and 
$\Delta_a$ the canonical dimension of the operator ${\cal O}_a(x)$. 

In order to eliminate the logarithmic ultraviolet divergences we replace the bare couplings by renormalized couplings plus counterterms:
\be
  \lambda_a = \mu^{d- \Delta_a} \bigg(  g_a + \sum_{k\ge 1}  \frac{B_a^{(k)}(\{g_b\})}{\epsilon^k} \bigg) \,,
\ee
where the counterterms of $\lambda_a$ can depend on all the renormalized couplings 
$\{g_b\}$. The counterterms can have finite parts in $\epsilon$, but do not have if we use minimal subtraction.
The renormalized action is obtained by substituting the bare couplings in terms of the renormalized ones and the infrared scale $\mu$:
\be 
S [\phi]= S_0[\phi] + S^{\rm int}[\phi] \;,\qquad
S^{\rm int}[\phi]=  \sum_a \mu^{d-\Delta_a} \bigg( g_a   + \sum_{k\ge 1} \frac{B_a^{(k)}(\{g_b\})}{\epsilon^k} \bigg) \int d^d x \; {\cal O}_a(x) \,,
\ee
and the counterterms are chosen such that the connected correlations:
\be\label{eq:n-point}
\Braket{\phi(x_1) \dots \phi(x_n)}_{c} =  \frac{1}{\int [d\phi]\; e^{-S(\phi)}}
\int [d\phi]   \;\phi(x_1) \dots \phi(x_n)  \; e^{-S[\phi]}
 - ( {\rm disconnected} ) \, ,
\ee
have no poles in $1/\epsilon$ when expressed in terms of the renormalized couplings.

\subsection{Operator mixing}

The (integrated) renormalized operator $\big[{\cal O}_a\big]$ is the derivative of the renormalized action with respect to the dimensionless renormalized coupling $g_a$:
\be
\int d^d x  \; \big[{\cal O}_a\big](x) \equiv 
 \mu^{- ( d-\Delta_a) } \,
\frac{\partial S^{\rm int}}{\partial g_a} \,,
\ee
and it is easy to check that when acting on a connected correlation the derivative with respect to $g_a$ brings down an $\big[{\cal O}_a\big]$ operator:
\be \label{eq:defOren}
 - \frac{\partial}{\partial g_a} \Braket{\phi(x_1) \dots \phi(x_n)}_{c} 
  =   \mu^{d-\Delta_a} \Braket{\phi(x_1) \dots \phi(x_n) \; \int d^d x  \; 
  \big[{\cal O}_a\big](x)  }_{c} \,.
\ee
As $\Braket{\phi(x_1) \dots \phi(x_n)}_{c}$ is finite, i.e.\ it has no poles in $1/\epsilon$, its derivative is also finite, with the possible exception of some critical points.

Observe that the renormalized operator $ \big[{\cal O}_a\big] ={\cal O}_a + \dots $ has the same dimension $\Delta_a$ as its bare counterpart $ {\cal O}_a $. 
The bare operators are linear combinations of the renormalized ones: 
\be
{\cal O}_a = \sum_b \mu^{\Delta_a} \, M_{ab} \, \mu^{-\Delta_b} \, \big[ {\cal O}_b\big] \,, \qquad 
\big[ {\cal O}_a\big] = \sum_b   \mu^{\Delta_a} \, Z_{ab} \, \mu^{ - \Delta_b} \,{\cal O}_b \;,
\ee 
where $M$ is the mixing matrix and $Z=M^{-1}$. The matrix elements $M_{ab}$ and $Z_{ab}$ are dimensionless. For a free theory, that is, neglecting all the radiative corrections, the mixing matrix is  $M_{ab} =  \delta_{ab} = Z_{ab}$. In the interacting case
$M$ is determined by observing that, as insertions in correlation functions, we have:
\be \label{eq:M-def}
\big[{\cal O}_a\big] \, = \, \mu^{- (d-\Delta_a)} \, \sum_{b} \frac{\partial \lambda_b}{\partial g_a} \, {\cal O}_b \, 
\Rightarrow\quad 
Z_{ab} = \frac{\partial \lambda_b}{\partial g_a}  \, \mu^{\Delta_b - d} \, = (M^{-1})_{ab} \, .
\ee
Denoting $\Delta$ the diagonal matrix with entries $\Delta_a$, and organizing the couplings $\lambda$ and the beta functions in (row) vectors,
we can obtain the $\beta$ functions in terms of the mixing matrix as:
\be \label{eq:M-inv}
  0 = \mu \frac{d}{d\mu} \lambda= \lambda \, (d-\Delta) + \beta \, Z \, \mu^{d-\Delta} \quad
 \Rightarrow \quad \beta \, = \, - \lambda  \, (d-\Delta) \, \mu^{\Delta-d} \, M \,.
\ee

Below we will deal with the energy-momentum tensor. The contribution of the interaction to the trace of the energy-momentum tensor is:
\be\label{eq:useful}
 \sum_{b} \lambda_b (d-\Delta_b) {\cal O}_b = \lambda (d-\Delta) \mu^{\Delta}  M
\mu^{-\Delta} [ {\cal O}] = - \beta \mu^{d-\Delta} [{\cal O}]  \;.
\ee

%%%%%%%%%%%%%%%%%%%%%%%%%%%%%%%%%%%%%%%%%%%%%%%%%%%%%%%%
\subsection{Stability matrix and scaling operators}
\label{sec:scaling-op}
%%%%%%%%%%%%%%%%%%%%%%%%%%%%%%%%%%%%%%%%%%%%%%%%%%%%%%%%
The renormalized $n$-point functions \eqref{eq:n-point} of our model satisfy, in the limit $\mu\to 0$, the usual Callan-Symanzik (CS) equation with zero anomalous dimension:
\be\label{eq:CSeq}
\left(\mu \partial_\mu + \sum_a \beta_a \partial_{g_a} \right) \Braket{\phi(x_1) \dots \phi(x_n)}_{c} = 0 \,.
\ee
Although this is a very familiar equation, some remarks are due. 
In the renormalized expansion any correlation is expressed as an explicit function of $\mu$ and $g$ with no poles in $1/\epsilon$. The left-hand side of the CS equation is just the total derivative with respect to $\mu$ of the correlation.

The right-hand side of the CS equation is exactly zero only if the fields acquire no anomalous dimension and if one uses a massless renormalization scheme in which the bare expansion is $\mu$ independent. 
This is  \emph{not} the case for our regularization scheme, where $\mu$ is an explicit infrared cutoff in the covariance of the bare action. When expressed in terms of bare constants, the correlations will depend on $\mu$ and the right-hand side of Eq.~\eqref{eq:CSeq}, rather than being zero, gives 
\be \Braket{-(\mu \partial_\mu S_0[\phi]) \phi(x_1) \dots \phi(x_n)}_{c} =-\zeta \mu^2 \int d^d x \Braket{ (\phi(\partial^2+\mu^2)^{\zeta-1} \phi)(x)\phi(x_1) \dots \phi(x_n)}_{c}  \;.
\ee
This is an $n$-point function with a two-valent vertex insertion, and it goes to zero faster than the left-hand side of \eqref{eq:CSeq} when the infrared regulator $\mu$ descends below the scale of the external momenta of the correlation.\footnote{Such situation is similar to that concerning the CS equation for massive theories, with $\mu$ replaced by the physical mass (see for example Ref.~\cite{Amit:1984ms}).} In other words, the CS equation with zero on the right-hand side holds in any schemes  in the deep infrared regime. This is due to the fact that the external momenta become the true infrared regulators and the correlation stops depending on $\mu$. The reader will note that this is in fact what we did in Section \ref{sec:2pt} in order to solve self-consistently for the full two point function in the case $\epsilon = 0$. 

Notice that there exists a well established renormalization scheme in which the bare functions are truly $\mu$ independent: the subtraction at momentum scale $\mu$, {\it \`a la} Gell-Mann and Low. In this case one does not introduce any infrared regulator in the bare action, but notices that the correlation functions are infrared convergent provided that the external momenta are non exceptional (i.e. no subset of external momenta adds up to zero). One eliminates the poles in $1/\epsilon$ by using a renormalization condition at scale $\mu$, such as $\Gamma^{(4)}(p)|_{p=\bar{p}} = \mu^{\epsilon} g$, with $\bar{p}_i\cdot\bar{p}_j = \frac{\mu^2}{4}(4\delta_{ij}-1)$, and solving for the bare coupling $\lambda = \mu^{\epsilon} (g  +\dots )$. As the $\beta$ functions at two loops are scheme independent, one can safely use the CS equation \eqref{eq:CSeq} up to this order in any scheme.

We now use the CS equation as stated in Eq.~\eqref{eq:CSeq} to study the scaling operators close to an infrared fixed point. Acting on the CS equation with a derivative $-\partial_{g_b}$, we obtain:
\be \label{eq:dgCS}
0 =\sum_a  \bigg[\delta_{ab}\, \mu \partial_\mu + (\partial_{g_b}\beta_a) + \beta_a \partial_{g_b} \bigg] \mu^{d-\Delta_a}\int d^d x   \Braket{\phi(x_1) \dots \phi(x_n)   \big[{\cal O}_a\big](x)  }_{c} \;.
\ee
At a fixed point we have $\beta_a|_{\star}=0$ hence the third term drops out. Defining the stability matrix $Y_{ba} = \partial_{ g_b} \beta_a|_{\star}$, the CS equation becomes, in matrix notation (aranging the operators $ \big[{\cal O}\big]$ in a column vector) and dropping the integral over $x$:
\be
0= \bigg[  \mu \partial_\mu+d  + Y \bigg] \mu^{-\Delta}  \;   \Braket{  \big[{\cal O}\big](x)  \, \phi(x_1) \dots \phi(x_n)    }_{c}\,.
\ee
Assuming that the stability matrix is diagonalizable, $Y = U \nu U^{-1}$, with $\nu$ being by definition the (diagonal matrix of) critical exponents, we find:
\be
0=  \big( \mu \partial_\mu +d + \nu \big) U^{-1}  \, \mu^{-\Delta} \big[{\cal O}\big](x) \,,
\ee
in the sense of operators inserted in correlation functions. This allows us to identify scaling operators at the fixed point, that is the operators which satisfy:
\be
\mu\frac{d}{d\mu} \bigg[ \big\{ {\cal O}_b \big\}(x)  \bigg] = 0 \;.
\ee 
In matrix notation, they write:
\be \label{eq:scal-op-def}
\big\{ {\cal O} \big\}(x) \, = \, \mu^{d + \nu}  \,  U^{-1} \, \mu^{-\Delta}\big[{\cal O}\big](x)  
\, = \, \mu^{d + \nu}  \,  U^{-1} \, Z \, \mu^{-\Delta} {\cal O} (x)
\, \equiv \, \mu^{\Delta^\star} \big\{ {\cal O} \big\}'(\mu x)  \,,
\ee
where the (inverse) mixing matrix $Z$ is evaluated at the fixed point, and we denoted by prime the dimensionless version of the operator.
The scaling operators are the eigenvectors of the dilatation, with eigenvalue $\Delta_b^{\star} = d+\nu_b $, which is the scaling dimension 
of $\big\{ {\cal O}_b \big\} $.
We write the fixed-point scaling dimensions in terms of canonical and anomalous dimensions as  $\Delta^\star_b = \Delta_b+\delta h_b$.
In the case of our quartic operators we have $\Delta_b=d$ at $\epsilon=0$, hence the critical exponent and the anomalous dimension coincide.

The stability matrix can be computed further from Eq.~\eqref{eq:M-def} and \eqref{eq:M-inv}. In matrix notation we get:
\be
 \begin{split}
   Y= \partial_g \beta = - Z\mu^{d-\Delta} (d-\Delta) \mu^{\Delta -d} Z^{-1} + \lambda ( d-\Delta) \mu^{\Delta -d} Z^{-1} (\partial_g Z) \,Z^{-1} \;.
 \end{split}
\ee
Observing that, as $Z$ does not depend explicitly on $\mu$, the matrix of anomalous dimensions can be written as:
\be
 \gamma =  - \mu \left( \frac{d}{d\mu} Z  \right) Z^{-1} = 
  -  \beta_a ( \partial_{g_a} Z ) Z^{-1} \;,
\ee
and using $\partial_{g_a} Z_{bc}= \partial_{g_b} Z_{ac}$, we conclude that:
\be
 d + Y = Z \Delta Z^{-1} + \gamma \;.
\ee

%%%%%%%%%%%%%%%%%%%%%%%%%%%%%%%%%%%%%%%%%%%%%%%%%%%%%%%%
\subsection{Action, mixing, and scaling operators at first order}
%%%%%%%%%%%%%%%%%%%%%%%%%%%%%%%%%%%%%%%%%%%%%%%%%%%%%%%%
Gathering the first order results in Eq.~\eqref{eq:lambda1_MS} and~\eqref{eq:lambda_m}, the minimally-subtracted action, with bare mass $m^{2\zeta} = m_{c}^{2\zeta} + \lambda_{\phi^2}$ and $m_c^{2\zeta}$ given by \eqref{eq:mbare}, and up to quadratic order in the couplings, is:
	\begin{align}
	\label{eq:one-loop-action}
		S &= \frac{1}{2} \int d^dxd^dy \;  \phi_{\mba} (x)  C_{\mu}^{-1}(x,y) \phi_{\mba}(y)
		+ \frac{1}{2} \bigg( \lambda_{\phi^2} - \lambda^2 \int d^d u \; C_{\mu}(u)^3 \bigg)  \int d^d x  \; \phi^2(x) \nonumber\\
		&\quad + \frac{1}{4} \int d^dx \left[ \lambda \phi^4_t(x) + \lambda_1 \phi^4_1(x) + \lambda_2 \bigg(  \phi^4_2(x) - 2 C_{\mu}(0) \; \phi^2(x) \bigg) \right] \, , 
	\end{align}
where $\Delta_{\phi} = \frac{d-\epsilon}{4}$ and the bare couplings at first order are:
\begin{align} \label{eq:couplings-renorm}
 &  \lambda_{\phi^2}  =  \mu^{d-2\Delta_{\phi}}g_{\phi^2} 
 \bigg( 1 + Q \, \frac{g_2 }{\epsilon} \bigg)  \,,
  & & \lambda  = \mu^{d-4\Delta_{\phi}} g \,,
    \\
& \lambda_1  = \mu^{d-4\Delta_{\phi}}
    \bigg(g_1 - Q \, 
    \frac{ g^2 }{ \epsilon }  + Q \, 
    \frac{ g^2_1 }{ \epsilon } \bigg) \,, & & \lambda_2  = \mu^{d-4\Delta_{\phi}}
    \bigg(g_2 - 3 Q \, 
    \frac{ g^2 }{ \epsilon } + Q \, 
    \frac{ g^2_2 }{ \epsilon }\bigg) \,, \nonumber
\end{align}
with:
\be
Q \equiv \frac{2}{ (4\pi)^{d/2} \Gamma(d/2) } \,.
\ee
Observe that in the second line of Eq.~\eqref{eq:one-loop-action} we can recognize the Wick ordered double-trace interaction at leading order in $N$ and up to a vacuum term\footnote{The standard definition of Wick ordering \cite{salmhofer:book} gives: 
\[
 :\phi^4_2(x): = \exp\bigg\{ 
-\frac{1}{2} \int d^dx d^dy \;  \partial_{\phi_{\mba}(x)} C_{\mu}(x,y) \partial_{\phi_{\mba}(y)} \bigg\} \phi^4_2(x) 
=
  \phi^4_2(x) - 2 (1+\frac{2}{N^3}) \phi^2(x)  \; C_{\mu}( 0) + 2 (N^3+2) C_{\mu}( 0)^2 \,.
\]
}
$
:\phi^4_2:\; \equiv \, \phi^4_2(x) - 2 C_{\mu}( 0) \; \phi^2(x) $.
At this order in $1/N$ it is superfluous to Wick order the other quartic interactions.

Contrary to the previous discussion, we have explicitly taken into account the counterterms necessary for eliminating the mass (power law) ultraviolet 
divergence\footnote{Using $C_{\mu}(x) = \mu^{2\Delta_{\phi}}C_1(\mu x)$, we have $ C_{\mu}(0) = \mu^{2\Delta_{\phi}}  C_{1}(0) , \; 
  \int d^du \; C_{\mu}(u)^3 = \mu^{6\Delta_{\phi}-d} \int d^du' \, C_1(u')
$.}. The renormalized operators are computed by taking the derivatives of the renormalized action with respect to the renormalized couplings:
%\be
 \begin{align}\label{eq:one-loop-mix}
  &  \big[\phi^2\big] (x) = Z_{\phi^2} \phi^2(x) \,, \qquad  \big[\phi^4_1\big](x)  = Z_{\phi^4_1} \phi^4_1(x) \,, \qquad
    \big[\phi^4_2\big](x)  = Z_{\phi^4_2} : \phi^4_2(x) : + Z_{\phi^4_2 ; \phi^2} \phi^2(x) \,, \\
 & \big[\phi^4_t\big] (x)  = \phi^4_t (x) + Z_{\phi^4_t;\phi^4_1} \phi^4_1 (x) + Z_{\phi^4_t;\phi^4_2} :\phi^4_2(x): + 
  Z_{\phi^4_t;\phi^2} \phi^2(x) \,, \nonumber
 \end{align}
%\ee
where, contrary to the previous section, we included the dimensions directly in the (inverse) mixing matrix elements:
 \be\label{eq:one-loop-mix1}
 \begin{split}
   &  Z_{\phi^2}  = 1 + Q \, \frac{g_2 }{\epsilon}  \,, \quad
 Z_{\phi^4_1} =   1 + 2 Q \, \frac{g_1 }{\epsilon}  \,, \quad Z_{\phi^4_2} = 1 + 2 Q \, \frac{g_2 }{\epsilon}  \,, \quad 
  Z_{\phi^4_2 ; \phi^2} =   \mu^{ 2\Delta_{\phi} } \, 2 Q \frac{ g_{\phi^2} }{\epsilon} 
  \,, \crcr
  & Z_{\phi^4_t;\phi^4_1} = -2Q \frac{g}{\epsilon} \,, \quad
  Z_{\phi^4_t;\phi^4_2} = -6 Q \frac{g}{\epsilon} \,,  
\quad   Z_{\phi^4_t;\phi^2} = 
     - \mu^{d-4\Delta_{\phi}} 4g \int d^du C(u)^3  \,.
 \end{split}
 \ee

Going back to our beta functions \eqref{eq:betas} and \eqref{eq:beta-mass}, we find the following diagonalizing transformation for the stability matrix, corresponding to a vector ordered as $\{g_{\phi^2},g,g_1,g_2\}$:
\be 
U^{-1} = 
\begin{pmatrix}
1 & 0 & 0 & 0\\
 0 & 1 & 1 & \sqrt{3} \\
 0 & 0 & 1 & 0\\
 0 & 0 & 0 & 1
\end{pmatrix} \,, 
\ee
and using \eqref{eq:scal-op-def} and \eqref{eq:one-loop-mix}, the scaling operators are:
\begin{align}\label{eq:one-loop-scale-op}
\{ \phi^2 \}(x) &= \mu^{ \delta h_{\phi^2}} Z_{\phi^2} \phi^2(x) \,, \crcr
 %\Psi_t (x)  
\{ \phi^4_t \}(x) &=  \phi^4_t (x) + \left(Z_{\phi^4_1} +Z_{\phi^4_t;\phi^4_1}\right) \phi^4_1 (x) \crcr
&\qquad + \left(\sqrt{3} Z_{\phi^4_2}+Z_{\phi^4_t;\phi^4_2}\right) :\phi^4_2(x):
+ \left(\sqrt{3} Z_{\phi^4_2 ; \phi^2}+Z_{\phi^4_t;\phi^2}\right) \phi^2(x) \,, \crcr 
 %\Psi_1(x) 
\{ \phi^4_1 \}(x) &= \mu^{\delta h_1}  Z_{\phi^4_1}  \phi^4_1(x)\,, \\
 %\Psi_2(x) 
\{ \phi^4_2 \}(x) &= \mu^{\delta h_2}  Z_{\phi^4_2} : \phi^4_2(x): \,.
  \nonumber
 \end{align}

%%%%%%%%%%%%%%%%%%%%%%%%%%%%%%%%%%%%%%%%%%%%%%%%%%%%%%%%
%%%%%%%%%%%%%%%%%%%%%%%%%%%%%%%%%%%%%%%%%%%%%%%%%%%%%%%%
\section{$D=d+p$ Dimensional Embedding}
\label{sec:embedding}
%%%%%%%%%%%%%%%%%%%%%%%%%%%%%%%%%%%%%%%%%%%%%%%%%%%%%%%%
%%%%%%%%%%%%%%%%%%%%%%%%%%%%%%%%%%%%%%%%%%%%%%%%%%%%%%%%
In this section, we give a proof of the conformal symmetry of the model (\ref{eq:action}) at the infrared fixed point,
following the discussion of Ref.~\cite{Paulos:2015jfa}.
The main idea is that the model (\ref{eq:action}) can be written with a standard short-range kinetic term by embedding it in $D=d+p$ (where $p=2-2\zeta$) dimensional space.
In this enlarged space, the action becomes:
	\begin{align} 
		 S[\Phi] \, &= \, \frac{1}{2} \int d^DX \, \Big( \partial_M \Phi_{\mba}(X) \Big)^2 \nonumber\\
		 & \ + \, \frac{1}{4} \int_{y=0} d^d x \,
		 \left[ \im \lambda \hat{\delta}^t_{\mba\mbb\mbc\mbd} + \lambda_1 \hat{P}^{(1)}_{\mba\mbb; \mbc\mbd} + \lambda_2 \hat{P}^{(2)}_{\mba\mbb; \mbc\mbd } \right]
		\Phi_{\mba}(x) \Phi_{\mbb}(x) \Phi_{\mbc}(x) \Phi_{\mbd}(x) \, ,
	\label{eq:action2}
	\end{align}
where the $D$-dimensional coordinates are labeled by $X^M=(x^{\mu}, y^m)$ and the original field is obtained by $\Phi|_{y \to 0}=\phi$.
In this $D$-dimensional space, one can write down a local energy-momentum tensor:
	\begin{align}
		T_{MN} \, &= \, \sum_{\mba} \left[ \partial_M \Phi_{\mba} \partial_N \Phi_{\mba} - \frac{1}{2} \delta_{MN} (\partial_K \Phi_{\mba})^2 \right] 
		\, - \, \frac{\im \lambda}{4} \, \delta_{MN}^{\parallel} \delta^p(y) \sum_{\mba,\mbb,\mbc,\mbd} \hat{\delta}^t_{\mba\mbb\mbc\mbd} \, 
		\Phi_{\mba} \Phi_{\mbb} \Phi_{\mbc} \Phi_{\mbd} \nonumber\\
		&\qquad - \,\frac{\delta_{MN}^{\parallel} \delta^p(y)}{4} \sum_{\mba,\mbb,\mbc,\mbd} 
		\left( \lambda_1 \hat{P}^{(1)}_{\mba\mbb; \mbc\mbd} + \lambda_2 \hat{P}^{(2)}_{\mba\mbb; \mbc\mbd } \right)
		\Phi_{\mba} \Phi_{\mbb} \Phi_{\mbc} \Phi_{\mbd} \, ,
	\end{align}
where $\delta_{MN}^{\parallel}=\delta_{\mu\nu}$ if both indices are in the $d$-dimensional space, and zero otherwise.
We also introduce the orthogonal projector  $\delta_{MN}^{\perp}=\delta_{MN}-\delta_{MN}^{\parallel}$.
Now we can write the divergence and trace of the energy-momentum tensor as:
	\begin{align}
		\partial^M T_{MN} \, &= \, - \, E_N \, + \, \delta^p(y) \, \delta_{MN}^{\perp} D^M  \, , \\
		\label{eq:Tclassic}
		T^M{}_M \, = &  \, - \, \Delta_{\phi} E \, + \, \left( \frac{1}{2} - \frac{D}{4} \right) \sum_{\mba} \partial_K^2 \Phi_{\mba}^2 \nonumber\\
		&\;  - \, \frac{\epsilon \, \delta^p(y)}{4} \sum_{\mba,\mbb,\mbc,\mbd} 
		\left(\im \lambda \, \hat{\delta}^t_{\mba\mbb\mbc\mbd} + \lambda_1 \hat{P}^{(1)}_{\mba\mbb;\mbc\mbd} + \lambda_2 \hat{P}^{(2)}_{\mba\mbb;\mbc\mbd} \right)
		\Phi_{\mba} \Phi_{\mbb} \Phi_{\mbc} \Phi_{\mbd} \, ,
	\end{align}
with 
	\begin{align}
		E \, &= \, \sum_{\mba} \Phi_{\mba} \left[ - \partial_K^2 \Phi_{\mba} \, + \, \delta^p(y) \sum_{\mbb,\mbc,\mbd}
		\left(\im \lambda \, \hat{\delta}^t_{\mba\mbb\mbc\mbd} + \lambda_1 \hat{P}^{(1)}_{\mba\mbb;\mbc\mbd} + \lambda_2 \hat{P}^{(2)}_{\mba\mbb;\mbc\mbd} \right)
		\Phi_{\mbb} \Phi_{\mbc} \Phi_{\mbd} \right] \, , \\
		E_N \, &= \, \sum_{\mba} \partial_N \Phi_{\mba} \left[ - \partial_K^2 \Phi_{\mba} \, + \, \delta^p(y) \sum_{\mbb,\mbc,\mbd}
		\left(\im \lambda \, \hat{\delta}^t_{\mba\mbb\mbc\mbd} + \lambda_1 \hat{P}^{(1)}_{\mba\mbb;\mbc\mbd} + \lambda_2 \hat{P}^{(2)}_{\mba\mbb;\mbc\mbd} \right)
		\Phi_{\mbb} \Phi_{\mbc} \Phi_{\mbd} \right] \, , \\
		D^N \, &= \, \sum_{\mba,\mbb,\mbc,\mbd}
		\left(\im \lambda \, \hat{\delta}^t_{\mba\mbb\mbc\mbd} + \lambda_1 \hat{P}^{(1)}_{\mba\mbb;\mbc\mbd} + \lambda_2 \hat{P}^{(2)}_{\mba\mbb;\mbc\mbd} \right)
		(\partial_N \Phi_\mba) \, \Phi_{\mbb} \Phi_{\mbc} \Phi_{\mbd}  %\, , \qquad ({\rm when}\ N = \nu) 
		\,.
	\end{align}	
We note that $E$ and $E_N$ are proportional to the equation of motion, so they vanish on-shell.
Therefore, on-shell the trace of the energy momentum tensor, Eq.~\eqref{eq:Tclassic} is equal to a double divergence, $\partial_K^2 \phi_{\mathbf{a}}^2$, plus a term proportional to $\epsilon$, and thus the theory is classically conformal invariant at $\epsilon=0$.
However, due to radiative corrections, the $\Phi^4|_{y=0}=\phi^4$ operators lead to infinities, hence they need to be renormalized as we discussed in the previous section.
Using Eq.~\ref{eq:useful} we arrive at the following expression for the trace of the energy-momentum tensor:
	\be\label{eq:trace}
	\begin{split}
		T^M{}_M \, &= \, - \, \Delta_{\phi} E \, + \, \left( \frac{1}{2} - \frac{D}{4} \right) \sum_{\mba} \partial_K^2 \Phi_{\mba}^2 \\
		&\quad \ + \, \frac{\mu^{\epsilon} \, \delta^p(y)}{4} \bigg[  \beta_t(g) \, \big[\phi_t^4\big](x)
		+ \beta_1(g, g_1)\, \big[\phi_1^4\big](x) + \beta_2(g, g_2)\, \big[\phi_2^4 \big](x) \bigg] \, ,
		\end{split}
		\ee
where we observe that in terms of the renormalized operators, the last terms survive at $\epsilon=0$, leading in general to a breaking of conformal invariance.
In order to simplify the notation we have omitted the bare mass term. We can include this term (restricted to $y=0$) in Eq.~\eqref{eq:action2} and tune the renormalized mass to zero. The trace of the energy momentum tensor still writes in terms of beta functions using Eq.~\eqref{eq:useful}. To keep things simple, we will continue ignoring this mass term and its mixing with the quartic operators.

The dilatation and special conformal transformation currents are constructed from the energy-momentum tensor as:
	\begin{equation}
		\mathcal{D}_M \, = \, T_{MN} X^N \, , \qquad \mathcal{C}_M{}^N \, = \, T_{MK} (2 X^K X^N - \delta^{KN} X^2 ) \, .
	\end{equation}
Their divergences are found as:
	\begin{align}
		\partial^M \mathcal{D}_M \,& = \, - X^M E_M + T^M{}_M \, , \\
		\partial^M \mathcal{C}_M{}^N \, &= \, - (2 X^K X^N - \delta^{KN} X^2 ) (E_K - \delta^p(y)\delta_{MK}^{\perp} D^M ) \, + \, 2 X^N T^M{}_M \, .
	\end{align}
Inserting these operators in a renormalized $n$-point function and integrating over the insertion point, up to a possible boundary term we obtain:
	\begin{gather}
		\int d^D X \Braket{ \Big( - X^M E_M(X) + T^M{}_M(X) \Big) \Phi(X_1) \cdots \Phi(X_n) } \, = \, 0 \, , \\
		\int d^D X \Braket{\bigg[ (2 X^K X^N - \delta^{KN} X^2 ) (E_K(X) - \delta^p(y) \delta_{MK}^{\perp} D^M(X)) \, - \, 2 X^N T^M{}_M \bigg] \Phi(X_1) \cdots \Phi(X_n) } \, = \, 0 \, .
	\end{gather}
The Schwinger-Dyson equations involving $E$ and $E_M$ are obtained from the path integral expression for the $n$-point function by the field redefinitions  $\Phi\to \Phi (1+\delta\Phi)$ and $\Phi\to \Phi +(\partial_M\Phi)\delta\Phi$, respectively. The result is \cite{Brown:1979pq}:
\begin{align} \label{eq:SD-eom}
		\Braket{ E(X)  \Phi(X_1) \cdots \Phi(X_n) }  \, &= \, \sum_{i=1}^n \, \delta(X-X_i) \,  \Braket{\Phi(X_1) \cdots \Phi(X_n) }  \, , \\
		 \label{eq:SD-eom2}
		 \Braket{E_M(X)  \Phi(X_1) \cdots \Phi(X_n) }  \, &= \, \sum_{i=1}^n \, \delta(X-X_i) \, \frac{\partial \ }{\partial X_i^M} \,  \Braket{\Phi(X_1) \cdots \Phi(X_n) }  \, .
\end{align}
Using these equations, together with the expression of the trace of the energy-momentum tensor (\ref{eq:trace}), 
we obtain the Ward identities for the dilatation and special conformal transformation currents.
Since the $n$-point function behaves continuously in the limit $y \to 0$ \cite{Paulos:2015jfa}, we can write down the Ward identities in the original $d$-dimensional space as:
	\begin{align}
		& \sum_{i=1}^n \Big[ x_i \cdot \partial_{x_i} + \Delta_{\phi} \Big] \Braket{\phi(x_1) \cdots \phi(x_n) }  \nonumber\\
		&\quad= \, \frac{\mu^{\epsilon}}{4} \int d^dx \Braket{ \bigg[  \beta_t(g) \,  \big[\phi_t^4\big](x)
		+ \beta_1(g, g_1)\, \big[\phi_1^4\big](x) + \beta_2(g, g_2)\, \big[\phi_2^4\big](x) \bigg] \phi(x_1) \cdots \phi(x_n) } \, ,
        \label{eq:scaleWI}
	\end{align}
and 
	\begin{align}
		& \sum_{i=1}^n \left[ (2 x_i^{\mu} x_i^{\nu} - \delta^{\mu\nu} x_i^2) \frac{\partial}{\partial x_i^{\mu}} + 2\Delta_{\phi} \, x_i^{\nu} \right] \Braket{\phi(x_1) \cdots \phi(x_n) }  \nonumber\\
		&\quad = \, \frac{\mu^{\epsilon}}{2} \int d^dx \, x^{\nu} \Braket{\bigg[  \beta_t(g) \, \big[\phi_t^4\big](x)
		+ \beta_1(g, g_1)\, \big[\phi_1^4\big](x) + \beta_2(g, g_2)\, \big[\phi_2^4\big](x) \bigg] \phi(x_1) \cdots \phi(x_n) }  \, .
	\end{align}

These Ward identities are well defined even at $\epsilon=0$  and the right-hand sides survives in this limit. We thus get in general a breaking of scale and special conformal invariance due to renormalization. At a fixed point, as $\beta_t(g)=\beta_1(g,g_1)=\beta_2(g,g_2)=0$,  the invariance is restored.
In fact, one should check that the integral on the right-hand side of the Ward identities does not blow up when we approach the fixed point as the inverse of the beta functions or worse. This subtlety is discussed in detail in Ref.~\cite{Paulos:2015jfa} and that discussion goes through in our case.\footnote{In our case, we can actually understand the absence of such singularity on the right-hand side of the scale Ward identity \eqref{eq:scaleWI} thanks to the presence of a small parameter, the tetrahedron coupling $g$. First, we can express the right-hand side of \eqref{eq:scaleWI} by means of \eqref{eq:defOren}, thus reducing the scale Ward identity to a Callan-Symanzik equation. Then, for infinitesimal $g$, we notice that $n$-point functions near the fixed point are essentially polynomials in $g$, hence their derivative at the fixed point is finite. A singularity might instead arise at finite $g$, and in particular at the critical value determined by the invertibility of the relation $g = \lambda Z(\lambda)^{-2}$, with $Z(\lambda)$ defined by the solution of Eq.~\eqref{eq:wf} (see Ref.~\cite{Benedetti:2019eyl}).}
The main point of repeating the derivation of the Ward identities here is to check that the structure of right-hand side  (beta function times operator insertion) generalizes to our model with multiple interaction terms.

We conclude that the $n$-point functions of fundamental fields are conformal invariant at the fixed point.
The next step would be to generalize the above result to correlators of composite operators. A possible strategy, starting from correlators of fundamental fields and using an operator-product expansion to generate the composite fields, is sketched in Ref.~\cite{Paulos:2015jfa}.
In the next section we will follow a different route, and compute perturbatively three-point functions of (quartic and quadratic) composite operators.

%%%%%%%%%%%%%%%%%%%%%%%%%%%%%%%%%%%%%%%%%%%%%%%%%%%%%%%%
\section{Correlation Functions}
\label{sec:3pt}
%%%%%%%%%%%%%%%%%%%%%%%%%%%%%%%%%%%%%%%%%%%%%%%%%%%%%%%%
In this section we will explicitly compute large-$N$ three-point functions among the $\phi^4$ and $\phi^2$ composite operators we discussed in the previous sections at the interacting IR fixed point. Since this fixed point depends parametrically on the exactly marginal coupling $g$, we will work at  small $g$, at lowest perturbative order in the couplings. 

\begin{figure}[htb]
	\begin{center}
		\scalebox{1}{\includegraphics{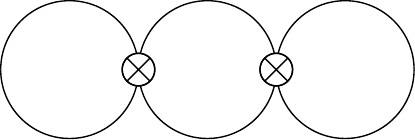}}
	\end{center}
	\caption{Would-be leading order diagram for $\Braket{\{ \phi_2^4\}\{ \phi_2^4\} }$.}
	\label{fig:2pt-2-tadpoles}
\end{figure}

%%%%%%%%%%%%%%%%%%%%%%%%%%%%%%%%%%%%%%
\paragraph{Large-$N$ and conformal limits.}
%%%%%%%%%%%%%%%%%%%%%%%%%%%%%%%%%%%%%%
Before proceeding, we stress one important aspect about the interplay between large-$N$ limit and conformal limit.
Namely, it should be kept in mind that in principle we should take the conformal limit (i.e.\ tune the theory to the fixed point) before or at least together with the large-$N$ limit, in order to keep non-trivial two-point functions in the limit. 
For example, if we were to take the large-$N$ limit first, away from criticality, the two-point function of double-trace operators would be dominated by the diagram in Fig.~\ref{fig:2pt-2-tadpoles}, rather than that of Fig.~\ref{fig:2pt-1}. Away from criticality, such diagram leads to a finite contribution (after renormalization) determining the two-point function at leading order of the large-$N$ expansion. However, at criticality its renormalized amplitude vanishes, and we would be left with a zero two-point function. The way out is to assume from the beginning that diagrams such as that in Fig.~\ref{fig:2pt-2-tadpoles} are zero because we are in the massless limit, and to rescale the operators in such a way that the first non-vanishing contribution to a two-point function at criticality is always normalized to order $N^0$.
Making sure that such a scaling exists in the case of tensor models is non-trivial, and it is what we want to address now.
%%%%%%%%%%%%%%%%%%%%%%%%%%%%%%%%%%%%%%%%%%%%%%%%%%%%%%%%
%%%%%%%%%%%%%%%%%%%%%%%%%%%%%%%%%%%%%%%%%%%%%%%%%%%%%%%%

Let us first review how correlators of invariants are treated in the case of matrix field theories.
The analogue of our interacting action \eqref{eq:action} is:
\be
S^{\rm int}[\phi] =  \int d^dx \, \left( \frac{ m^{2\zeta}}{2}  \Tr[ \phi(x)^2 ] + \frac{\lambda_s}{4 N}  \Tr[ \phi(x)^4] + \frac{\lambda_d}{4 N^2}  (\Tr[ \phi(x)^2])^2   \right) \, .
\ee
It is convenient to rescale $\phi\to \sqrt{N} \phi$, so that the action becomes:
\be
S[\phi] = N \int d^dx \, \left( \frac12 \phi_{ab}(x)(   - \partial^2)^{\zeta}\phi_{ba}(x) + \frac{ m^{2\zeta}}{2}  \Tr[ \phi(x)^2 ] + \frac{\lambda_s}{4}  \Tr[ \phi(x)^4] + \frac{\lambda_d}{4 N}  (\Tr[ \phi(x)^2])^2   \right) \, .
\ee
In such rescaled variables, the functional:
\be
W[J] =\frac{1}{N^2} \log \int [d\phi] e^{-S[\phi] + \sum_{t\geq 1} N^{2-t}  \sum_i \int J^{(t)}_i {\cal O}^{(t)}_i } \, ,
\ee
admits a large-$N$ limit, in the perturbative sense. Here, ${\cal O}^{(t)}_i$ stands for an invariant built from $t$ traces of products of $\phi$ fields and contains \emph{no explicit factors} of $N$.
The functional derivatives of $ W[J]$ with respect to the sources $J_i^{(t)}$ are proportional to the connected correlators of ${\cal O}^{(t)}_i$:
\be
\frac{\delta^n W}{\delta { J}^{(t_1)}_{i_1}(x_1) \cdots \delta { J}^{(t_n)}_{i_n}(x_n)} = N^{\sum_j (2-t_j)}\Braket{{\cal O}^{(t_1)}_1(x_1) 
\dots {\cal O}^{(t_n)}_1 (x_n)  }_{\rm connected} \,,
\ee
and have an expansion in $1/N$ starting at order $N^0$. Restricting to single-trace operators, the leading order graphs are connected planar ribbon graphs (or 3-colored graphs in our notation). Including multi-trace operators, the leading order graphs are nodal surfaces, i.e.\ trees of bubbles (or cacti) diagrams with bubbles made of planar graphs.
Disconnected correlators, obtained from $Z[J]=e^{N^2 W[J]}$, have leading-order contributions of order $N^{2c}$, with $c$ the number of connected components.

In applications to conformal field theories it is customary to consider instead correlators of ${\cal O}^{(t)}_i $, without explicit factors of $N$. They are obtained by rescaling appropriately the derivatives of $W$, or equivalently, by taking derivatives with respect to ${\tilde J}^{(t)}_i = N^{2-t} J^{(t)}_i$ of the functional ${\tilde W}[ \{ {\tilde J}^{(t)}_i \} ]=N^2 W[\{ N^{t-2} {\tilde J}^{(t)}_i \} ]$. In this normalization, the connected two-point functions of single-trace operators at leading-order are of order $N^0$, while their higher-point functions are suppressed in $1/N$.
For multi-trace operators, a subtlety rarely emphasized appears, namely, their leading-order two-point functions naively seem to be of higher order than $N^0$, and higher-point functions seem to have even higher powers of $N$. In fact, we have:
\be
\frac{\delta^n {\tilde W}}{\delta {\tilde J}^{(t_1)}_{i_1}(x_1) \cdots \delta {\tilde J}^{(t_n)}_{i_n}(x_n)}
 = N^{2+ \sum_j (t_j-2)} \frac{\delta^n W}{\delta { J}^{(t_1)}_{i_1}(x_1) \cdots \delta { J}^{(t_n)}_{i_n}(x_n)} \, ,
\ee
with a leading order contribution scaling as $N^{2+\sum_j (t_j-2)}$. For $n=2$ and $t_j=1$, we get a correlator of order $N^0$, as anticipated. On the other hand, if for example all the $t_i$ are greater than two, we naively seem to get an arbitrarily large power of $N$. However, the (naive) leading-order contributions for such multi-trace operators come from cactus diagrams, as the one of Fig.~\ref{fig:2pt-2-tadpoles}, which always contain a factor corresponding to the one-point function of a single-trace component of the multi-trace invariant (the end leaves of the cactus). In the conformal limit such contributions vanish, and a factorization into two-point functions of single-trace components dominates.\footnote{The factorization property is often stated in a slightly different fashion in the literature \cite{Aharony:1999ti}. Namely, considering only single trace operators, one observes that
\be\nonumber
\Braket{ {\cal O}^{(1)}_1(x_1) \cdots {\cal O}^{(1)}_n(x_n) }_{\rm conn.}  = \sum_{g\geq 0} N^{2-2g -n} F_g(\lambda) \, ,
\ee
from which it is deduced that higher-point functions are suppressed with respect to two-point functions. However, as one-point functions are more favored than two-point functions, in order to have a factorization in two-point functions it must be assumed that one-point functions are zero, as guaranteed by scale invariance.
}
Notice that since two-point functions of single-trace operators are normalized to one, by factorization also the two-point-functions of multi-trace operators are normalized to one.
Summarizing, one is typically interested in:
\be
{\tilde W}[{\tilde J}] =\log \int [d\phi] e^{-S_{\rm CFT}[\phi] + \sum_{t\geq 1}  \sum_i \int {\tilde J}^{(t)}_i {\cal O}^{(t)}_i } \, 
\ee
which gives two-point functions normalized to one, and higher-point functions suppressed in $1/N$.

We now go back to our tensor field theory. In analogy to the matrix case, we rescale the field as:
\be \label{eq:phi-rescale}
\phi_{\mba}\to N^{3/4} \phi_{\mba} \,,
\ee
so that the action becomes:
\be
\begin{split}
    S[\phi]  & = N^{3/2}  \frac{1}{2} \int d^dx \, \phi_{\mba}(x) (   - \partial^2)^{\zeta}\phi_{\mba}(x) + S^{\rm int}[\phi]  \,,  \\
    S^{\rm int}[\phi]  & =
    N^{3/2} \bigg[
    \frac{ m^{2\zeta}}{2} \int d^dx \, \phi_{\mba}(x) \phi_{\mba}(x)  \\
		& \qquad + \frac{1}{4} \int d^d x \, \left( \im\lambda \delta^t_{\mba \mbb\mbc\mbd} + \lambda_p N^{-1/2} \delta^p_{\mba\mbb; \mbc\mbd}
		+ \lambda_d N^{-3/2} \delta^d_{\mba\mbb; \mbc\mbd } \right)
		\phi_{\mba}(x) \phi_{\mbb}(x)  \phi_{\mbc}(x) \phi_{\mbd}(x) \bigg]  \, ,
	\end{split}
	\ee
where we reinstated the pillow and double trace couplings. Observe that the operators $\delta^t, \delta^p$ and $\delta^d$, defined in Eq.~\eqref{eq:deltas-nohat}, merely contract indices and do not contain explicit factors of $N$.

According to Appendix \ref{app:1/N},
the generating functional of connected correlators admitting  a large-$N$ limit in the perturbative sense is now:
\be
W[J] =\frac{1}{N^3} \log \int [d\phi] e^{-S[\phi] + \sum_{b} N^{\frac32-\rho_b} \int J_b {\cal O}_b } \, ,
\ee
with $\rho_b\geq 0$ chosen according to the optimal scaling defined in Ref.~\cite{Carrozza:2015adg}:
\be
\rho_b=\frac{F_b-3}{2} \,,
\ee
where $F_b$ counts the total number of cycles of alternating colors $i$ and $j$ with $i,j ~\in \lbrace 1,2,3\rbrace$.
Invariants with $\rho_b=0$ are called \emph{maximally single-trace} (MST) \cite{Ferrari:2017jgw}.

Defining again correlators as derivatives with respect to non-rescaled sources ${\tilde J}_b = N^{\frac32-\rho_b} J_b$ of the functional ${\tilde W}[ \{ {\tilde J}_b \} ]=N^3 W[\{ N^{\rho_b-\frac32} {\tilde J}_b \} ]$, we have the following expansion:
\be \label{eq:Wexp}
\frac{\delta^n {\tilde W}}{\delta {\tilde J}_{b_1}(x_1) \cdots \delta {\tilde J}_{b_n}(x_n)}
 = \sum_{\omega\geq 0} N^{3-\omega - \sum_i (\frac32-\rho_{b_i}) } F_{\omega}(\lambda) \, .
\ee
For MST operators we thus have an analogous result as for single-trace matrix operators, that is, their two-point functions are of order one, and higher-point functions are suppressed. Assuming again that one-point functions are zero, for operators which are products of MST operators, which we will call \emph{maximally multi-trace} (MMT), we obtain a factorization property as for the matrix multi-trace operators (see Appendix~\ref{app:1/N} for more details).

For non-MMT operators with $\rho_b>0$ we seem to have again two-point functions with a higher scaling than $N^0$, and for those with $\rho_b>3/2$ we seem to have again a possible arbitrarily large power of $N$. However, we conjecture that, as for multi-trace matrix operators, the leading order diagrams (those with $\omega=0$) vanish in the conformal case, and the first non-vanishing order has $\omega \geq 3-\sum_i (\frac32-\rho_{b_i})$. We hope to return to this conjecture in full generality in a future publication. For the pillow invariant which is contained in $\phi^4_1$, and which has $\rho_p=1/2$, we can explicitly check that for example the two-point function at leading order is of order $N^0$ (see App.~\ref{app:pillow}).

Now we proceed to the perturbative computation of the two and three-point connected correlations of the MST operators $\phi^4_t$ and $\phi^2$. The correlations involving the double trace $\phi^4_2$ are obtained at leading order using the large $N$ factorization. We do not study here the correlations involving $\phi^4_1$, which due to the pillow operator is neither MST nor MMT, but we will briefly discuss them in App.~\ref{app:pillow}. As explained above, a connected correlation function of $n$ MST operators has an a priori scaling $N^{3-\frac{3}{2}n}$, that is $N^{0}$ for two-point correlations and $N^{- 3/2}$ for the three-point ones. Whenever a correlation scales less than that, we consider it suppressed in the large $N$ limit. 

%%%%%%%%%%%%%%%%%%%%%%%%%%%%%%%%%%%%%%
\paragraph{Rescaled action and renormalized operators.}
%%%%%%%%%%%%%%%%%%%%%%%%%%%%%%%%%%%%%%
In order to regularize the UV divergences we set $\epsilon > 0$. We compute the correlations up to first order in the couplings in the critical theory, $\lambda_{\phi^2}=0$. At the relevant order the action in Eq.~\eqref{eq:one-loop-action} becomes:
\begin{equation}\label{eq:new-action}
 S = N^{3/2}  \int  \bigg[ \frac{1}{2}  \, 
\phi_{\mba} C_{\mu}^{-1} \phi_{\mba} + \frac{\lambda}{4}  \, \phi^4_t + \frac{\lambda_1}{4 N^{1/2}} \, \phi^4_1  + \frac{\lambda_2}{4 N^{3/2}} : \phi^4_2: \bigg] \,,
\end{equation}
where the operators are defined as before, except for the factors of $N$ due to the rescaling \eqref{eq:phi-rescale} (remember the distinction between hatted and un-hatted deltas in Eq.~\ref{eq:deltas}):
\be\label{eq:quartinv-2}
\begin{split}
 & \phi^4_t(x) \equiv \im {\delta}^t_{\mba \mbb\mbc\mbd} \, \phi_{\mba}(x) \phi_{\mbb}(x) \phi_{\mbc}(x) \phi_{\mbd }(x) \,, \\
 & \phi^4_1(x) \equiv 3 ({\delta}^p_{\mba\mbb;\mbc\mbd} - \frac{1}{N}{\delta}^d_{\mba\mbb;\mbc\mbd})\, \phi_{\mba}(x) \phi_{\mbb}(x) \phi_{\mbc}(x) \phi_{\mbd }(x) \,, \\
 & \phi^4_2(x) \equiv {\delta}^{d}_{\mba\mbb; \mbc\mbd} \, \phi_{\mba}(x) \phi_{\mbb}(x) \phi_{\mbc}(x) \phi_{\mbd }(x) \, ,
\end{split}
\ee
where, in spite of the new scaling in $N$ we maintain the same notation for the quartic invariants.

At this order we have $\lambda = \mu^{d-4\Delta_{\phi}} g, \, \lambda_1 = \mu^{d-4\Delta_{\phi}} g_1, \, \lambda_2 = \mu^{d-4\Delta_{\phi}} g_2$ with $g,g_1,g_2$ the dimensionless couplings. 
Below we will use $\lambda$'s for the perturbative expressions at $\epsilon>0$, and $g$'s for the final expressions in dimensionless variables.
At linear order, the renormalized operators are:\footnote{Notice that to obtain from Eq.~\eqref{eq:M-def} the renormalized operators at linear order, we need to use the couplings at quadratic order in Eq.~\eqref{eq:couplings-renorm}.}
\be
 \begin{split}\label{eq:new-one-loop-mix}
  &  \big[\phi^2\big]  = Z_{\phi^2} \phi^2 \,, \qquad
   \big[\phi^4_1\big]  = Z_{\phi^4_1} \phi^4_1 \,, \qquad
   \big[\phi^4_2\big]  = Z_{\phi^4_2} : \phi^4_2 : \,, \crcr
 & \qquad Z_{\phi^2}  = 1 + Q \, \frac{g_2 }{\epsilon}  \,, \quad
 Z_{\phi^4_1} =   1 + 2 Q \, \frac{g_1 }{\epsilon}  \,, \quad Z_{\phi^4_2} = 1 + 2 Q \, \frac{g_2 }{\epsilon}  \,, \crcr
 & \big[\phi^4_t\big]  = \phi^4_t + N^{-1/2} Z_{\phi^4_t;\phi^4_1} \phi^4_1 + N^{-3/2} Z_{\phi^4_t;\phi^4_2} :\phi^4_2: + 
  Z_{\phi^4_t;\phi^2} \phi^2 \,, \crcr
  & \qquad Z_{\phi^4_t;\phi^4_1} = -2Q \frac{g}{\epsilon} \,, \quad
  Z_{\phi^4_t;\phi^4_2} = -6 Q \frac{g}{\epsilon} \,,  
\quad   Z_{\phi^4_t;\phi^2} = 
     - 4 \mu^{ d-4\Delta_{\phi}} g \int d^du \, C(u)^3  \,,
 \end{split}
 \ee
with $Q = 2  (4\pi)^{ - d/2} \Gamma( d / 2 )^{-1}$. All our correlations will be written up to terms of order $g^2$ which from now we omit. 
We will fix the pillow and double-trace couplings at their IR fixed point, which at the same order in $N$ and $g$ reads  $ g_1^{\star} =  g^{\star},  g_2^{\star} = \sqrt{3}  g^{\star}$.

The scaling operators are obtained as in Sec.~\ref{sec:scaling-op}, but with some different factors of $N$. In particular, from the above discussion, we have the following modification of Eq.~\eqref{eq:dgCS}:
\be
\begin{split}
0 &= -\partial_{g_b} \left(\mu \partial_\mu + \beta_a \partial_{g_a} \right) \Braket{\phi(x_1) \dots \phi(x_n)}_{c} \\
&= \sum_a  \left(\delta_{ab}\, \mu \partial_\mu + (\partial_{g_b}\beta_a) \right) \mu^{d-\Delta_a} N^{\frac32-\rho_a} \int d^d x   \Braket{\phi(x_1) \dots \phi(x_n)   \big[{\cal O}_a\big](x)  }_{c} \\
& \qquad - \sum_a \mu^{2d-\Delta_a-\Delta_b} \, N^{3-\rho_a-\rho_b}\, \beta_a \int d^d x d^d y   \Braket{\phi(x_1) \dots \phi(x_n)   \big[{\cal O}_a\big](x) \big[{\cal O}_b\big](y)  }_{c} \,,
\end{split}
\ee
from which, going at the fixed point and multiplying by $N^{-\frac32+\rho_b}$, we arrive at the same expression for the scaling operators as in Eq.~\eqref{eq:scal-op-def}, but with the replacement $(U^{-1})_{ba}\to N^{\rho_b} (U^{-1})_{ba} N^{-\rho_a}$.
The scaling operators are then:
 \begin{align}\label{eq:new-one-loop-scale-op}
\{ \phi^2 \}(x) &= \mu^{ \delta h_{\phi^2}} Z_{\phi^2} \phi^2(x) \,, \crcr
\{ \phi^4_t \}(x) &=  \phi^4_t (x) + N^{-\frac12}\left(Z_{\phi^4_1} +Z_{\phi^4_t;\phi^4_1}\right) \phi^4_1 (x) \crcr
&\quad +N^{-\frac32} \left(\sqrt{3} Z_{\phi^4_2}+Z_{\phi^4_t;\phi^4_2}\right) :\phi^4_2(x):
+ \left(N^{-\frac32}\sqrt{3} Z_{\phi^4_2 ; \phi^2}+Z_{\phi^4_t;\phi^2}\right) \phi^2(x) \,, \crcr 
\{ \phi^4_1 \}(x) &= \mu^{\delta h_1}  Z_{\phi^4_1}  \phi^4_1(x)\,, \\
\{ \phi^4_2 \}(x) &= \mu^{\delta h_2}  Z_{\phi^4_2} : \phi^4_2(x): \,.
  \nonumber
 \end{align}

%%%%%%%%%%%%%%%%%%%%%%%%%%%%%%%%%%%%%%
\paragraph{Contact terms.}
%%%%%%%%%%%%%%%%%%%%%%%%%%%%%%%%%%%%%%
Because of the melonic convolution in Eq.~\eqref{eq:convolution}, we will encounter ``contact'' terms proportional to $(\partial^{2})^n \delta(x_{ij})$.
This should not come as a surprise, as such terms are for example expected in the three-point function of exactly marginal operators \cite{Seiberg:1988pf}.
One should distinguish them from terms like $\lim_{\epsilon \to 0} \frac{1}{|x|^{d-\epsilon}} \sim \frac{1}{\epsilon}  \delta(x)$, which require regularization and lead to an anomaly  \cite{Freedman:1991tk,Bzowski:2015pba}.
Contact terms with finite coefficients are compatible with conformal transformations. The difference can be reformulated by the observation \cite{Todorov:1985xs} that in $d$ dimensions the only homogeneous distribution of dimension $d$ is the Dirac delta, while any generalized function which coincides with $\frac{1}{|x|^d}$ for $|x|\neq 0$ is an associate homogeneous distribution \cite{Gelfand}, that is, under scale transformations it transforms with an in homogeneous contact term.

%%%%%%%%%%%%%%%%%%%%%%%%%%%%%%%%%%%%%%%%%%%%%%%%%%%%%%%%
\subsection{Two-point functions}
%%%%%%%%%%%%%%%%%%%%%%%%%%%%%%%%%%%%%%%%%%%%%%%%%%%%%%%%

We denote by prime dimensionless positions, $x' = \mu x$ and so on. We first compute the relevant two-point functions.

\begin{figure}[htb]
	\begin{center}
		\scalebox{1}{\includegraphics{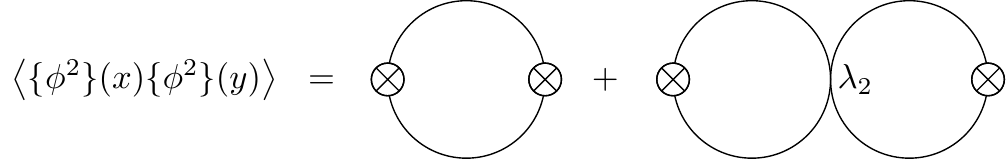}}
	\end{center}
	\caption{Leading order diagrams for $\Braket{ \{ \phi^2 \} \{ \phi^2 \} }$.}
	\label{fig:phi2phi2}
\end{figure}

%%%%%%%%%%%%%%%%%%%%%%%%%%%%%%%%%%%%%%%%%%%%%%%%%%%%%%%%
\paragraph{The $\Braket{ \{ \phi^2\} \{ \phi^2 \}}$ correlation.}
%%%%%%%%%%%%%%%%%%%%%%%%%%%%%%%%%%%%%%%%%%%%%%%%%%%%%%%%
Up to the first order in the coupling constant, at leading order in $N$ only the two diagrams represented in  Fig.~\ref{fig:phi2phi2} contribute. They give:
	\begin{equation}
		\Braket{ \{\phi^2 \}(x) \{\phi^2 \} (y) } \, =  \, \mu^{2 \delta h_{\phi^2}}\, 2 \left[ Z_{\phi^2}^2 \, C(x-y)^2 \, - \, \lambda_2 \int d^dz \, C(x-z)^2 C(z-y)^2 \,  \right] \, ,
	\end{equation}
where $Z_{\phi^2}$ is the renormalization constant of the $\phi^2$ operator in Eq.~\eqref{eq:new-one-loop-mix}.
%%%%%%%%%%%%%%%%%%%%%%%%%%%%%%%%%%%%%%%%%%%%%%%%%%%%%%%%
%%%%%%%%%%%%%%%%%%%%%%%%%%%%%%%%%%%%%%%%%%%%%%%%%%%%%%%%

Computing the integral at first order in $\epsilon$ (see appendix \ref{app:conf-int}) we obtain: 
\be
\Braket{ \{\phi^2 \} (x) \{\phi^2 \} (y) }
		\, = \,\mu^{4\Delta_{\phi}+2\delta h_{\phi^2}} 
		\frac{2 \, c(\zeta)^2}{ |x'-y'|^{4\Delta_{\phi} }} \left[ Z_{\phi^2}^2 \, - \,2 \, Q \, g_2 
		\left( \frac{1}{\epsilon} + \log|x'-y'| - \log 2 - \psi(\tfrac{d}{4}) + \mathcal{O}(\epsilon) \right)  \right] \, .
\ee
As expected, the $1/\epsilon$ pole cancels as, according to Eq.~\eqref{eq:new-one-loop-mix}, $Z_{\phi^2} \, = \,  1 \, + \, Q \frac{ g_2 }{\epsilon}$. 
We now take $\epsilon$  to $0$. The $g_2 \log|x'-y'|$ term combines with the constant term to give a correction to the scaling law, hence at first order in $g$ the two-point function is:
	\begin{equation}
		\Braket{ \{\phi^2 \}(x) \{\phi^2 \}(y) }\, = \, 
		%\mu^{4\Delta^{\star}_{\phi} + 2\delta h_{\phi^2}}
		 \frac{2 \, c(d/4)^2}{ |x-y|^{4\Delta^{\star}_{\phi} +2\delta h_{\phi^2}}}
		\left[ 1 + 2\, Q \, g_2^{\star} \Big( \log 2 + \psi(d/4) \Big)  \right] \, ,
	\end{equation}
where $\Delta_{\phi}^{\star}=d/4$ is the dimension of the field at $\epsilon=0$. 
The dimension of $\phi^2$ is
$\Delta_{\phi^2}^{\star} = 2\Delta_{\phi}^{\star} + \delta h_{\phi^2}$ with 
$\delta h_{\phi^2} \, = \, Q\, g_2^{\star} $ which reproduces the anomalous dimension found in Eq.~\eqref{eq:phi2-andim}.

%%%%%%%%%%%%%%%%%%%%%%%%%%%%%%%%%%%%%%%%%%%%%%%%%%%%%%%%
\paragraph{The $\Braket{\{\phi^4_t \}\{\phi^4_t\}}$ correlation.}
%%%%%%%%%%%%%%%%%%%%%%%%%%%%%%%%%%%%%%%%%%%%%%%%%%%%%%%%
Next, we consider the two-point functions of $\phi_t^4$ operators.
As the operator is complex we take the two-point function of $\phi_t^4$ with the hermitian conjugate $(\phi_t^4 )^{\dagger} = -\phi^4_t$. 

\begin{figure}[htb]
	\begin{center}
		\scalebox{1}{\includegraphics{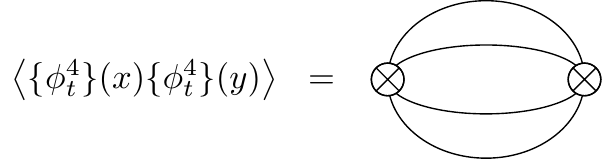}}
	\end{center}
	\caption{Leading order diagram for $\Braket{ \{ \phi_t^4 \} \{ \phi_t^4 \} }$.}
	\label{fig:2pt}
\end{figure}

At first order in the couplings and at leading order in $N$ only one diagram, depicted in Fig.~\ref{fig:2pt}, contributes, yielding:
	\begin{equation}
\Braket{ \{\phi_t^4\}(x) \{(\phi_t^4 )^{\dagger} \}(y) } \, 
		= 4 C(x-y)^4 = \, %\mu^{8\Delta_{\phi}} 
		\frac{ 4 \, c(\zeta)^4}{|x-y|^{8\Delta_{\phi}}} \, .
	\end{equation}
The $\epsilon \to 0$ limit is trivial.
%%%%%%%%%%%%%%%%%%%%%%%%%%%%%%%%%%%%%%%%%%%%%%%%%%%%%%%%
%%%%%%%%%%%%%%%%%%%%%%%%%%%%%%%%%%%%%%%%%%%%%%%%%%%%%%%%
This is the standard form for the two-point function of an exactly marginal operator, leading to a conformal anomaly in even dimensions \cite{Gomis:2015yaa}. The appearance of an anomaly can be understood in the spirit of our earlier comment on contact terms by noticing that for $d=2n$ we have $1/|x|^{2d-\epsilon} \sim \frac{1}{\epsilon} (\partial^{2})^n \delta(x)$.
Since in our model we have assumed $d<4$ from the beginning, the anomaly only concerns $d=2$.

\begin{figure}[htb]
	\begin{center}
		\scalebox{1}{\includegraphics{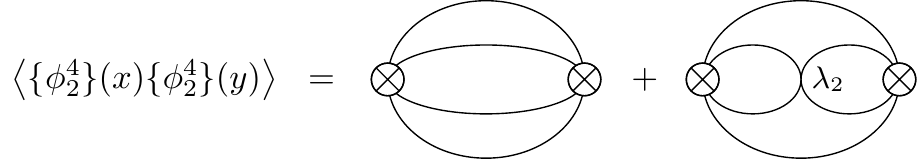}}
	\end{center}
	\caption{Leading order diagrams for $\Braket{ \{ \phi_2^4 \} \{ \phi_2^4 \} }$.}
	\label{fig:2pt-1}
\end{figure}

%%%%%%%%%%%%%%%%%%%%%%%%%%%%%%%%%%%%%%%%%%%%%%%%%%%%%%%%
\paragraph{The $\Braket{ \{ \phi^4_2 \}\{ \phi^4_2 \} }$ correlation.}
%%%%%%%%%%%%%%%%%%%%%%%%%%%%%%%%%%%%%%%%%%%%%%%%%%%%%%%%
At the appropriate order we have two graphs (see Fig.\ref{fig:2pt-1}),
%%%%%%%%%%%%%%%%%%%%%%%%%%%%%%%%%%%%%%%%%%%%%%%%%%%%%%%%
%%%%%%%%%%%%%%%%%%%%%%%%%%%%%%%%%%%%%%%%%%%%%%%%%%%%%%%%
where the quartic vertex is $:\phi^4_2:$, yielding:
\be
 \Braket{ \{ \phi^4_2 \} (x) \{ \phi^4_2 \}(y)  } \, = \, \mu^{2\delta h_2} \;8 \bigg[ Z_{\phi^4_2}^2 \, C(x-y)^4  - 2 \, \lambda_2 \,  C(x-y)^2 \int d^dz \, C(x-z)^2 C(z-y)^2   \bigg] \,.
\ee
Following the same steps as before we get, after cancellation of the pole and taking the limit $\epsilon \to 0$:
\be
 \Braket{ \{ \phi^4_2 \} (x) \{ \phi^4_2 \}(y) } =  
% \mu^{8\Delta_{\phi}^{\star} +2\delta h_2} 
 \frac{8\, c(d/4)^4}{ |x-y|^{8\Delta_{\phi}^{\star} + 2\delta h_2 }}
 \bigg[1 + 4 \, Q \, g_2^{\star} \bigg(  \log 2 + \psi(\tfrac{d}{4}) \bigg)  \bigg] \,,
\ee
with $\delta h_2 = 2 \, Q \,  g_2^{\star}$ reproducing the perturbative
result Eq.~\eqref{eq:dim-an-quart}. As expected for a multitrace operator we have 
\be
\Braket{ \{ \phi^4_2 \} (x) \{  \phi^4_2 \}(y) } = 
2 \Braket{ \{ \phi^2 \} (x) \{ \phi^2 \} (y) } 
\Braket{\{ \phi^2 \} (x) \{ \phi^2\}(y) } \,.
\ee

\begin{figure}[htb]
	\begin{center}
		\scalebox{1.2}{\includegraphics{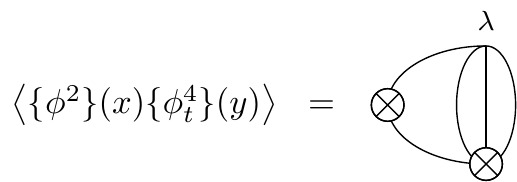}}
	\end{center}
	\caption{Leading order diagram for $\Braket{  \{ \phi^2 \} \{ \phi^4_t \} }$.}
	\label{fig:bare}
\end{figure}

%%%%%%%%%%%%%%%%%%%%%%%%%%%%%%%%%%%%%%%%%%%%%%%%%%%%%%%%
 \paragraph{The $\Braket{ \{ \phi^2 \} \{ \phi^4_t \} }$ correlation.}
%%%%%%%%%%%%%%%%%%%%%%%%%%%%%%%%%%%%%%%%%%%%%%%%%%%%%%%%
For a conformal field theory we expect that the two-point functions of operators with different dimensions are zero. 
We check this for $\Braket{ \{ \phi^2 \} \{ \phi^4_t \} }$.
The leading-order contribution to this correlator is depicted in Fig.~\ref{fig:bare}
(and we get the appropriate counterterm subtraction from the mixing of $\phi^4_t$ with $\phi^2$) which yields:
	\begin{equation}
  \Braket{ \{ \phi^2\} (x) \{  \phi^4_t \} (y) }  \,  \sim  \, \lambda \, C(x-y) \int d^dz \, C(x-z) \bigg( C(z-y)^3  - \delta_{yz} \int d^du \; C^3(u) \bigg)  \, .
	\label{eq:phi^2phi^4_t}
	\end{equation}
%%%%%%%%%%%%%%%%%%%%%%%%%%%%%%%%%%%%%%%%%%%%%%%%%%%%%%%%
%%%%%%%%%%%%%%%%%%%%%%%%%%%%%%%%%%%%%%%%%%%%%%%%%%%%%%%%

Using the convolution (\ref{eq:convolution}), one can evaluate the $z$-integral and obtain:
\be
 \Braket{ \{ \phi^2 \} (x) \{ \phi^4_t \} (y) }  \, \sim \,  4 \, \lambda \frac{c(\zeta)^4c(4\zeta-d)}{c(3\zeta-d)} \frac{1}{|x-y|^{10\Delta_{\phi} -d}}  \, ,
\ee
which is zero in the $\epsilon\to 0$ limit.
We conclude that $\Braket{ \{ \phi^2 \} (x) \{ \phi^4_t \} (y) } \lesssim N^{-1} $, hence it is suppressed at large $N$.

%%%%%%%%%%%%%%%%%%%%%%%%%%%%%%%%%%%%%%%%%%%%%%%%%%%%%%%%
\subsection{Three-point functions}
%%%%%%%%%%%%%%%%%%%%%%%%%%%%%%%%%%%%%%%%%%%%%%%%%%%%%%%%
We now compute the three-point functions at first order in the couplings and leading order in large $N$. 
 
\begin{figure}[htb]
	\begin{center}
		\vspace{10pt} \scalebox{1}{\includegraphics{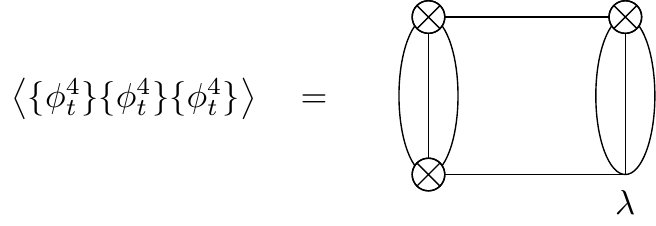}} \vspace{-8pt}
	\end{center}
	\caption{Leading order diagram for $\Braket{ \{ \phi_t^4 \} \{ \phi_t^4 \}  
	\{ \phi_t^4 \}  }$. }
	\label{fig:ttt-pert}
\end{figure}

%%%%%%%%%%%%%%%%%%%%%%%%%%%%%%%%%%%%%%%%%%%%%%%%%%%%%%%%
\paragraph{The $\Braket{ \{ \phi_t^4 \}\{ \phi_t^4 \} \{ \phi_t^4 \}}$ correlation.}
\label{sec:ttt}
%%%%%%%%%%%%%%%%%%%%%%%%%%%%%%%%%%%%%%%%%%%%%%%%%%%%%%%%
We consider the three-point function $\Braket{ \{ \phi_t^4\} (x) 
\{\phi_t^4 \} (y) \{ \phi_t^4 \} (z) }$. Up to the first order in the coupling constant, the only relevant Feynman diagram at order $N^{-3/2}$ is depicted in Fig.~\ref{fig:ttt-pert}  (and we get the appropriate counterterm subtraction from the mixing of $\phi^4_t$ with $\phi^2$).
%%%%%%%%%%%%%%%%%%%%%%%%%%%%%%%%%%%%%%%%%%%%%%%%%%%%%%%%
%%%%%%%%%%%%%%%%%%%%%%%%%%%%%%%%%%%%%%%%%%%%%%%%%%%%%%%%
Its contribution is given by:
\be
\begin{split}	\label{eq:ttt}
    \Braket{ \{ \phi_t^4 \} (x) \{ \phi_t^4 \} (y) \{ \phi_t^4 \}(z) } 
    &= \,N^{-\frac{3}{2}}\, 4^2 \, \lambda C(x-y)^3 C(y-z) 
    \\ &\qquad \times 
    \int d^du \, \bigg[ C(z-u)^3 - \delta_{zu} \int d^d v \,  C(v)^3 \bigg]C(u-x)  \\
     & \quad  + \, ({\rm permutations}) \\
     & = \,N^{-\frac{3}{2}}\, 4^2 \, \lambda\frac{c(d/4)^7}{c(-d/4)} \frac{1}{|x-y|^{2d}} \delta(x-z) + \, ({\rm permutations})
        \,  \, ,
\end{split}
\ee
where we used again Eq.~\eqref{eq:convolution}.
This is of the expected form for the three-point function of exactly marginal operators \cite{Seiberg:1988pf}, and
up to contact terms with finite coefficients, it is consistent with the conclusion of Ref.~\cite{Bzowski:2015pba}, according to which a marginal operator should have vanishing three-point function with itself.

%%%%%%%%%%%%%%%%%%%%%%%%%%%%%%%%%%%%%%%%%%%%%%%%%%%%%%%%
\paragraph{The $\Braket{ \{\phi^2 \} \{ \phi^2 \} \{ \phi^2 \}}$ correlation.}
%%%%%%%%%%%%%%%%%%%%%%%%%%%%%%%%%%%%%%%%%%%%%%%%%%%%%%%%

\begin{figure}[htb]
	\begin{center}
		\hspace{-20pt}\scalebox{1.2}{\includegraphics{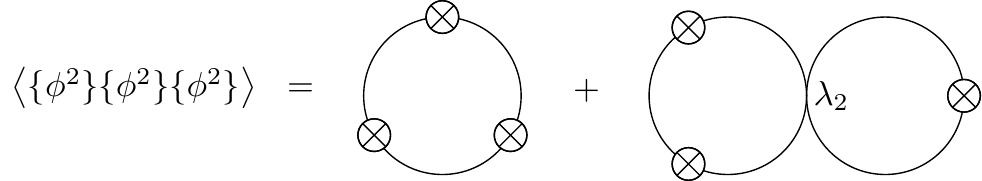}}
	\end{center}
	\caption{Leading order diagrams for $\Braket{ \{ \phi^2 \} \{ \phi^2 \} 
	\{\phi^2 \}  }$.}
	\label{fig:222}
\end{figure}

Up to the first order in the coupling we have the diagrams depicted in 
Fig.~\ref{fig:222}.
%%%%%%%%%%%%%%%%%%%%%%%%%%%%%%%%%%%%%%%%%%%%%%%%%%%%%%%%
%%%%%%%%%%%%%%%%%%%%%%%%%%%%%%%%%%%%%%%%%%%%%%%%%%%%%%%%

They yield:
	\begin{align}
     & \Braket{ \{ \phi^2 \} (x) \{ \phi^2 \} (y) \{ \phi^2 \}(z) }
     \, =  \,\mu^{3\delta h_{\phi^2}}\, 8N^{-\frac{3}{2}} \bigg[ Z_{\phi^2}^3 \,C(x-y) C(y-z) C(z-x) \\
     &\qquad \qquad \qquad
     - \, \lambda_2 \, C(x-y) \int d^du \, C(x-u) C(y-u) C(z-u)^2 
      - (x \leftrightarrow z) - (y \leftrightarrow z)
     \bigg]  \, . \nonumber
	\end{align}
Using Appendix \ref{app:conf-int}, the right and side of the above equation becomes:
\be
8N^{-\frac{3}{2}} c(\zeta)^3 \mu^{6\Delta_{\phi}+3\delta h_{\phi^2}}
 \frac{  Z_{\phi^2}^3 - Qg_2 \left[ \frac{3}{\epsilon} \, + \,  \log\big( |x'-y'| |y'-z'| |z'-x'| \big)  %+ \cdots 
  \right] }
		{ |x'-y'|^{2\Delta_{\phi} } |y'-z'|^{2\Delta_{\phi} } 
		|z'-x'|^{2\Delta_{\phi} } } \,,
\ee
where the dots denote some finite part. As expected, the pole cancels and in the $\epsilon \to 0$ limit we get:
\be
		\Braket{ \{  \phi^2 \} (x) \{ \phi^2 \} (y) \{ \phi^2 \}(z) }
		\, = \, 
		N^{-\frac{3}{2}} \,% \mu^{6\Delta_{\phi}^{\star}+3\delta h_{\phi^2}} \,
		\frac{ 8  \, c(d/4)^3 \, \big(1 - Qg_2^{\star} %+\cdots 
		\big) }
		{|x-y|^{2\Delta_{\phi}^{\star} + \delta h_{\phi^2}} \, |y-z|^{2\Delta_{\phi}^{\star} + \delta h_{\phi^2}} \, |z-x|^{2\Delta_{\phi}^{\star} + \delta h_{\phi^2}}}  \, ,
\ee
which is the right form of the conformal three-point function with the correct anomalous dimension given in Eq.~\eqref{eq:phi2-andim}.

\begin{figure}[htb]
	\begin{center}
		\hspace{-20pt}\scalebox{1.4}{\includegraphics{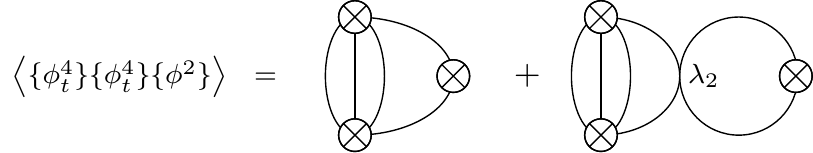}}
	\end{center}
	\caption{Leading order diagrams for $\Braket{ \{ \phi_t^4 \} \{ \phi_t^4 \}   \{ \phi^2 \}  }$.}
	\label{fig:tt2}
\end{figure}

%%%%%%%%%%%%%%%%%%%%%%%%%%%%%%%%%%%%%%%%%%%%%%%%%%%%%%%%
\paragraph{The $\Braket{ \{ \phi_t^4 \} \{ \phi_t^4 \} \{ \phi^2 \} }$ correlation.}
%%%%%%%%%%%%%%%%%%%%%%%%%%%%%%%%%%%%%%%%%%%%%%%%%%%%%%%%
For the three-point function
$\Braket{ \{ \phi_t^4\} \{ \phi_t^4 \} \{ \phi^2 \} } $
two graphs, represented in Fig.~\ref{fig:tt2} contribute.
%%%%%%%%%%%%%%%%%%%%%%%%%%%%%%%%%%%%%%%%%%%%%%%%%%%%%%%%
%%%%%%%%%%%%%%%%%%%%%%%%%%%%%%%%%%%%%%%%%%%%%%%%%%%%%%%%

We have:
\be
\begin{split}
& \Braket{ \{ \phi_t^4 \} (x) \{ \phi_t^4 \} (y) \{  \phi^2 \} (z) }= \,\mu^{\delta h_{\phi^2}}\,32\,   N^{-\frac{3}{2}} 
        C(x-y)^3 \Bigg[   Z_{\phi^2} C(x-z) C(y-z) \, \crcr
       & \qquad\qquad \qquad  - \, \lambda_2 \int d^du \, C(x-u) C(y-u) C(u-z)^2
		 \Bigg]  \,.
\end{split}
\ee
Using the conformal integral given in Appendix \ref{app:conf-int}, the right hand side above writes:
	\begin{equation}
	32 N^{-\frac{3}{2}} \,  c(\zeta)^5 \, \mu^{10\Delta_{\phi} + \delta h_{\phi^2}}
		\frac{  Z_{\phi^2} - Qg_2 \left[ \frac{1}{\epsilon} \, + \,  
		\log\frac{ |y'-z'| |z'-x'| }{|x'-y'|} %+ \cdots 
		\right]}
		{  |x'-y'|^{6\Delta_{\phi} } |y'-z'|^{ 2\Delta_{\phi} } 
		|z'-x'|^{ 2\Delta_{\phi} }} \, .
	\end{equation}
Again the pole cancels and in the $\epsilon \to 0$ limit we get:
	\begin{equation}
      \Braket{ \{ \phi_t^4 \} (x) \{ \phi_t^4 \}(y) 
		\{ \phi^2 \} (z)  } 
		\, = \, N^{-\frac{3}{2}} \,%\mu^{10\Delta_{\phi}^{\star}+\delta h_{\phi^2}} \, 
		\frac{	 32 \, c(d/4)^5 \,  \big(1 - Qg_2^{\star} %+\cdots
		 \big) } 
		{|x-y|^{6\Delta_{\phi}^{\star} -\delta h_{\phi^2}}
		\, |y-z|^{2\Delta_{\phi}^{\star} + \delta h_{\phi^2}} \, 
		|z-x|^{2\Delta_{\phi}^{\star} + \delta h_{\phi^2}}} \, ,
	\end{equation}
which is of the correct conformal form.

%%%%%%%%%%%%%%%%%%%%%%%%%%%%%%%%%%%%%%%%%%%%%%%%%%%%%%%%
\paragraph{The $\Braket{ \{  \phi_t^4 \} \{  \phi^2 \}  \{ \phi^2 \}}$ correlation.}
%%%%%%%%%%%%%%%%%%%%%%%%%%%%%%%%%%%%%%%%%%%%%%%%%%%%%%%%
The leading order contributions to this correlation are depicted in Fig.~\ref{eq:new-one-loop-scale-op}.

%%%%%%%%%%%%%%%%%%%%%%%%%%%%%%%%%%%%%%%%%%%%%%%%%%%%%%%%
\begin{figure}[htb]
	\begin{center}
		\hspace{-20pt}\scalebox{1.2}{\includegraphics{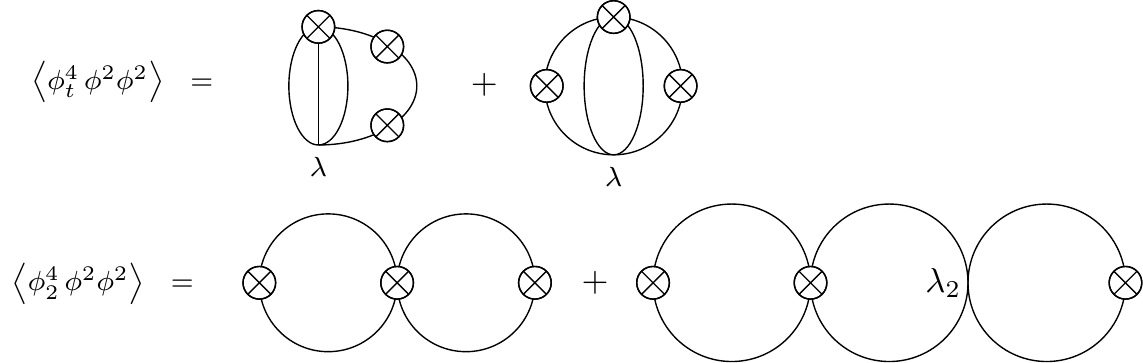}}
	\end{center}
	\caption{Leading diagrams for $\Braket{ \{ \phi_t^4 \} \{ \phi^2 \} \{ \phi^2 \} }$.}
	\label{fig:t22}
\end{figure}
%%%%%%%%%%%%%%%%%%%%%%%%%%%%%%%%%%%%%%%%%%%%%%%%%%%%%%%%
Taking into account the mixing and the scaling operators we get:
\begin{equation}
\begin{split}
    & \Braket{ \{\phi^4_t\} (x) \{ \phi^2 \} (y) \{  \phi^2 \} (z) } \crcr
    & \;\; = \, \mu^{2\delta h_{\phi^2}} \bigg(  \Braket{ \phi_t^4 (x)  \phi^2(y) \phi^2(z) } \, +    \, N^{-3/2} (\sqrt{3} Z_{\phi^4_2}+Z_{\phi^4_t;\phi^4_2}) Z_{\phi^2}^2 \, \Braket{ \phi_2^4 (x)  \phi^2(y) \phi^2(z) } \bigg) \crcr
    &\;\; = \, \mu^{2\delta h_{\phi^2}}  \, N^{-\frac{3}{2}} \bigg[ 16 \, \lambda \, C(x-y)C(y-z) \int d^du \, C(x-u)^3 C(u-z) \, + \, (y \leftrightarrow z) \crcr
		&\qquad \qquad \qquad + \, 48 \,  \lambda \, C(x-y)C(x-z) \int d^du \, C(x-u)^2 C(y-u) C(z-u) \crcr
		&\qquad \qquad\qquad + \, 8 \, \left( \sqrt{3} (1+ 4 Q \, \frac{g_2}{\epsilon}) - 6 Q \, \frac{g}{\epsilon} \right) \, C(x-y)^2 C(x-z)^2 \crcr
		&\qquad \qquad \qquad - 8 \sqrt{3} \lambda_2 C(x-y)^2 \int d^du \, C(x-u)^2 C(u-z)^2 \, + \, (y \leftrightarrow z) \bigg] \, ,
\end{split}
\end{equation}
where the terms correspond to the diagrams in Fig.\ref{fig:t22}. 
In order to obtain this equation we note that at order $N^{-3/2}$ and up to linear order in the couplings the term coming from the mixing with $\phi_1^4$ vanishes and $ \Braket{ \phi_t^4\phi^2 \phi^2 } $ starts at order $g$, hence we can ignore the renormalization of $\phi^2$ in this correlation. However in 
$\Braket{ \phi_2^4\phi^2 \phi^2 } $ we need to take it into account. We also omit the subtraction term coming from the mixing of $\phi^4_t$ with $\phi^2$.
We use once more  Eq.~\eqref{eq:convolution} and obtaining contact terms proportional to $\delta(x-z)/|x-y|^d$. In conjunction with Appendix.\ref{app:conf-int}, we find at first order in the coupling constant: 
	\begin{equation}
	 \begin{split}
    \Braket{ \{\phi^4_t\} (x) \{ \phi^2 \} (y) \{  \phi^2 \} (z) }  
     \, & = \, N^{-\frac{3}{2}} %\mu^{8\Delta_{\phi}^{\star}}
     \mu^{ 2\delta h_{\phi^2} } 
      \frac{16 \, c(d/4)^5 \, g^{\star}}{c(-d/4)}
		\left[ \frac{\delta(x-z)}{|x-y|^{4\Delta_{\phi}^{\star}}} \, + \, \frac{\delta(x-y)}{|x-z|^{4\Delta_{\phi}^{\star}}} \right] \crcr
    &\quad  + \, \, %\mu^{8\Delta_{\phi}^* + 2\delta h_{\phi^2} } \,
     N^{-\frac{3}{2}} \frac{8\sqrt{3} \, c(d/4)^4 \left[ 1 + 2\sqrt{3} Q g_\star (\kappa - 2\log 2 -2 \psi(d/4)) \right]}
		{|x-y|^{4\Delta_{\phi}^{\star}} |x-z|^{4\Delta_{\phi}^{\star}} |y-z|^{2\delta h_{\phi^2}}} \, ,
    	 \end{split}
	\end{equation}
where $\delta h_{\phi^2} =  Q\, g_2^{\star} = \sqrt{3} \, Q\, g^{\star}$ and $\kappa$ is defined in Eq.(\ref{eq:kappa}).
Notice that in the first term of the above correlator there is no correction to scaling associated to $\delta h_{\phi^2}$, because such term is proportional to $g^{\star}$, hence the scaling correction will appear at next order.

%%%%%%%%%%%%%%%%%%%%%%%%%%%%%%%%%%%%%%%%%%%%%%%%%%%%%%%%
%%%%%%%%%%%%%%%%%%%%%%%%%%%%%%%%%%%%%%%%%%%%%%%%%%%%%%%%
\section*{Acknowledgments}
%%%%%%%%%%%%%%%%%%%%%%%%%%%%%%%%%%%%%%%%%%%%%%%%%%%%%%%%
%%%%%%%%%%%%%%%%%%%%%%%%%%%%%%%%%%%%%%%%%%%%%%%%%%%%%%%%
We are grateful to Guillaume Bossard and Anastasios Petkou for helpful discussion.
We would also like to thank Sabine Harribey for her collaboration at the early stage of this project.

The work of DB and RG is supported by the European Research Council (ERC) under the European Union's Horizon 2020 research and innovation program (grant agreement No818066).
The work of KS is supported by the European Research Council (ERC) under the European Union's Horizon 2020 research and innovation program (grant agreement No758759). 
This work was partly supported by Perimeter Institute for Theoretical Physics.

\clearpage
\appendix

%%%%%%%%%%%%%%%%%%%%%%%%%%%%%%%%%%%%%%%%%%%%%%%%%%%%%%%%
%%%%%%%%%%%%%%%%%%%%%%%%%%%%%%%%%%%%%%%%%%%%%%%%%%%%%%%%
\section{Integrals}
\label{app:int}
%%%%%%%%%%%%%%%%%%%%%%%%%%%%%%%%%%%%%%%%%%%%%%%%%%%%%%%%
%%%%%%%%%%%%%%%%%%%%%%%%%%%%%%%%%%%%%%%%%%%%%%%%%%%%%%%%

We compute in this appendix several integrals we encountered in this paper.

%%%%%%%%%%%%%%%%%%%%%%%%%%%%%%%%%%%%%%%%%%%%%%%%%%%%%%%%
\paragraph{The subtracted melon integral.}
%%%%%%%%%%%%%%%%%%%%%%%%%%%%%%%%%%%%%%%%%%%%%%%%%%%%%%%%
We first compute the subtracted melon integral:
\be
      \int_{0}^{\infty} da_1da_2da_3 \; \; \frac{(a_1a_2a_3)^{\zeta-1}}{
  (a_1a_2+a_1a_3+a_2a_3)^{d/2}}\; \bigg(1- e^{ - \frac{a_1a_2a_3}{a_1a_2+a_1a_3+a_2a_3} }  \bigg) 
  \,.
\ee
Using a Taylor expansion with integral rest we have:
\be
\begin{split}
   & \int_0^1 dt 
  \int_{0}^{\infty} da \; 
  \frac{ (a_1a_2a_3)^{\zeta-1} }{
  \left( a_1a_2+a_1a_3+a_2a_3 \right)^{d/2}}
   \left( -   \frac{d}{dt} \;  e^{ -t \frac{ a_1 a_2a_3  }{a_1a_2 +a_1a_3 + a_2a_3 }  }
\right)\crcr
 &  \qquad =    \int_0^1 dt \; t^{ d - 3 \zeta -1 }
  \int_{0}^{\infty} da \; 
  \frac{ (a_1a_2a_3)^{\zeta} }{
  \left( a_1a_2 + a_1a_3+a_2a_3 \right)^{d/2+1}} 
 e^{ - \frac{a_1a_2a_3}{ a_1a_2+a_1a_3+a_2a_3 } } 
\,.
\end{split}
 \ee
The integral over $t$ converges for $d>3\zeta$. Let us compute a slight generalization of this integral to $q$ parameters 
$\alpha$. Changing variables to $\beta = a^{-1}$ and integrating out $t$ yields:
\begin{align}
   \; \frac{1}{  \frac{d}{2}(q-1)  -\zeta q  } \int_0^{\infty} d\beta \; 
 \frac{\prod_{i=1}^{q } \beta_i^{d/2 - \zeta-1} }{ \left( \sum_{i=1}^{q} \beta_i\right)^{d/2+1}}   e^{-\frac{1}{\sum_{i=1}^{q} \beta_i}} \,.
\end{align}
Introducing $x =  \sum_i \beta_i $ and $\beta_i =s_i x$ the integral becomes:
\begin{align}
 &    \frac{1}{  \frac{d}{2}(q-1)  -\zeta q }
\int_0^{\infty} d x \; x^{ q(d/2-\zeta) - d/2 -1 }  e^{-\frac{1}{x}}  \int_0^1 \frac{ds_1}{s_1}  s_1^{\frac{d}{2}-\zeta}
     \int_0^{1-s_1}  \frac{ds_2}{s_2} s_2^{\frac{d}{2} -\zeta } \dots \crcr
  & \qquad \qquad   \int_{0}^{1-s_1 -\dots -s_{q-2}} ds_{q-1} \;  
     s_{q-1}^{\frac{d}{2}-\zeta-1} (1 - s_1 - \dots s_{q-1})^{\frac{d}{2}-\zeta-1} \,.
\end{align}
We now use:
\be
 \int_0^{1-x} ds \; s^{u-1} (1-x-s)^{v -1} =
 (1-x)^{u + v-1} \frac{\Gamma(u)\Gamma(v)}{\Gamma(u+v)} \,,
\ee
and we finally obtain the subtracted integral:
 \be   
  \frac{\Gamma\bigg[1 - [\frac{d}{2}(q-1)  -\zeta q ] \bigg] }{ \frac{d}{2}(q-1)  -\zeta q  } \; 
  \frac{ \Gamma\left( \frac{d}{2} -\zeta \right)^{q} }{  \Gamma\left[   \left( \frac{d}{2} -\zeta \right) q   \right]} \,.
 \ee
 
%%%%%%%%%%%%%%%%%%%%%%%%%%%%%%%%%%%%%%%%%%%%%%%%%%%%%%%%
\paragraph{The $D$ integral.}
%%%%%%%%%%%%%%%%%%%%%%%%%%%%%%%%%%%%%%%%%%%%%%%%%%%%%%%%
We will repeatedly use below the integral\footnote{
We have:
\be
 \int_{0}^{\infty}[da] \;  
 \frac{ (a_1 a_2)^{ u -1}}{(a_1+a_2)^{ \gamma }} \; 
 e^{- (a_1+a_2) }   =  \int_{0}^{\infty} ds \; s^{-\gamma} e^{-s } \int_0^s da_1\; a_1^{u-1}(s-a_1)^{u-1} = \int_{0}^{\infty} ds \; s^{2u-\gamma-1} e^{-s } \int_0^1 dx\; x^{u-1}(1-x)^{u-1}   
 \,. \ee
}:
\be\label{eq:I}
  \int_{0}^{\infty}[da] \;  
 \frac{ (a_1 a_2)^{ u -1}}{(a_1+a_2)^{ \gamma }} \;
 e^{- (a_1+a_2) } =  \frac{\Gamma(u)^2 \Gamma(2u-\gamma)}{\Gamma(2u)} \,,
\ee
which is convergent for $2{\rm Re}( u ) > {\rm Re}(\gamma)$ and ${\rm Re}(u)>0$. In the the particular case $u = \zeta, \gamma  = d/2 $ we get:
\begin{align}
\label{eq:singtildeD}
  D  =  
  \frac{ \Gamma(\zeta)^2 }{\Gamma(2\zeta)} 
  \Gamma ( \epsilon /2 ) \,.
\end{align}
Denoting $\psi$ the digamma function (the logarithmic derivative of $\Gamma$) we have:
\be
D = \frac{2}{\epsilon} \;
 \frac{\Gamma(d/4+ \epsilon/4)^2}{\Gamma(d/2 + \epsilon/2) } 
 \Gamma(1+\epsilon /2) =  
 \frac{2}{\epsilon} \;
 \frac{\Gamma(d/4)^2}{\Gamma(d/2) } 
 + \frac{\Gamma(d/4)^2}{\Gamma(d/2) }  
 \bigg[ \psi(d/4) -  \psi(d/2) +  \psi(1)  \bigg] + O(\epsilon)
\ee
 
%%%%%%%%%%%%%%%%%%%%%%%%%%%%%%%%%%%%%%%%%%%%%%%%%%%%%%%%
\paragraph{The $ S_1$ integral.}
%%%%%%%%%%%%%%%%%%%%%%%%%%%%%%%%%%%%%%%%%%%%%%%%%%%%%%%%
The next integral we want to compute is:
\be
  S_1 =   \int_{0}^{\infty} [dadb] \; \frac{(a_1 a_2 b_1 b_2)^{\zeta-1}}{ \big[(a_1 + a_2)(b_1 + b_2 ) + b_1 b_2 \big] ^{d/2 }} \;e^{-(a_1+a_2+b_1+b_2)} \,.
\ee 
We use Mellin parameters to write:
\be
 \frac{1}{\big[(a_1 + a_2)(b_1 + b_2 ) + b_1 b_2 \big]^{d/2}}
  =  \int_{0^--\im \infty}^{0^- + \im \infty} \frac{dz}{2\pi \im} \; \frac{
  \Gamma(d/2 +z)}{\Gamma( d/2 )} \Gamma(-z) 
   \;  \frac{ (b_1b_2)^z }{(b_1+b_2)^{z+d/2}} \;\frac{1}{(a_1+a_2)^{z+d/2}} \,,
\ee
and Eq.~\eqref{eq:I} allows us to integrate $a$ and $b$. We thus obtain:
\be
   S_1 = \frac{\Gamma(\zeta)^2}{\Gamma(2\zeta)\Gamma(d/2) } 
\int_{0^--\im \infty}^{0^- + \im \infty} \frac{dz}{2\pi \im} \; 
\Gamma( d/2 + z)  \; 
\frac{\Gamma(\zeta+z )^2}{ \Gamma(2\zeta +2z) } 
\Gamma\left(\frac{\epsilon }{2} + z \right) 
\Gamma(-z) \Gamma\left(\frac{\epsilon}{2} - z \right) \,.
\ee
In the right half complex plane the integrand has poles at 
$z = n, n + \epsilon/2 $. Only the poles in $0,\epsilon/2$ have large residues at $\epsilon \to 0$, hence:
\begin{align}\label{eq:S1full}
  S_1 = & \frac{\Gamma(\zeta)^4}{\Gamma(2\zeta)^2}
  \Gamma(\epsilon/2)  \Gamma(\epsilon/2) + 
  \; \frac{ \Gamma(\zeta)^2 }{ \Gamma(2\zeta) \Gamma(d/2) }  
  \Gamma(d/2 + \epsilon/2)\frac{\Gamma(\zeta + \epsilon/2)^2}{ \Gamma(2\zeta +\epsilon ) } \Gamma(\epsilon) \Gamma(-\epsilon/2)   \crcr
   & + \frac{\Gamma(\zeta)^2}{\Gamma(2\zeta)\Gamma(d/2) } \int_{1^--\im \infty}^{1^- + \im \infty} \frac{dz}{2\pi \im} \; 
\Gamma(d/2  + z)  \; 
\frac{\Gamma(\zeta+z)^2}{ \Gamma(2\zeta +2z ) } \Gamma\left(\frac{\epsilon }{2} + z \right) 
\Gamma(-z) \Gamma\left(\frac{\epsilon}{2} - z \right)\,,
\end{align}
and the last line is finite in the $\epsilon \to 0$ limit.
At small $\epsilon$ we get:
\be
	S_1 \, = \, \frac{2}{\epsilon^2} \; \frac{\Gamma(d/4)^4}{\Gamma(d/2)^2}
	\, + \, \frac{1}{\epsilon} \;\frac{\Gamma(d/4)^4}{\Gamma(d/2)^2} \bigg[ 3\psi(1) - \psi(d/2) \bigg] + O(\epsilon^0) \, , 
\ee
In particular, we have
\be
2 S_1 - D^2 \, = \, \frac{2}{\epsilon} \, \frac{\Gamma(d/4)^4}{\Gamma(d/2)^2} \bigg[ \psi(1) + \psi(d/2) - 2 \psi(d/4) \bigg] + O(\epsilon^0) \, .
\ee

%%%%%%%%%%%%%%%%%%%%%%%%%%%%%%%%%%%%%%%%%%%%%%%%%%%%%%%%
\paragraph{The melon integral with momentum insertion.}
%%%%%%%%%%%%%%%%%%%%%%%%%%%%%%%%%%%%%%%%%%%%%%%%%%%%%%%%

We are interested in evaluating the coefficient of $p^{2n}$ in the Taylor expansion of the integral:
\be
    \int\frac{d^dq_1}{(2\pi)^d} \frac{d^dq_2}{(2\pi)^d}
   \frac{1}{[ (p+q_1+q_2)^2+ \mu^2]^{\zeta}} 
    \frac{1}{(q_2^2+\mu^2)^{\zeta}}  \frac{1}{(q_1^2+\mu^2)^{\zeta}} 
    (q_1^2)^{n}         \frac{1}{(q_1^2+\mu^2)^{\zeta}}    \,.
\ee
Using the parametric representation, and observing that $(q_1^2)^{n} e^{-a_1 q_1^2} = (-\partial_{a_1})^n e^{-a_1 q_1^2} $, the momentum integrals can be computed, yielding:
\be
 \frac{1}{ (4\pi)^d\Gamma(\zeta)^4}
  \int_{a,b} \; (a_1a_2b_1b_2)^{\zeta-1}  e^{-\mu^2(a_1+a_2+b_1+b_2)}
   \left( -\frac{ \partial }{ \partial a_1 }  \right)^n
    \bigg[ \frac{
     e^{- p^2\frac{ b_1b_2 (a_1+a_2)}{  (a_1+a_2)(b_1+b_2)+ b_1b_2 }}
    }{[(a_1+a_2)(b_1+b_2)+ b_1b_2]^{d/2}}\bigg] \,.
\ee
The coefficient of $p^{2n}$ in the Taylor expansion of this integral is 
$\frac{\mu^{-2\epsilon}}{ (4\pi)^d\Gamma(\zeta)^4} S_1^{(n)}$ with:
\be  S_1^{(n)}=\frac{1}{n!}
  \int_{a,b} \; (a_1a_2b_1b_2)^{\zeta-1}  e^{-(a_1+a_2+b_1+b_2)}
   \left( \frac{ \partial }{ \partial a_1 }  \right)^n
    \bigg[ \frac{  [ b_1b_2 (a_1+a_2) ]^n
    }{[(a_1+a_2)(b_1+b_2)+ b_1b_2]^{d/2+n }}\bigg] \,.
\ee
In order to compute the leading divergence of $S_1^{(n)}$, we note that:
\be
 \frac{1}{\big[(a_1 + a_2)(b_1 + b_2 ) + b_1 b_2 \big]^{d/2+n}}
  =  \int_{0^--\im \infty}^{0^- + \im \infty} \frac{dz}{2\pi \im} \; \frac{
  \Gamma(d/2 + n +z)}{\Gamma( d/2 + n )} \Gamma(-z) 
   \;   \frac{ (a_1+a_2)^z(b_1+b_2)^z }{ (b_1b_2)^{ d/2+ n + z} } \,,
\ee
and, as we encounter no singularities, we move the integration contour to 
$z = -d/2 + \im \mathbb{R}$. We thus get:
\be
\begin{split}
& S_1^{(n)}=\frac{1}{n!} \int_{-\im \infty}^{ \im \infty} \frac{dz}{2\pi \im} \; \frac{
  \Gamma( n  +z)}{\Gamma( d/2 +  n )} \Gamma(d/2-z) 
  \int_{b} \; \frac{  (b_1b_2)^{\zeta - z -1 }  }{(b_1+b_2)^{d/2-z}} e^{-(b_1+b_2)} \crcr
& \qquad \qquad   \int_{a} \; \frac{(a_1a_2)^{\zeta-1}}{(a_1+a_2)^{d/2-z}} 
e^{-(a_1+a_2)} \frac{\Gamma(n +1 -d/2 +z )}{\Gamma(n-d/2+z)} \,,
   \end{split}
\ee
and using Eq.~\eqref{eq:I} to integrate out the $a$'s and $b$'s we obtain:
\be
\begin{split}
 S_1^{(n)}=& \frac{1}{n!} \int_{-\im \infty}^{ \im \infty} \frac{dz}{2\pi \im} \; \frac{
  \Gamma( n  +z)}{\Gamma( d/2 +n )} \Gamma(d/2-z) \crcr
  & \qquad
   \frac{\Gamma(\zeta-z)^2}{\Gamma(2\zeta -2z )}
   \Gamma\big(\frac{\epsilon}{2} -z \big) \; 
   \frac{\Gamma(n+1 -d/2 +z )}{\Gamma(n-d/2+z)}\;
   \frac{\Gamma(\zeta)^2}{\Gamma(2\zeta)}\Gamma\big(\frac{\epsilon}{2} + z \big) \,.
\end{split}
\ee
The only pole of the integrand in the right-half complex plane with residue of order $1/\epsilon$ is located at $z = \epsilon/2$. Moving the contour across the pole we get:
\be
 S_1^{(n)} =  \frac{1}{\epsilon} \;  \frac{ \Gamma(d/4)^4 }{ \Gamma(d/2) } 
  \frac{\Gamma(n+1-d/2) }{n \Gamma(d/2+n) \Gamma(1-d/2)}
  + O(\epsilon^0) \,.
\ee
 
%%%%%%%%%%%%%%%%%%%%%%%%%%%%%%%%%%%%%%%%%%%%%%%%%%%%%%%%
\subsection{The conformal integrals}
\label{app:conf-int}
%%%%%%%%%%%%%%%%%%%%%%%%%%%%%%%%%%%%%%%%%%%%%%%%%%%%%%%%

We work at $\epsilon>0$, that is $\Delta_{\phi} = ( d-\epsilon)/ 4$. In the main text we encounter the following integrals involving bare propagators:
\be\label{eq:I_d(2)} 
\int d^du \, C(x-u)^2 C(y-u)^2  \,  = \, \frac{2 \, Q \, c(\zeta)^2}{ |x-y|^{4\Delta_{\phi} }} \, \left[ \frac{1}{\epsilon} + \log|x-y| - \log 2 - \psi(\tfrac{d}{4})
		\, + \, \mathcal{O}(\epsilon) \right] \, , 	
\ee
and 
\be \label{eq:I_d(3)}
\begin{split}
 		& \int d^du \, C(x-u) C(y-u) C(z-u)^2 \, =\\
        & \qquad \qquad \qquad  = \, \frac{Q \, c(\zeta)^2}{ |y-z|^{2\Delta_{\phi} } |x-z|^{2\Delta_{\phi} } } \,\bigg[ \frac{1}{\epsilon} +
		\log \left( \frac{ |z-x||z-y|}{|x-y|} \right) + \kappa \, + \, \mathcal{O}(\epsilon) \bigg] \, ,
\end{split}
\ee
with
	\begin{equation}
		\kappa \, = \, \frac{1}{2} \Big[ \psi(\tfrac{d}{2}) - 4 \psi(\tfrac{d}{4}) - \gamma -2\log 2 \Big] \, .
	\label{eq:kappa}
	\end{equation}

These integrals are computed using two conformal integrals. First we have:
	\begin{equation}
		I_d(\nu_1, \nu_2) \, = \, \int \frac{d^d u}{|x-u|^{2\nu_1} |y-u|^{2\nu_2}}
		\, = \, \frac{\pi^{\frac{d}{2}} \Gamma(d/2-\nu_1)\Gamma(d/2-\nu_2)\Gamma(\nu_1+\nu_2-d/2)}{\Gamma(\nu_1)\Gamma(\nu_2)\Gamma(d-\nu_1-\nu_2) \; \; |x-y|^{2\nu_1+2\nu_2-d} }\, ,
	\end{equation}
which follows from the Fourier transform (\ref{eq:fourier}).
In particular for $\nu_1=\nu_2=2\Delta_\phi=\frac{d-\epsilon}{2}$ we get:
	\begin{equation}
		I_d(2\Delta_\phi, 2\Delta_\phi) \, = \, \frac{2(2\pi)^d Q}{|x-y|^d} \left[ \frac{1}{\epsilon} \, + \, 2\log|x-y| \, + \, \mathcal{O}(\epsilon) \right] \, ,
	\end{equation}
and multiplying $c(\zeta)^4$ and rearranging the coefficient we obtain Eq.~\eqref{eq:I_d(2)}.

In order to prove Eq~.\eqref{eq:I_d(3)}, we start from:
	\begin{equation}
		I_d(\nu_1, \nu_2, \nu_3) \, = \, \int \frac{d^d u}{|x-u|^{2\nu_1} |y-u|^{2\nu_2} |z-u|^{2\nu_3}} \, .
	\end{equation}
In particular, we are interested in the case
$\nu_1  =  \nu_2 = \Delta_{\phi}  =( d-\epsilon)/4 , 
\, \nu_3 =  2 \Delta_{\phi} = (d-\epsilon)/2$.
Using the Mellin-Barnes representation \cite{Davydychev:1995mq}, we rewrite the integral as:
	\begin{align}
		I_d(\nu_1, \nu_2, \nu_3) \, &= \, \frac{\pi^{\frac{d}{2}} |x-y|^{d-2\sum \nu_i}}{\Gamma(d-\sum \nu_i) \prod\Gamma(\nu_i)}
		\int_{c-\im \infty}^{c+\im \infty} \frac{ds}{2\pi \im} \int_{c-\im \infty}^{c+\im \infty} \frac{dt}{2\pi \im} \, \alpha^s \beta^t \, \Gamma(-s) \Gamma(-t) \nonumber\\
		&\quad \times \, \Gamma(\tfrac{d}{2} - \nu_2 - \nu_3 -s) \Gamma(\tfrac{d}{2} - \nu_1 - \nu_3 -t) \Gamma(\nu_3 + s + t) \Gamma\left(\sum \nu_i - \tfrac{d}{2} +s +t\right) \, ,
	\end{align}
where we defined $\alpha = (y-z)^2/ (x-y)^2,  \, \beta  = (z-x)^2/ (x-y)^2$.
The integration contour (i.e. the constant $c$) is chosen to separate all poles of the first four Gamma functions from the poles of the last two Gamma functions.
We close the contour to the right so that we pick up all poles of the first four Gamma functions, but none of the poles of the last two Gamma functions. The relevant poles are located at: 
	\begin{align}
		&s \, = \, n_1 \, , \qquad s \, = \, \frac{d}{2} \, - \, \nu_2 \, - \, \nu_3 \, + \, n_2 \, , \nonumber\\
		&t \, = \, m_1 \, , \qquad t \, = \, \frac{d}{2} \, - \, \nu_1 \, - \, \nu_3 \, + \, m_2 \, , 
	\end{align}
with $n_{1,2}, m_{1,2}=0,1,2, \cdots$.
The complete answer for the integral is given by the sum of all of these pole contributions.
This is a daunting task to complete, so we look at the singular contribution in the limit $\epsilon \to 0$ with the 
choice $\nu_1  =  \nu_2 = \Delta_{\phi} , \, \nu_3 =  2 \Delta_{\phi}$.
We note that for this choice one of the Gamma functions in the overall coefficient becomes:
	\begin{equation}
		\Gamma\Big(d-\sum_i \nu_i \Big) \, = \, \Gamma(\epsilon) \, = \, \frac{1}{\epsilon} \, + \, \mathcal{O}(\epsilon^0) \, .
	\end{equation}
The leading behavior of the integral is $\mathcal{O}(\epsilon^{-2})$, coming from the poles at $n_2=m_2=0$.
Namely, the poles of the third and fourth gamma functions at $s=t=(3\epsilon-d)/4$ lead to a $\mathcal{O}(\epsilon^{-1})$ contribution from each of the fifth and sixth gamma functions.
Overall we get:
	\begin{equation}
		I_d\Big( \Delta_{\phi}, \Delta_{\phi}, 2 \Delta_{\phi} \Big) 
		\, = \, \frac{(2\pi)^d Q}{|y-z|^{\frac{d}{2}} |z-x|^{\frac{d}{2}}}
		\left[ \frac{1}{\epsilon} \, + \, \log\frac{|y-z|^{3/2}|z-x|^{3/2}}{ |x-y| } + \kappa'  + \mathcal{O}(\epsilon) \right] \, ,
	\end{equation}
where
	\begin{equation}
		\kappa' \, = \, \frac{1}{2} \Big[ \psi(\tfrac{d}{2}) - 2 \psi(\tfrac{d}{4}) - \gamma \Big] \, .
	\end{equation}
Multiplying by $c(\zeta)^4$, using $\Delta_{\phi}  = ( d-\epsilon)/4$, and rearranging the coefficient we obtain
Eq.~\eqref{eq:I_d(3)}.

%%%%%%%%%%%%%%%%%%%%%%%%%%%%%%%%%%%%%%%%%%%%%%%%%%%%%%%%
%%%%%%%%%%%%%%%%%%%%%%%%%%%%%%%%%%%%%%%%%%%%%%%%%%%%%%%%
\section{The Bilinear Operators}
\label{app:bilinear}
%%%%%%%%%%%%%%%%%%%%%%%%%%%%%%%%%%%%%%%%%%%%%%%%%%%%%%%%
%%%%%%%%%%%%%%%%%%%%%%%%%%%%%%%%%%%%%%%%%%%%%%%%%%%%%%%%
The spin-zero bilinear operators of the type $\phi(-\partial^2)^n \phi$ can be treated similarly to the $\phi^2$ perturbation.
We start by including a bare perturbation:
\be
 - \frac{1}{2}  \lambda_{(n)} \int d^dx \, \phi ( - \partial^2)^n \phi \,,
 \label{eq:perturbation}
\ee
and we evaluate the Taylor coefficient of $p^{2n}$ in the one-particle irreducible two-point function, which we denote $\Gamma^{R}_{(n)}$. As the tadpole is local, only the melon with one bi-valent vertex $ \lambda_{(n)}$ inserted on one of its edges contributes:
\be
\begin{split}
\Gamma^{R}_{(n)} & =  \lambda_{(n)} - 3\lambda^2  \lambda_{(n)}
 \bigg[ \int\frac{d^dq_1}{(2\pi)^d} \frac{d^dq_2}{(2\pi)^d}\crcr
& \qquad \qquad
   \frac{1}{[ (p+q_1+q_2)^2+ \mu^2]^{\zeta}} 
    \frac{1}{(q_2^2+\mu^2)^{\zeta}}  \frac{1}{(q_1^2+\mu^2)^{\zeta}} 
    (q_1^2)^{n}         \frac{1}{(q_1^2+\mu^2)^{\zeta}} \bigg]_{p^{2n}} \,,
\end{split}
\ee
where the subscript $p^{2n}$ signifies that we are only interested in the coefficient of $p^{2n}$ in the Taylor expansion of the integral. Using appendix \ref{app:int}, the bare expansion becomes:
\be
  \Gamma^{R}_{(n)} = \lambda_{(n)} - 3\lambda_{(n)} \frac{\lambda^2 }{(4\pi)^d\Gamma(\zeta)^4} \mu^{-2\epsilon}\; S_1^{(n)}  \,,\qquad
   S_1^{(n)} =  \frac{1}{\epsilon} \;  \frac{ \Gamma(d/4)^4 }{ \Gamma(d/2) } 
  \frac{\Gamma(n+1-d/2) }{n \Gamma(d/2+n) \Gamma(1-d/2)}
  + O(\epsilon^0) \,.
\ee
Similar to the mass parameter, we obtain:
\be
 \tilde \lambda_{(n)} = \mu^{d - \Delta_{(n)}} \bigg( \tilde g_{(n)} + 3 \tilde g^2 \tilde g_{(n)} S_1^{(n)} \bigg)
 \,,\qquad \beta_{(n)} = -(d-\Delta_{(n)}) \tilde g_{(n)} + 6 \tilde g_{(n)} \tilde g^2 \epsilon S_1^{(n)} \,,
\ee
where $\Delta_{(n)} =2\Delta_{\phi}+2n$ is the classical dimension
of the operator $\phi (-\partial^2)^n \phi$. We note that $\tilde g_{(n)}=0$ 
is always a fixed point of this equation, and that at $\epsilon=0$ the beta function simplifies to:
\be
 \beta_{(n)} = -\bigg[d- \bigg(2\Delta_{\phi}^{\star} +2n \bigg) \bigg] \tilde g_{(n)} +
 6\frac{ \Gamma(d/4)^4 }{ \Gamma(d/2) } 
  \frac{\Gamma(n+1-d/2) }{n \Gamma(d/2+n) \Gamma(1-d/2)} \tilde g_{(n)} \tilde g^2  \,.
\ee
At the fixed point $\tilde g^{\star}$ the operator $\phi(-\partial^2)^n \phi$ acquires an anomalous dimension
\be
 \delta h_{ (n) } = 6\frac{ \Gamma(d/4)^4 }{ \Gamma(d/2) } 
  \frac{\Gamma(n+1-d/2) }{n \Gamma(d/2+n) \Gamma(1-d/2)}  \tilde g^{\star}{}^2 \,, \qquad
  \Delta_{(n)}^{\star} = 2\Delta_{\phi}^{\star} + 2n + \delta h_{ (n) } \,,
\ee
reproducing the results derived in \cite{Benedetti:2019eyl} by diagonalizing the four-point kernel.

%%%%%%%%%%%%%%%%%%%%%%%%%%%%%%%%%%%%%%%%%%%%%%%%%%%%%%%%
%%%%%%%%%%%%%%%%%%%%%%%%%%%%%%%%%%%%%%%%%%%%%%%%%%%%%%%%
\section{The $1/N$ Expansion Revisited}
\label{app:1/N}
%%%%%%%%%%%%%%%%%%%%%%%%%%%%%%%%%%%%%%%%%%%%%%%%%%%%%%%%
%%%%%%%%%%%%%%%%%%%%%%%%%%%%%%%%%%%%%%%%%%%%%%%%%%%%%%%%

In the main body of the paper we are interested in tensors of rank $D=3$. However, the discussion below applies to any rank $D$. 

A \emph{$D$--colored graph} \cite{RTM,review,color} is a graph such that:
\begin{itemize}
 \item all the vertices are $D$--valent 
 \item the edges have a color $1,\dots D$ and at any vertex we have exactly one incident edge for each color
\end{itemize}

A $D$--colored graph is connected if any two vertices are joined by a path of (colored) edges such that two consecutive edges in the path share a vertex. 
For the $D$--colored graph $h$  we denote $V(h)$, $E(h)$, $C(h)$ and $F(h)$ the numbers of
vertices, edges, connected components and faces (i.e. bi colored cycles) of $h$.
We also denote $E^c(h)$ the number or edges of color $c$ and $F^c(h)$ the number of faces which contain the color $c$.

\paragraph{Invariants and Feynman graphs.} 

The $O(N)^D$ invariants $\Tr_{b}(T)$ are $D$--colored graphs \cite{RTM} 
$b$. The vertices of $b$ are associated to the tensors $T$ and the edges (colored $1,\dots D$) are associated to the contractions of indices:
\be
 \Tr_{b}(T) = \prod_{v\in b} T_{a^1_v a^2_v a^3_v}
  \prod_{e^c = (v,w)} \delta_{a^c_v a^c_w} \,,
\ee
where $v$ runs over the vertices of $b$ and $e^c$ over its edges ($c$ denotes the color of the edge $e^c$). The tetrahedral graph in $D=3$ corresponds to $
 \delta^t_{\mba\mbb\mbc\mbd}\, T_{\mba} T_{\mbb } T_{\mbc}T_{\mbd} $. 
 We call $b$ the \emph{bubbles}. 
 
 We are interested in the partition function:
\be
 W = \frac{1}{N^D} \ln \bigg(\int [dT] \;e^{-N^{D/2} \big(T_{\mba} T_{\mba} +
 \sum_{b} \lambda_{b} \, N^{-\rho_{b}} \Tr_{b}(T) \big) } \bigg) \,,
\ee
where $\rho_b\ge 0$ are scalings chosen such that the large $N$ limit of $W$ exists. Observe that, contrary to~\cite{Carrozza:2015adg}, we allow the bubbles $b$ to have several connected components. This is for instance the case of the double trace interaction bubble $\delta^d_{\mba\mbb;\mbc\mbd}\, T_{\mba} T_{\mbb } T_{\mbc}T_{\mbd} $. Somewhat abusively, we some times call a bubble with several connected components a ``multi-trace'' interaction.

The generating function $W$ is a sum over connect Feynman graphs $G$ which have a new color $0$ for the Wick contractions (propagators). As the propagators represent pairings of tensors, they connect vertices and $G$ is a $(D+1)$--colored graph. Denoting $G^{\hat 0}$ the graph obtained from $G$ by erasing the edges of color $0$ we have:
\be\label{eq:original}
W = \sum^{G^{\hat 0} = \cup b}_{{\rm i-connected} \; G }
\left( \prod_b - \lambda_{b} \right)
  N^{- D - \frac{D}{2} E^0(G) + \sum_b \big(\frac{D}{2} -\rho_{b} \big) + F^0(G)}
   \,.
\ee

Due to the disconnected bubbles (multi trace interactions), the notion of connectivity in equation \eqref{eq:original} subtle, hence the notation ``i-connected'' in the sum. The graph $G$ is i-connected if any two interaction bubbles are joined by a path of edges of color $0$ such that any two consecutive edge in the path are incident to the same interaction bubble. However, the graph $G$ can be disconnected as a colored graph, $C(G)>1$, because the edges in this path can be incident to different connected components in the bubbles. An example of an i-connected graph $G$ which has $C(G)>1$ is a double trace interaction decorated by two tadpole edges  $\delta^d_{\mba\mbb;\mbc\mbd}\, \braket{T_{\mba} T_{\mbb }}
\braket{T_{\mbc}T_{\mbd}} $.

It is a standard result \cite{review,RTM} that the total number of faces of a $(D+1)$--colored graph $G$ is:
\be
 F(G) = D  \, C(G) + \frac{D(D-1)}{4} V(G) - \bar \omega(G) \,,\qquad 
 \bar \omega(G) = \frac{1}{2(D-1)!} \sum_{\pi} k(\pi) \ge 0 \,,
\ee
where $\pi$ runs over the $D!$ jackets of $G$ (that is the embedding of $G$ corresponding to cycles over the colors) and $k(\pi)$ is the non orientable genus of the jacket $\pi$.
The non negative half integer $ \bar \omega(G)$ is the \emph{degree} of $G$.
The degree of a disconnected graph is the sum of the degrees of its connect components. The bubbles $b$ have only $D$ colors therefore:
\be
 F(b) = (D-1) \, C(b) + \frac{(D-1)(D-2)}{4}V(b) - \bar \omega(b) \,.
\ee

The crucial property of the degree is that for any $(D+1)$--colored graph $G$:
\be\label{eq:ineq1}
\bar \omega(G) \ge \frac{D}{D-1} \; \bar \omega(G^{\hat 0}) \,.
\ee
This is a bit subtle. As $G$ has $D+1$ colors, $G^{\hat 0}$ has only $D$ colors. There is a $D$ to $1$ correspondence between the jackets $\pi$ of $G$ and the jackets $\pi^{\hat 0}$ of $G^{\hat 0}$ consisting in deleting the edges of color $0$ in the jacket. As the non orientable genus can not increase by deleting edges we have $k(\pi)\ge k(\pi^{\hat 0})$ and consequently 
$\sum_{\pi} k(\pi) \ge D \sum_{\pi^{\hat 0}} k(\pi^{\hat 0})$. 

A $(D+1)$--colored graph $G$ has at most $\sum_{b\in G^{\hat 0} } C(b)$ connected components. If $G$ is i-connected, then it posses a tree of edges of color $0$ connecting all the bubbles. Each edge in this tree joins two connected components on two different bubbles, hence decrease the maximal number of connect components of $G$ by $1$. Overall we get an upper bound on the number of connected components of $G$:
\be\label{eq:ineq2}
  1 + \sum_{b\in G^{\hat 0}} \big[ C(b) -1 \big] \ge C(G) \,.
\ee

Among the invariants (i.e. bubbles), an interesting subclass consists in the 
\emph{maximally single trace} (MST) ones. They are those bubbles with only one face for each couple of colors. They are obviously connected and have exactly $D(D-1)/  2$ faces hence maximal possible degree:
\be
 \bar \omega(b) = -\frac{(D-1)(D-2)}{2} + \frac{(D-1)(D-2)}{4} V(b) \,,
\ee
at fixed number of vertices.

\paragraph{D=3.} Let us fix the ideas for $D=3$.  The bubbles $b$ are $3$ colored graphs. As such they are embedded graphs (ribbon graph, combinatorial map) with $   V(b) - E(b) + F(b) = 2C(b)- k(b) $, where $k(b)$ is the non orientable genus of $b$. Every $b$ admits two jackets, $(123)$ and $(132)$, which are identical up to orientation and have non orientable genus $k(b)$. The degree of $b$ is its non orientable genus $\bar \omega(b) = k(b)$. The MST invariants have three faces and non orientable genus $ k(b) = -1 + V(b) / 2 $.
For instance the tetrahedron is MST and has non orientable genus $1$. The wheel sextic interaction \cite{Benedetti:2019rja} is also MST and has non orientable genus $2$. The Feynman graphs $G$ have 4 color and 6 jackets, which are in $3\to 1$ correspondence with the jackets of $G^{\hat 0}$:
\be
  \begin{cases}
                   (0123) \\  
                    (0 2 3 1) \\
                    (0 3 1 2) 
                  \end{cases}  \to (123) \qquad
  \begin{cases}
                    (0 1 3 2)  \\
                    (0 2 1 3) \\
                    (0 3 2 1 )
                     \end{cases} \to (132) \,, \qquad
                     \omega(G) = \frac{1}{4} \sum_{6 \, {\rm cycles} \, \pi} k(\pi) \ge \frac{3}{2}  k(G^{\hat 0}) \,.
\ee

\paragraph{The $1/N$ series.} As $E^0(G) =\sum_b V(b) /2 $ and $F^0(G) = F(G) -\sum_b F(b)$ the scaling with $N$ of a Feynman graph $G$:
 \be
  - D - \frac{D}{4} \sum_b V(b) + \sum_b \bigg(\frac{D}{2} -\rho_{b} \bigg) + 
  F(G) - \sum_b F(b) \,.
 \ee
 
We now chose to scale all the invariants by the ``optimal scaling'' introduced in  \cite{Carrozza:2015adg,Ferrari:2017jgw}:
\be
 \rho_b = \frac{F(b) }{D-1} -  \frac{D}{2} \ge 0\,.
\ee
With this optimal scaling Eq.~\eqref{eq:original} becomes:
\be
W = \sum^{G^{\hat 0} = \cup b}_{{\rm i-connected} \; G }
\left( \prod_b - \lambda_{b} \right)
  N^{ -D \bigg(1 + \sum_b [C(b)-1 ]  -C(G )\bigg)  - \bigg(  \bar \omega(G) - \frac{D}{D-1} \sum_b \bar \omega(b) \bigg)}
   \,,
\ee
which, due to the inequalities~\eqref{eq:ineq1}~\eqref{eq:ineq2}, is a series in $1/N$ indexed by:
\be
 \omega(G) = D \bigg(1 + \sum_b [C(b)-1 ]  -C(G )\bigg)  + \bigg(  \bar \omega(G) - \frac{D}{D-1} \sum_b \bar \omega(b) \bigg) \ge 0 \,.
\ee

The optimal scaling leads to a good large $N$ limit. For some classes of interaction bubbles (like the MST or the melonic ones) the optimal scaling is the minimal scaling which still leads to a large $N$ limit.
It should be stressed however that this is not true in general: finding the minimal $\rho_b$ which still leads to a large $N $ limit for an arbitrary interaction is a difficult open question \cite{Bonzom:2016dwy}.
 
 \paragraph{Correlations.} The connected correlation of $p$ bubbles $b_1,\dots b_p$ is 
 \be
 \begin{split}
    \Braket{\Tr_{b_1}(T) \dots \Tr_{b_p}(T)}_{\rm connected} & = 
   N^{D - \sum_i \big( \frac{D}{2} -\rho_{b_i} \big) } \frac{\delta^p W}{ \delta \lambda_{b_1} \dots \delta \lambda_{b_p}}  \crcr
  &  = \sum^{G^{\hat 0} \supset \cup b_i}_{{\rm i-connected} \; G }
     N^{D - \sum_i \big( \frac{D}{2} -\rho_{b_i}  \big) -\omega(G) }
      \left( (-1)^p \prod^{b\neq b_i}_{b \subset G^{\hat 0} } - \lambda_{b} \right)
   \,.
 \end{split}
 \ee
For MST invariants we have:
\be
  \Braket{\Tr_{b_1}(T) \dots \Tr_{b_p}(T)}_{\rm connected} \lesssim N^{D- \frac{D}{2}p}
  \,,
\ee
which leads to a large $N$ factorization of the expectations. Any correlation factors into connected correlations:
\be
 \Braket{\Tr_{b_1}(T) \dots \Tr_{b_p}(T) } =
  \sum_{P} \prod_{B\in P}  \Braket{ \prod_{j\in B} \Tr_{b_j}(T) }_{\rm connected}
  \,,
\ee
where the sum runs over the partitions $P$ of $\{1, \dots p\}$, $B$ runs over the the blocks in the partition $P$ and $j$ over the elements in the block $B$. In this decomposition the partitions with the larges number of blocks will dominate. As the one point functions are zero in a CFT, the dominant partition will have either only connected two-point functions or at most one connected three-point function. 
Furthermore, if $O$ is a product of single trace operators and $[ \Tr_{b_1}(T) \Tr_{b_1}(T) ]$ is a double trace one we have:
 \be
 \begin{split}
 & \Braket{ O [ \Tr_{b_1}(T) \Tr_{b_1}(T) ] }_{\rm connected}
   =   \Braket{O \Tr_{b_1}(T) \Tr_{b_2}(T) }_{\rm connected} \crcr
& \qquad \qquad   +  \Braket{O \Tr_{b_1}(T)  }_{\rm connected}
     \Braket{  \Tr_{b_2}(T) }_{\rm connected}
  +   \Braket{O \Tr_{b_2}(T)  }_{\rm connected}
  \Braket{  \Tr_{b_1}(T) }_{\rm connected} \,,
 \end{split}
 \ee
and the last two terms are zero in a CFT.
 
%%%%%%%%%%%%%%%%%%%%%%%%%%%%%%%%%%%%%%%%%%%%%%%%%%%%%%%%
%%%%%%%%%%%%%%%%%%%%%%%%%%%%%%%%%%%%%%%%%%%%%%%%%%%%%%%%
\section{Comments on the Correlation Functions of $\phi^4_1$}
\label{app:pillow}
%%%%%%%%%%%%%%%%%%%%%%%%%%%%%%%%%%%%%%%%%%%%%%%%%%%%%%%%
%%%%%%%%%%%%%%%%%%%%%%%%%%%%%%%%%%%%%%%%%%%%%%%%%%%%%%%%

The $\phi^4_1$ operator is neither MST nor MMT, as the pillow is a connected invariant with $\rho_b=1/2$.
In order to get a general idea of how it contributes to $n$-point functions we will consider first a simplified with only pillow operators of a single type, and then we will explicitly compute the two-point function of $\phi^4_1$ at leading order.

Let us first consider a general correlator of $n$ pillow operators with an arbitrary number of perturbative pillow vertices. 
We restrict to a single type of pillow operator, with single lines of color one: ${\cal O}_p = \phi_{a_1 b_1 c_1} \phi_{a_2 b_1 c_1} \phi_{a_1 b_2 c_2} \phi_{a_2 b_2 c_2} $.
The intermediate field representation, known also as Hubbard-Stratonovich transformation, amounts to replacing it in the path integral by the integral over an auxiliary real symmetric $N\times N$ matrix field (the intermediate field), with ultralocal free covariance proportional to $N$ (from the scaling of the pillow in the original action) and which couples to the composite matrix $\phi_{a_1 b_1 c_1} \phi_{a_2 b_1 c_1}$ (see for example \cite{Benedetti:2015ara,Benedetti:2017fmp}). 
The original field appears then only quadratically in the new action, and thus it forms $V_q$ loop-vertices of valency $q$, for $q\geq 1$, each containing two faces of the tensor model. Denoting by $E$ the number of intermediate field propagators, and by $F$ the number of faces that the intermediate field forms, we thus have that the connected $n$-point function of pillows scales as
\be
N^{\sum_{q\geq 1}(2-\frac{3}{2} q) V_q + F +E -n} = N^{\sum_{q\geq 1}(1-\frac{q}{2}) V_q + 2-2g -n} \,,
\ee
where the factor $n$ is due to the fact that the inserted operators, unlike the perturbative vertices, carry no factor $N$.
The amplitude would therefore be dominated by an intermediate field graph which is planar and which maximizes the number of univalent loop-vertices, that is, a usual cactus diagram. However, assuming that univalent loop-vertices (tadpoles in the original representation) have zero amplitude, we are left with dominant graphs being made of two-valent loop-vertices, joined in a planar way. Their amplitude scales like $N^{2-n}$. 
Comparing with Eq.~\eqref{eq:Wexp}, this means that such dominant graphs have $\omega=1$.
Therefore, we conclude that also pillows at criticality have two-point functions of order $N^0$, and higher-point functions suppressed in $1/N$.

%%%%%%%%%%%%%%%%%%%%%%%%%%%%%%%%%%%%%%%%%%%%%%%%%%%%%%%%
\paragraph{Two-point function of $\phi^4_1$.}
%%%%%%%%%%%%%%%%%%%%%%%%%%%%%%%%%%%%%%%%%%%%%%%%%%%%%%%%
We now study the two-point function $\Braket{ \{ \phi^4_1 \} \{ \phi^4_1 \}  }$.
The same two graphs as for the correlation $\Braket{ \{ \phi^4_2 \} \{  \phi^4_2 \} }$ contribute with replacing $\lambda_2$ by $\lambda_1$.
The global scaling is $N^0$ (as for the other two-point functions) and we get:
\be
 \Braket{ \{ \phi^4_1 \} (x) \{\phi^4_1 \} (y) } = 
  \mu^{2\delta h_1} Z_{\phi^4_1}^2 \bigg( 4 C(x-y)^4  - 8 \lambda_1 C(x-y)^2 \int d^dz \, C(x-z)^2 C(z-y)^2  \bigg) \,.
\ee
Recalling that $Z_{\phi^4_1} = 1+2 Q \frac{g_1}{\epsilon}$, the $1/\epsilon$ pole cancels and in the $\epsilon \to 0$ limit we obtain:
\be
 \Braket{ \{ \phi^4_1  \} (x) \{  \phi^4_1 \} (y)  } =  
 \mu^{8\Delta_{\phi}^{\star} +2\delta h_1} 
 \frac{4 \, c(d/4)^4}{ |x'-y'|^{8\Delta_{\phi}^{\star} + 2\delta h_1 }}
 \bigg[1 + 4 \, Q \, g_1^{\star} \bigg(  \log 2 + \psi(d/4) \bigg)  \bigg] \,,
\ee
with $\delta h_1 = 2 \, Q \, g_1^{\star}= 2 \, Q \, g^{\star}$ reproducing the perturbative computation in Eq.~\eqref{eq:dim-an-quart}.

\newpage

%----- Bibliography ----------------------

%---------------------------------------------

\addcontentsline{toc}{section}{References}

%---------------------------------------------

\providecommand{\href}[2]{#2}\begingroup\raggedright\endgroup


\begin{thebibliography}{10}

\bibitem{Benedetti:2019eyl}
D.~Benedetti, R.~Gurau and S.~Harribey, \emph{Line of fixed points in a bosonic
  tensor model}, \href{https://doi.org/10.1007/JHEP06(2019)053}{JHEP {\bfseries
  06} (2019) 053} [\href{https://arxiv.org/abs/1903.03578}{{\tt
  arXiv:1903.03578}}].

\bibitem{Benedetti:2019ikb}
D.~Benedetti, R.~Gurau, S.~Harribey and K.~Suzuki, \emph{{Hints of unitarity at
  large $N$ in the $O(N)^3$ tensor field theory}}, \href{https://doi.org/10.1007/JHEP02(2020)072}{JHEP {\bfseries
  02} (2020) 072} 
  \href{https://arxiv.org/abs/1909.07767}{{\tt arXiv:1909.07767}}.

\bibitem{Poland:2018epd}
D.~Poland, S.~Rychkov and A.~Vichi, \emph{The conformal bootstrap: Theory,
  numerical techniques, and applications},
  \href{https://doi.org/10.1103/RevModPhys.91.015002}{Rev.Mod.Phys. {\bfseries
  91} (2019) 015002} [\href{https://arxiv.org/abs/1805.04405}{{\tt
  arXiv:1805.04405}}].

\bibitem{Nakayama:2013is}
Y.~Nakayama, \emph{Scale invariance vs conformal invariance},
  \href{https://doi.org/https://doi.org/10.1016/j.physrep.2014.12.003}{Physics
  Reports {\bfseries 569} (2015) 1 }
  [\href{https://arxiv.org/abs/1302.0884}{{\tt arXiv:1302.0884}}].

\bibitem{Wilson:1972cf}
K.~G. Wilson, \emph{Quantum field theory models in less than four-dimensions},
  \href{https://doi.org/10.1103/PhysRevD.7.2911}{Phys.Rev.D {\bfseries 7}
  (1973) 2911}.

\bibitem{Bonzom:2011zz}
V.~Bonzom, R.~Gurau, A.~Riello and V.~Rivasseau, \emph{{Critical behavior of
  colored tensor models in the large {$N$} limit}},
  \href{https://doi.org/10.1016/j.nuclphysb.2011.07.022}{Nucl. Phys. {\bfseries
  B853} (2011) 174} [\href{https://arxiv.org/abs/1105.3122}{{\tt
  arXiv:1105.3122}}].

\bibitem{RTM}
R.~Gurau, \emph{{Random Tensors}}. Oxford University Press, Oxford, 2016.

\bibitem{Klebanov:2018fzb}
I.~R. Klebanov, F.~Popov and G.~Tarnopolsky, \emph{{TASI Lectures on Large $N$
  Tensor Models}}, \href{https://doi.org/10.22323/1.305.0004}{PoS {\bfseries
  TASI2017} (2018) 004} [\href{https://arxiv.org/abs/1808.09434}{{\tt
  arXiv:1808.09434}}].

\bibitem{Ferrari:2017jgw}
F.~Ferrari, V.~Rivasseau and G.~Valette, \emph{{A New Large $N$ Expansion for
  General Matrix?Tensor Models}},
  \href{https://doi.org/10.1007/s00220-019-03511-7}{Commun. Math. Phys.
  {\bfseries 370} (2019) 403} [\href{https://arxiv.org/abs/1709.07366}{{\tt
  arXiv:1709.07366}}].

\bibitem{Prakash:2019zia}
S.~Prakash and R.~Sinha, \emph{Melonic dominance in subchromatic sextic tensor
  models},  \href{https://arxiv.org/abs/1908.07178}{{\tt arXiv:1908.07178}}.

\bibitem{Moshe:2003xn}
M.~Moshe and J.~Zinn-Justin, \emph{{Quantum field theory in the large N limit:
  A Review}}, \href{https://doi.org/10.1016/S0370-1573(03)00263-1}{Phys. Rept.
  {\bfseries 385} (2003) 69} [\href{https://arxiv.org/abs/hep-th/0306133}{{\tt
  arXiv:hep-th/0306133}}].

\bibitem{DiFrancesco:1993nw}
P.~Di~Francesco, P.~H. Ginsparg and J.~Zinn-Justin, \emph{{{2-D} {Gravity} and
  random matrices}}, \href{https://doi.org/10.1016/0370-1573(94)00084-G}{Phys.
  Rept. {\bfseries 254} (1995) 1}
  [\href{https://arxiv.org/abs/hep-th/9306153}{{\tt arXiv:hep-th/9306153}}].

\bibitem{Giombi:2017dtl}
S.~Giombi, I.~R. Klebanov and G.~Tarnopolsky, \emph{{Bosonic tensor models at
  large {$N$} and small {$\epsilon$}}},
  \href{https://doi.org/10.1103/PhysRevD.96.106014}{Phys. Rev. {\bfseries D96}
  (2017) 106014} [\href{https://arxiv.org/abs/1707.03866}{{\tt
  arXiv:1707.03866}}].

\bibitem{Prakash:2017hwq}
S.~Prakash and R.~Sinha, \emph{{A Complex Fermionic Tensor Model in $d$
  Dimensions}}, \href{https://doi.org/10.1007/JHEP02(2018)086}{JHEP {\bfseries
  02} (2018) 086} [\href{https://arxiv.org/abs/1710.09357}{{\tt
  arXiv:1710.09357}}].

\bibitem{Benedetti:2017fmp}
D.~Benedetti, S.~Carrozza, R.~Gurau and A.~Sfondrini, \emph{{Tensorial
  Gross-Neveu models}}, \href{https://doi.org/10.1007/JHEP01(2018)003}{JHEP
  {\bfseries 01} (2018) 003} [\href{https://arxiv.org/abs/1710.10253}{{\tt
  arXiv:1710.10253}}].

\bibitem{Giombi:2018qgp}
S.~Giombi, I.~R. Klebanov, F.~Popov, S.~Prakash and G.~Tarnopolsky,
  \emph{{Prismatic Large $N$ Models for Bosonic Tensors}},
  \href{https://doi.org/10.1103/PhysRevD.98.105005}{Phys. Rev. {\bfseries D98}
  (2018) 105005} [\href{https://arxiv.org/abs/1808.04344}{{\tt
  arXiv:1808.04344}}].

\bibitem{Benedetti:2018ghn}
D.~Benedetti and N.~Delporte, \emph{{Phase diagram and fixed points of
  tensorial Gross-Neveu models in three dimensions}},
  \href{https://doi.org/10.1007/JHEP01(2019)218}{JHEP {\bfseries 01} (2019)
  218} [\href{https://arxiv.org/abs/1810.04583}{{\tt arXiv:1810.04583}}].

\bibitem{Popov:2019nja}
F.~K. Popov, \emph{Supersymmetric tensor model at large $n$ and small
  $\epsilon$}, \href{https://doi.org/10.1103/PhysRevD.101.026020}{Phys.Rev.D
  {\bfseries 101} (2020) 026020} [\href{https://arxiv.org/abs/1907.02440}{{\tt
  arXiv:1907.02440}}].

\bibitem{Ambjorn:1990ge}
J.~Ambjorn, B.~Durhuus and T.~Jonsson, \emph{{Three-dimensional simplicial
  quantum gravity and generalized matrix models}},
  \href{https://doi.org/10.1142/S0217732391001184}{Mod. Phys. Lett. {\bfseries
  A6} (1991) 1133}.

\bibitem{Sasakura:1990fs}
N.~Sasakura, \emph{{Tensor model for gravity and orientability of manifold}},
  \href{https://doi.org/10.1142/S0217732391003055}{Mod. Phys. Lett. {\bfseries
  A6} (1991) 2613}.

\bibitem{color}
R.~Gurau, \emph{{Colored Group Field Theory}},
  \href{https://doi.org/10.1007/s00220-011-1226-9}{Commun. Math. Phys.
  {\bfseries 304} (2011) 69} [\href{https://arxiv.org/abs/0907.2582}{{\tt
  arXiv:0907.2582}}].

\bibitem{review}
R.~Gurau and J.~P. Ryan, \emph{{Colored tensor models - a review}},
  \href{https://doi.org/10.3842/SIGMA.2012.020}{SIGMA {\bfseries 8} (2012) 020}
  [\href{https://arxiv.org/abs/1109.4812}{{\tt arXiv:1109.4812}}].

\bibitem{Oriti:2011jm}
D.~Oriti, \emph{{The microscopic dynamics of quantum space as a group field
  theory}},  in \emph{{Proceedings, Foundations of Space and Time: Reflections
  on Quantum Gravity}}, pp.~257--320, 2011,
  \href{https://arxiv.org/abs/1110.5606}{{\tt arXiv:1110.5606}},
  \href{https://inspirehep.net/record/942719/files/arXiv:1110.5606.pdf}{https://inspirehep.net/record/942719/files/arXiv:1110.5606.pdf}.

\bibitem{Witten:2016iux}
E.~Witten, \emph{An {SYK}-like model without disorder},
  \href{https://doi.org/10.1088/1751-8121/ab3752}{J. Phys. {\bfseries A52}
  (2019) 474002} [\href{https://arxiv.org/abs/1610.09758}{{\tt
  arXiv:1610.09758}}].

\bibitem{Gurau:2016lzk}
R.~Gurau, \emph{{The complete {$1/N$} expansion of a SYK--like tensor model}},
  \href{https://doi.org/10.1016/j.nuclphysb.2017.01.015}{Nucl. Phys. {\bfseries
  B916} (2017) 386} [\href{https://arxiv.org/abs/1611.04032}{{\tt
  arXiv:1611.04032}}].

\bibitem{Klebanov:2016xxf}
I.~R. Klebanov and G.~Tarnopolsky, \emph{Uncolored random tensors, melon
  diagrams, and the {SYK} models},
  \href{https://doi.org/10.1103/PhysRevD.95.046004}{Phys. Rev. {\bfseries D95}
  (2017) 046004} [\href{https://arxiv.org/abs/1611.08915}{{\tt
  arXiv:1611.08915}}].

\bibitem{Peng:2016mxj}
C.~Peng, M.~Spradlin and A.~Volovich, \emph{{A Supersymmetric SYK-like Tensor
  Model}}, \href{https://doi.org/10.1007/JHEP05(2017)062}{JHEP {\bfseries 05}
  (2017) 062} [\href{https://arxiv.org/abs/1612.03851}{{\tt
  arXiv:1612.03851}}].

\bibitem{Krishnan:2016bvg}
C.~Krishnan, S.~Sanyal and P.~N. Bala~Subramanian, \emph{{Quantum Chaos and
  Holographic Tensor Models}},
  \href{https://doi.org/10.1007/JHEP03(2017)056}{JHEP {\bfseries 03} (2017)
  056} [\href{https://arxiv.org/abs/1612.06330}{{\tt arXiv:1612.06330}}].

\bibitem{Krishnan:2017lra}
C.~Krishnan, K.~V. Pavan~Kumar and D.~Rosa, \emph{Contrasting {SYK}-like
  models}, \href{https://doi.org/10.1007/JHEP01(2018)064}{JHEP {\bfseries 01}
  (2018) 064} [\href{https://arxiv.org/abs/1709.06498}{{\tt
  arXiv:1709.06498}}].

\bibitem{Bulycheva:2017ilt}
K.~Bulycheva, I.~R. Klebanov, A.~Milekhin and G.~Tarnopolsky, \emph{Spectra of
  operators in large {$N$} tensor models},
  \href{https://doi.org/10.1103/PhysRevD.97.026016}{Phys. Rev. {\bfseries D97}
  (2018) 026016} [\href{https://arxiv.org/abs/1707.09347}{{\tt
  arXiv:1707.09347}}].

\bibitem{Choudhury:2017tax}
S.~Choudhury, A.~Dey, I.~Halder, L.~Janagal, S.~Minwalla and R.~Poojary,
  \emph{{Notes on melonic $O(N)^{q-1}$ tensor models}},
  \href{https://doi.org/10.1007/JHEP06(2018)094}{JHEP {\bfseries 06} (2018)
  094} [\href{https://arxiv.org/abs/1707.09352}{{\tt arXiv:1707.09352}}].

\bibitem{Halmagyi:2017leq}
N.~Halmagyi and S.~Mondal, \emph{Tensor models for black hole probes},
  \href{https://doi.org/10.1007/JHEP07(2018)095}{JHEP {\bfseries 07} (2018)
  095} [\href{https://arxiv.org/abs/1711.04385}{{\tt arXiv:1711.04385}}].

\bibitem{Klebanov:2018nfp}
I.~R. Klebanov, A.~Milekhin, F.~Popov and G.~Tarnopolsky, \emph{{Spectra of
  eigenstates in fermionic tensor quantum mechanics}},
  \href{https://doi.org/10.1103/PhysRevD.97.106023}{Phys. Rev. {\bfseries D97}
  (2018) 106023} [\href{https://arxiv.org/abs/1802.10263}{{\tt
  arXiv:1802.10263}}].

\bibitem{Carrozza:2018psc}
S.~Carrozza and V.~Pozsgay, \emph{{SYK-like tensor quantum mechanics with
  $\mathrm{Sp}(N)$ symmetry}},
  \href{https://doi.org/10.1016/j.nuclphysb.2019.02.012}{Nucl. Phys. {\bfseries
  B941} (2019) 28} [\href{https://arxiv.org/abs/1809.07753}{{\tt
  arXiv:1809.07753}}].

\bibitem{Klebanov:2019jup}
I.~R. Klebanov, P.~N. Pallegar and F.~K. Popov, \emph{Majorana fermion quantum
  mechanics for higher rank tensors},
  \href{https://doi.org/10.1103/PhysRevD.100.086003}{Phys.Rev.D {\bfseries 100}
  (2019) 086003} [\href{https://arxiv.org/abs/1905.06264}{{\tt
  arXiv:1905.06264}}].

\bibitem{Ferrari:2019ogc}
F.~Ferrari and F.~I. Schaposnik~Massolo, \emph{Phases of melonic quantum
  mechanics}, \href{https://doi.org/10.1103/PhysRevD.100.026007}{Phys.Rev.D
  {\bfseries 100} (2019) 026007} [\href{https://arxiv.org/abs/1903.06633}{{\tt
  arXiv:1903.06633}}].

\bibitem{Delporte:2018iyf}
N.~Delporte and V.~Rivasseau, \emph{The tensor track {V}: Holographic tensors},
   in \emph{{Proceedings, 17th Hellenic School and Workshops on Elementary
  Particle Physics and Gravity (CORFU2017)}}, SISSA, 2018,
  \href{https://arxiv.org/abs/1804.11101}{{\tt arXiv:1804.11101}},
  \href{https://pos.sissa.it/318/}{https://pos.sissa.it/318/}.

\bibitem{Sachdev:1992fk}
S.~Sachdev and J.~Ye, \emph{{Gapless spin fluid ground state in a random,
  quantum Heisenberg magnet}},
  \href{https://doi.org/10.1103/PhysRevLett.70.3339}{Phys. Rev. Lett.
  {\bfseries 70} (1993) 3339}
  [\href{https://arxiv.org/abs/cond-mat/9212030}{{\tt
  arXiv:cond-mat/9212030}}].

\bibitem{Kitaev2015}
A.~Kitaev, \emph{{A simple model of quantum holography}}, {KITP strings seminar
  and Entanglement 2015 (Feb. 12, April 7, and May 27, 2015) }.

\bibitem{Maldacena:2016hyu}
J.~Maldacena and D.~Stanford, \emph{{Remarks on the Sachdev-Ye-Kitaev model}},
  \href{https://doi.org/10.1103/PhysRevD.94.106002}{Phys. Rev. {\bfseries D94}
  (2016) 106002} [\href{https://arxiv.org/abs/1604.07818}{{\tt
  arXiv:1604.07818}}].

\bibitem{Polchinski:2016xgd}
J.~Polchinski and V.~Rosenhaus, \emph{{The spectrum in the Sachdev-Ye-Kitaev
  model}}, \href{https://doi.org/10.1007/JHEP04(2016)001}{JHEP {\bfseries 04}
  (2016) 001} [\href{https://arxiv.org/abs/1601.06768}{{\tt
  arXiv:1601.06768}}].

\bibitem{Jevicki:2016bwu}
A.~Jevicki, K.~Suzuki and J.~Yoon, \emph{Bi-local holography in the {SYK}
  model}, \href{https://doi.org/10.1007/JHEP07(2016)007}{JHEP {\bfseries 07}
  (2016) 007} [\href{https://arxiv.org/abs/1603.06246}{{\tt
  arXiv:1603.06246}}].

\bibitem{Gross:2016kjj}
D.~J. Gross and V.~Rosenhaus, \emph{{A generalization of Sachdev-Ye-Kitaev}},
  \href{https://doi.org/10.1007/JHEP02(2017)093}{JHEP {\bfseries 02} (2017)
  093} [\href{https://arxiv.org/abs/1610.01569}{{\tt arXiv:1610.01569}}].

\bibitem{Carrozza:2015adg}
S.~Carrozza and A.~Tanasa, \emph{{$O(N)$} random tensor models},
  \href{https://doi.org/10.1007/s11005-016-0879-x}{Lett. Math. Phys. {\bfseries
  106} (2016) 1531} [\href{https://arxiv.org/abs/1512.06718}{{\tt
  arXiv:1512.06718}}].

\bibitem{Bonzom:2012hw}
V.~Bonzom, R.~Gurau and V.~Rivasseau, \emph{Random tensor models in the large n
  limit: Uncoloring the colored tensor models},
  \href{https://doi.org/10.1103/PhysRevD.85.084037}{Phys.Rev.D {\bfseries 85}
  (2012) 084037} [\href{https://arxiv.org/abs/1202.3637}{{\tt
  arXiv:1202.3637}}].

\bibitem{Fisher:1972zz}
M.~E. Fisher, S.-k. Ma and B.~Nickel, \emph{Critical exponents for long-range
  interactions},
  \href{https://doi.org/10.1103/PhysRevLett.29.917}{Phys.Rev.Lett. {\bfseries
  29} (1972) 917}.

\bibitem{Sak:1973}
J.~Sak, \emph{Recursion relations and fixed points for ferromagnets with
  long-range interactions}, \href{https://doi.org/10.1103/PhysRevB.8.281}{Phys.
  Rev. B {\bfseries 8} (1973) 281}.

\bibitem{Brydges:2002wq}
D.~C. Brydges, P.~K. Mitter and B.~Scoppola, \emph{{Critical (Phi**4)(3,
  epsilon)}}, \href{https://doi.org/10.1007/s00220-003-0895-4}{Commun. Math.
  Phys. {\bfseries 240} (2003) 281}
  [\href{https://arxiv.org/abs/hep-th/0206040}{{\tt arXiv:hep-th/0206040}}].

\bibitem{Abdesselam:2006qg}
A.~Abdesselam, \emph{A complete renormalization group trajectory between two
  fixed points},
  \href{https://doi.org/10.1007/s00220-007-0352-x}{Commun.Math.Phys. {\bfseries
  276} (2007) 727} [\href{https://arxiv.org/abs/math-ph/0610018}{{\tt
  arXiv:math-ph/0610018}}].

\bibitem{Brezin:2014}
E.~Brezin, G.~Parisi and F.~Ricci-Tersenghi, \emph{The crossover region between
  long-range and short-range interactions for the critical exponents},
  \href{https://doi.org/10.1007/s10955-014-1081-0}{J. Stat. Phys. {\bfseries
  157} (2014) 855} [\href{https://arxiv.org/abs/1407.3358}{{\tt
  arXiv:1407.3358}}].

\bibitem{Defenu:2014}
N.~Defenu, A.~Trombettoni and A.~Codello, \emph{{Fixed-point structure and
  effective fractional dimensionality for $O(N)$ models with long-range
  interactions}}, \href{https://doi.org/10.1103/physreve.92.052113}{Phys. Rev.
  {\bfseries E92} (2015) 052113} [\href{https://arxiv.org/abs/1409.8322}{{\tt
  arXiv:1409.8322}}].

\bibitem{Paulos:2015jfa}
M.~F. Paulos, S.~Rychkov, B.~C. van Rees and B.~Zan, \emph{Conformal invariance
  in the long-range ising model},
  \href{https://doi.org/10.1016/j.nuclphysb.2015.10.018}{Nucl.Phys.B {\bfseries
  902} (2016) 246} [\href{https://arxiv.org/abs/1509.00008}{{\tt
  arXiv:1509.00008}}].

\bibitem{Behan:2017emf}
C.~Behan, L.~Rastelli, S.~Rychkov and B.~Zan, \emph{A scaling theory for the
  long-range to short-range crossover and an infrared duality},
  \href{https://doi.org/10.1088/1751-8121/aa8099}{J.Phys.A {\bfseries 50}
  (2017) 354002} [\href{https://arxiv.org/abs/1703.05325}{{\tt
  arXiv:1703.05325}}].

\bibitem{Fisher:1978pf}
M.~Fisher, \emph{Yang-lee edge singularity and phi**3 field theory},
  \href{https://doi.org/10.1103/PhysRevLett.40.1610}{Phys.Rev.Lett. {\bfseries
  40} (1978) 1610}.

\bibitem{Cardy:1985yy}
J.~L. Cardy, \emph{Conformal invariance and the yang-lee edge singularity in
  two-dimensions},
  \href{https://doi.org/10.1103/PhysRevLett.54.1354}{Phys.Rev.Lett. {\bfseries
  54} (1985) 1354}.

\bibitem{Brown:1979pq}
L.~S. Brown, \emph{Dimensional regularization of composite operators in scalar
  field theory}, \href{https://doi.org/10.1016/0003-4916(80)90377-2}{Annals
  Phys. {\bfseries 126} (1980) 135}.

\bibitem{Caffarelli}
L.~Caffarelli and L.~Silvestre, \emph{An extension problem related to the
  fractional laplacian},
  \href{https://doi.org/10.1080/03605300600987306}{Communications in Partial
  Differential Equations {\bfseries 32} (2007) 1245}
  [\href{https://arxiv.org/abs/math/0608640}{{\tt arXiv:math/0608640}}].

\bibitem{Osborn:1993cr}
H.~Osborn and A.~Petkou, \emph{Implications of conformal invariance in field
  theories for general dimensions},
  \href{https://doi.org/10.1006/aphy.1994.1045}{Annals Phys. {\bfseries 231}
  (1994) 311} [\href{https://arxiv.org/abs/hep-th/9307010}{{\tt
  arXiv:hep-th/9307010}}].

\bibitem{Bzowski:2015pba}
A.~Bzowski, P.~McFadden and K.~Skenderis, \emph{Scalar 3-point functions in
  {CFT}: renormalisation, beta functions and anomalies},
  \href{https://doi.org/10.1007/JHEP03(2016)066}{JHEP {\bfseries 03} (2016)
  066} [\href{https://arxiv.org/abs/1510.08442}{{\tt arXiv:1510.08442}}].

\bibitem{Todorov:1985xs}
I.~Todorov, \emph{Local field representations of the conformal group and their
  applications},  in \emph{{Streit, L. (Ed.), Mathematics + Physics. Lectures
  On Recent Results, Vol. 1}}, pp.~195--338, 1985.

\bibitem{Seiberg:1988pf}
N.~Seiberg, \emph{Observations on the moduli space of superconformal field
  theories}, \href{https://doi.org/10.1016/0550-3213(88)90183-6}{Nucl. Phys.
  {\bfseries B303} (1988) 286}.

\bibitem{Nakayama:2019mpz}
Y.~Nakayama, \emph{Conformal contact terms and semi-local terms},
  \href{https://arxiv.org/abs/1906.07914}{{\tt arXiv:1906.07914}}.

\bibitem{Gomis:2015yaa}
J.~Gomis, P.-S. Hsin, Z.~Komargodski, A.~Schwimmer, N.~Seiberg and S.~Theisen,
  \emph{Anomalies, conformal manifolds, and spheres},
  \href{https://doi.org/10.1007/JHEP03(2016)022}{JHEP {\bfseries 03} (2016)
  022} [\href{https://arxiv.org/abs/1509.08511}{{\tt arXiv:1509.08511}}].

\bibitem{Gurdogan:2015csr}
O.~Gurdogan and V.~Kazakov, \emph{New integrable 4d quantum field theories from
  strongly deformed planar {$\mathcal N = 4$} supersymmetric {Yang-Mills}
  theory}, \href{https://doi.org/10.1103/PhysRevLett.117.201602}{Phys.Rev.Lett.
  {\bfseries 117} (2016) 201602} [\href{https://arxiv.org/abs/1512.06704}{{\tt
  arXiv:1512.06704}}].

\bibitem{Kazakov:2018qbr}
V.~Kazakov and E.~Olivucci, \emph{Biscalar integrable conformal field theories
  in any dimension},
  \href{https://doi.org/10.1103/PhysRevLett.121.131601}{Phys.Rev.Lett.
  {\bfseries 121} (2018) 131601} [\href{https://arxiv.org/abs/1801.09844}{{\tt
  arXiv:1801.09844}}].

\bibitem{Grabner:2017pgm}
D.~Grabner, N.~Gromov, V.~Kazakov and G.~Korchemsky, \emph{Strongly
  $\gamma$-deformed $\mathcal{N}=4$ supersymmetric yang-mills theory as an
  integrable conformal field theory},
  \href{https://doi.org/10.1103/PhysRevLett.120.111601}{Phys.Rev.Lett.
  {\bfseries 120} (2018) 111601} [\href{https://arxiv.org/abs/1711.04786}{{\tt
  arXiv:1711.04786}}].

\bibitem{Gromov:2017cja}
N.~Gromov, V.~Kazakov, G.~Korchemsky, S.~Negro and G.~Sizov,
  \emph{Integrability of conformal fishnet theory},
  \href{https://doi.org/10.1007/JHEP01(2018)095}{JHEP {\bfseries 01} (2018)
  095} [\href{https://arxiv.org/abs/1706.04167}{{\tt arXiv:1706.04167}}].

\bibitem{Diaz:2018eik}
P.~Diaz and J.~Rosabal, \emph{Spontaneous symmetry breaking in tensor
  theories}, \href{https://doi.org/10.1007/JHEP01(2019)094}{JHEP {\bfseries 01}
  (2019) 094} [\href{https://arxiv.org/abs/1809.10153}{{\tt
  arXiv:1809.10153}}].

\bibitem{Benedetti:2019sop}
D.~Benedetti and I.~Costa, \emph{{$SO(3)$-invariant phase of the $O(N)^3$
  tensor model}},  \href{https://arxiv.org/abs/1912.07311}{{\tt
  arXiv:1912.07311}}.

\bibitem{zinnjustin}
J.~Zinn-Justin, \emph{Quantum field theory and critical phenomena}. Clarendon
  Press, 1996.

\bibitem{Amit:1984ms}
D.~Amit,
\emph{Field Theory, the Renormalization Group, and Critical Phenomena}. World Scientific, 1984.


\bibitem{salmhofer:book}
M.~Salmhofer, \emph{Renormalization: An Introduction}, Theoretical and
  Mathematical Physics. Springer-Verlag Berlin Heidelberg, 1999,
  \href{https://doi.org/10.1007/978-3-662-03873-4}{10.1007/978-3-662-03873-4}.

\bibitem{Aharony:1999ti}
O.~Aharony, S.~S. Gubser, J.~M. Maldacena, H.~Ooguri and Y.~Oz, \emph{Large {N}
  field theories, string theory and gravity},
  \href{https://doi.org/10.1016/S0370-1573(99)00083-6}{Phys.Rept. {\bfseries
  323} (2000) 183} [\href{https://arxiv.org/abs/hep-th/9905111}{{\tt
  arXiv:hep-th/9905111}}].

\bibitem{Freedman:1991tk}
D.~Z. Freedman, K.~Johnson and J.~I. Latorre, \emph{Differential regularization
  and renormalization: A new method of calculation in quantum field theory},
  \href{https://doi.org/10.1016/0550-3213(92)90240-C}{Nucl.Phys.B {\bfseries
  371} (1992) 353}.

\bibitem{Gelfand}
I.~M. Gelfand and G.~E. Shilov, \emph{{Generalized functions, Vol.1}}, AMS
  Chelsea Publishing. Academic Press, New York, NY, 1964.

\bibitem{Davydychev:1995mq}
A.~I. Davydychev and J.~Tausk, \emph{A magic connection between massive and
  massless diagrams},
  \href{https://doi.org/10.1103/PhysRevD.53.7381}{Phys.Rev.D {\bfseries 53}
  (1996) 7381} [\href{https://arxiv.org/abs/hep-ph/9504431}{{\tt
  arXiv:hep-ph/9504431}}].

\bibitem{Benedetti:2019rja}
D.~Benedetti, N.~Delporte, S.~Harribey and R.~Sinha, \emph{Sextic tensor field
  theories in rank $3$ and $5$},  \href{https://arxiv.org/abs/1912.06641}{{\tt
  arXiv:1912.06641}}.

\bibitem{Bonzom:2016dwy}
V.~Bonzom, \emph{Large {$N$} limits in tensor models: Towards more universality
  classes of colored triangulations in dimension {$d\geq 2$}},
  \href{https://doi.org/10.3842/SIGMA.2016.073}{SIGMA {\bfseries 12} (2016)
  073} [\href{https://arxiv.org/abs/1603.03570}{{\tt arXiv:1603.03570}}].

\bibitem{Benedetti:2015ara}
D.~Benedetti and R.~Gurau, \emph{{Symmetry breaking in tensor models}},
  \href{https://doi.org/10.1103/PhysRevD.92.104041}{Phys. Rev. {\bfseries D92}
  (2015) 104041} [\href{https://arxiv.org/abs/1506.08542}{{\tt
  arXiv:1506.08542}}].

\end{thebibliography}
\end{document}